\documentclass{aastex}
\usepackage{graphicx,epsfig,emulateapj5}

\newcommand{\etal}{{et al.~}}

\newcommand{\bq}{\begin{equation}}
\newcommand{\eq}{\end{equation}}

\def\gtsim{\lower.5ex\hbox{$\buSildrel > \over\sim$}}
\def\ltsim{\lower.5ex\hbox{$\buildrel < \over\sim$}}

\def\arcmin{^\prime}

\def\farcs{\hbox{$.\!\!^{\prime\prime}$}}

\def\apjl{ApJL}
\def\apj{ApJ}
\def\apjs{ApJS}
\def\mnras{MNRAS}

\def\aj{AJ}
\def\aap{A\&A}

\def\nat{Nature}
\def\pasp{PASP}
\slugcomment{Accepted by the Astrophysical Journal} 
\shorttitle{Galaxy Interactions and Star Formation over 7 Gyr}
\shortauthors{Jogee, Miller, Penner, and the GEMS collaboration}
\received{xxxx}

\begin{document}
\title
{History of Galaxy Interactions and their Impact on Star Formation over the Last 7 Gyr from GEMS}
\author{
Shardha Jogee \altaffilmark{1},
Sarah H. Miller \altaffilmark{1,14},
Kyle Penner \altaffilmark{1,16},
Rosalind E. Skelton \altaffilmark{2},
Christopher J. Conselice \altaffilmark{3},
Rachel S.  Somerville \altaffilmark{2},
Eric F.Bell  \altaffilmark{2},
Xian Zhong Zheng   \altaffilmark{4},
Hans-Walter Rix \altaffilmark{2},
Aday R. Robaina \altaffilmark{2},
Fabio D. Barazza \altaffilmark{5},
Marco Barden \altaffilmark{6},
Andrea Borch  \altaffilmark{2},
Steven V.W. Beckwith \altaffilmark{7},
John A. R. Caldwell \altaffilmark{8},
Chien Y. Peng \altaffilmark{9},
Catherine Heymans \altaffilmark{10},
Daniel H. McIntosh \altaffilmark{11},
Boris H{\"a}u{\ss}ler \altaffilmark{2},
Knud Jahnke \altaffilmark{2},
Klaus Meisenheimer  \altaffilmark{2},
Sebastian F. Sanchez \altaffilmark{12},
Lutz Wisotzki \altaffilmark{13},
Christian Wolf \altaffilmark{14},
Casey Papovich \altaffilmark{15}
} 
\email{sj@astro.as.utexas.edu}
\altaffiltext{1}
{Department of Astronomy, University of Texas at Austin, 1 University Station C1400, Austin, TX 78712-0259}
\altaffiltext{2}
{Max-Planck-Institut f\"{u}r Astronomie, K\"{o}nigstuhl 17, D-69117,Heidelberg, Germany}
\altaffiltext{3}
{School of Physics and Astronomy, The University of Nottingham, University Park, Nottingham NG7 2RD, UK}
\altaffiltext{4}
{Purple Mountain Observatory, West Beijing Road 2, Nanjing, 210008, China}
\altaffiltext{5}
{Laboratoire d'Astrophysique, \'Ecole Polytechnique F\'ed\'erale de Lausanne (EPFL), Observatoire, 1290 Sauverny, Switzerland}
\altaffiltext{6}
{Institute for Astro- and Particle Physics, University of Innsbruck, 
Technikerstr. 25/8, A-6020 Innsbruck,   Austria}
\altaffiltext{7}
{Department of Physics and Astronomy, Johns Hopkins University,  Charles and 4th Street, Baltimore, MD 21218}
\altaffiltext{8}
{University of Texas, McDonald Observatory, Fort Davis TX, 79734 USA}
\altaffiltext{9}
{NRC Herzberg Institute of Astrophysics, Victoria, Canada}
\altaffiltext{10}
{SUPA, Institute for Astronomy, University of Edinburgh, Royal Observatory, Edinburgh EH9 3HJ}
\altaffiltext{11}
{Department of Astronomy, University of Massachusetts, 710 North Pleasant Street, Amherst, MA 01003, USA}
\altaffiltext{12}
{Centro Astronómico Hispano Alemán, Calar Alto, CSIC-MPG, C/Jesús Durbán Remón 2-2, E-04004 Almeria, Spain}
\altaffiltext{13}
{Astrophysikalisches Institut Potsdam, An der Sternwarte 16, D-14482 Potsdam, Germany}
\altaffiltext{14}
{Astrophysics, University of Oxford, Keble Road, Oxford OX1 3RH, U.K.}
\altaffiltext{15}
{George P. and Cynthia W. Mitchell Institute for Fundamental Physics and Astronomy, and
Department of Physics, Texas A\&M University, College Station, TX,77843-4242}
\altaffiltext{16}
{Department of Astronomy, University of Arizona, Steward Observatory, 933 N. Cherry Avenue, Tucson, AZ 85721}

\begin{abstract}
We perform a comprehensive estimate of the frequency of galaxy mergers and their impact on star formation over  $z \sim$~0.24--0.80 (lookback  time $T_{\rm b}\sim$~3--7 Gyr)  using  $\sim$~3600 ($M \ge$~$1 \times  10^{9}$ $M_{\odot}$) galaxies with GEMS $HST$, COMBO-17, and Spitzer data. Our  results are: (1)~ Among $\sim$~790 high  mass ($M \ge$~$2.5 \times  10^{10}$ $M_{\odot}$) galaxies, the visually-based merger fraction over $z\sim$~0.24--0.80, ranges from  9\% $\pm$~5\%  to 8\% $\pm$~2\%. Lower limits on the major merger and minor merger fraction over this interval range from  1.1\% to 3.5\% , and 3.6\%  to 7.5\%, respectively. This is the first, albeit approximate, empirical estimate of the frequency of  minor mergers over the last 7 Gyr. Assuming a  visibility timescale of $\sim$~0.5 Gyr, it follows that over $T_{\rm b}\sim$~3--7 Gyr, $\sim$~68\% of  high  mass systems have undergone a merger  of mass ratio $>$~1/10, with $\sim$~16\%, 45\%, and 7\%  of these corresponding respectively to major, minor, and ambiguous `major or minor' mergers. The average merger rate is $\sim$~ a few $\times 10^{-4}$ galaxies Gyr$^{-1}$  Mpc$^{-3}$. Among $\sim$~2840 blue cloud galaxies of mass $M \ge$~$1.0 \times 10^{9}$ $M_{\odot}$, similar results hold. (2)~We compare the empirical merger fraction and merger rate for high mass galaxies to  three $\Lambda$CDM-based models:  halo occupation distribution  models, semi-analytic models, and hydrodynamic SPH simulations. We find  qualitative  agreement  between observations and models such that the (major+minor) merger fraction or rate from different models bracket the observations, and show a factor of five dispersion. Near-future improvements can now start to rule out certain merger scenarios. (3)~Among $\sim$~3698 $M \ge$~$1.0 \times 10^{9}$ $M_{\odot}$ galaxies, we find that the mean SFR of visibly merging systems is only modestly enhanced compared to non-interacting galaxies over $z \sim$~0.24--0.80. Visibly merging systems only account for a small fraction ($<$ 30\%) of the cosmic SFR density over $T_{\rm b}\sim$~3--7 Gyr. This  complements the results of Wolf et al. (2005) over a shorter time interval of $T_{\rm b}\sim$~6.2--6.8 Gyr, and suggests that the behavior of the cosmic SFR density over the last 7 Gyr is predominantly shaped by non-interacting galaxies.
\end {abstract}


\section{Introduction}\label{scintro}

Hierarchical  $\Lambda$ cold dark matter ($\Lambda$CDM)  models provide a successful 
paradigm for the growth of dark matter on large scales.
The evolution of galaxies  within $\Lambda$CDM  cosmogonies
depends on the baryonic merger history,  the star formation (SF) history,
the nature and level of feedback from supernovae and AGN,  the 
redistribution of angular  momentum  via bars or mergers, 
and other aspects of the baryonic physics. 
Empirical constraints on the fate  of the baryonic component, in particular
their  merger and  SF history, are key  for  developing a coherent 
picture of  galaxy evolution and testing  galaxy evolution models 
 (e.g., Kauffmann et al. 1993;  Somerville \& Primack 1999; Navarro \& Steinmetz 2000; 
Murali \etal  2002; Governato et al. 2004; Springel et al. 2005a,b; Maller \etal  2006).
Such constraints  can  help to resolve 
several major areas of discord between observations and 
$\Lambda$CDM models of galaxy evolution, such   as the 
angular momentum crisis,  the problem of bulgeless and low bulge-to-total ($B/T$) ratio 
galaxies  (Navarro \& Benz 1991;   D'Onghia \& Burkert \&  2004; 
Kautsch \etal 2006; Barazza, Jogee, \&  Marinova 2008; Weinzirl \etal  2009), 
and the substructure  problem.


The merger history of galaxies impacts the mass assembly, star 
formation history, AGN activity and structural evolution of galaxies.  
Yet, the merger rate has  proved hard to robustly measure for a 
variety of reasons.  Initially, small samples hindered the first efforts 
to measure merger rates (Patton et al. 2000; Le Fevre et al. 2000; 
Conselice et al. 2003).  Later studies drew from larger samples, and have 
used a variety of methods to characterize the interaction history of 
galaxies at $z<1$.  
Studies based  on close pairs report a major merger fraction 
of $\sim$~2\% to 10\% over z$\sim$~0.2 to 1.2  (Lin 2004;  Kartaltepe 
et al. 2007;  de Ravel et al. 2008) and  
$\sim$~5\% for massive galaxies at  z$\sim$~0.4 to 0.8 (Bell  et al. 2006).  

Among these studies, the estimated fraction of galaxies, which have 
experienced a major merger since  $z\sim$~1.2  
differ.  Lin et  al. (2004) find that $\sim$~9\% of  
the massive early type galaxies experienced a major merger  since 
$z\sim$~1.2, while Bell et al. (2006) report that $\sim$~50\% of all 
galaxies with present-day stellar masses above $5 \times  10^{10}$ $M_{\odot}$ 
have undergone a major merger since $z\sim$~0.8. This difference of a 
factor of five between the two studies is addressed by Bell et al. (2006; 
see their footnote 9) and is traced primarily to differences in the following 
factors: the dataset used, the redshift integration method, the way in which 
the authors handle the relative number density of the parent population 
from which pairs are drawn and that of the remnant population, the assumed 
visibility timescale, and the fraction of pairs estimated to be true 
gravitationally bound pairs.

Studies based on Gini-M20 coefficients report a fairly constant fraction 
($\sim$ 7 $\pm$ 0.2 \%) of  disturbed galaxies over z$\sim$~0.2 to 1.2  
among bright galaxies (Lotz et al. 2008) in the AEGIS survey. 
Similar trends are found from early results based on visual classification 
and asymmetry parameters of high mass galaxies (e.g., Jogee et al. 2008). 
The study by Cassata et al. 
(2005) based on both pairs and asymmetries report a mild increase in the 
merger rate  with redshift up to  z$\sim$~1, with the caveat of a small 
sample size.
The merger rate/fraction at $z>1$ remains highly uncertain, owing to 
relatively modest volumes and bandpass shifting effects, but there is 
general trend towards  higher merger fractions at higher redshifts (e.g., 
Conselice et al. 2003; Cassata et al. 2005).

Studies to date have  brought important insights but face several limitations.
In the case of studies based on  close (separation $\sim$~5 to 
30 kpc) pairs, the translation of the pair frequency into 
a merger rate is somewhat uncertain due to several factors. The uncertainties in the 
spectrophotometric redshifts  for some of the galaxies in pairs can cause us
to overestimate or  underestimate the true close pair fraction, with the latter
effect being more likely. Corrections for this effect are uncertain and depend on 
the shape of the spectrophotometric redshift errors (e.g., see Bell et al. 2006 
for discussion).  Secondly, even pairs with members at the same redshift may not 
be gravitationally bound, and  may therefore not evolve into a  merger 
in the future: this effect causes the close pair fraction to be upper limits for 
the merger fraction. Thirdly, gravitationally bound pairs captured by this method sample 
different phases of an interaction depending on the separation, and any merger rate 
inferred depends on the separation, orbital eccentricity, and orbital geometry.

In the case of studies, which use automated parameters, such as CAS asymmetry $A$ and 
clumpiness parameters  (Conselice et al 2000; Conselice 2003)  or Gini-M${\rm 20}$ 
coefficients (e.g., Lotz et al. 2004)  to identify merging galaxies 
can fail to pick  stages of both major and minor  mergers where  
distortions do not dominate the total light  contribution  ($\S$~\ref{sccas1}). 
Comparison with simulations suggest that the CAS criterion ($A>$~0.35 and $A>S$; Conselice 2003) 
capture major mergers about 1/3 of the time, while the eye
is sensitive to major  merger features over twice as long (e.g., Conselice 2006; 
$\S$~\ref{sccas1}). To complicate matters, automated asymmetry parameters can also capture  
non-interacting galaxies  hosting small-scale asymmetries that are produced by stochastic
star formation ($\S$~\ref{scasts}). Thus, it is important to use several methods to 
assess the robustness of results and identify the  systematics.

In this paper, we present a complementary study of the frequency of mergers and 
their impact  on the SF activity of galaxies over  $z \sim$~0.24--0.80  
(lookback times $T_{\rm b}$ of 3--7 Gyr\footnote{
We assume a flat cosmology with $\Omega_{\rm m, 0}$ = 1-
$\Omega_{\lambda}$ = 0.3 and H$_{0}$~=~70 km s$^{-1}$ Mpc$^{-1}$
throughout.})  using $HST$ ACS, COMBO-17, and Spitzer 24~$\mu$m 
data of $\sim$~3600 galaxies in the GEMS survey.  The outline of the paper is 
given below and describes how this study complements existing work:

\begin{enumerate}
\vspace{-3mm} 
\item
We use a large sample of $\sim$~3600 ($M \ge$~$1 \times  10^{9}$ $M_{\odot}$)  galaxies 
to get robust number statistics for the merger fraction among 
$\sim$~790 high mass  ($M \ge$~$2.5 \times 10^{10}$ $M_{\odot}$) galaxies and 
$\sim$~2840 $M \ge$~$1 \times  10^{9}$ $M_{\odot}$ blue cloud galaxies 
($\S$~\ref{scdatas};  Table~\ref{tvclas1}; Table~\ref{tvclas2}).

\item
Two independent methods  are used to identify merging galaxies:  a 
physically-driven visual classification system complemented with spectrophotometric 
redshifts and stellar masses  ($\S$~\ref{scvis1} to $\S$~\ref{scvis2}), 
as well as  automated  CAS asymmetry and clumpiness parameters ($\S$~\ref{sccas1}). 
This allows one of the most extensive comparisons to date between CAS-based and visual 
classification results ($\S$~\ref{scasts}).

\vspace{-3mm} 
\item 
We design the visual classification system in a way that allows merger 
fractions and rates from observations and theoretical models 
to be readily compared. We classify as mergers those systems that show 
evidence of having  experienced  a merger of mass ratio $ >$~1/10 
within the last visibility timescale.
Throughout this paper,  we use the standard definition whereby  major mergers 
are defined to have stellar mass ratio (1/4 $< M1/M2 \le$ 1/1),  while  minor 
mergers  have (1/10 $< M1/M2 \le$ 1/4). 
We set lower limits on the major and minor merger fraction ($\S$~\ref{scvctes}). 
To our knowledge, this is the first, albeit approximate, empirical 
estimate of the frequency of  minor mergers over the last 7 Gyr.
While many earlier studies focused on major mergers, it is important to constrain 
minor mergers as well, since they dominate the merger rates in $\Lambda$CDM models,
and play an important role in building the bulges of massive galaxies (e.g., Weinzirl
et al. 2009).

\vspace{-3mm} 
\item 
We compare the empirical  merger fraction and  rate
to  a suite of  $\Lambda$CDM-based simulations of  galaxy evolution, including 
halo occupation distribution models, semi-analytic models, and hydrodynamic SPH simulations 
($\S$~\ref{scmodel}). To our knowledge, these extensive comparisons have not been attempted 
to date, and  are long overdue.

\vspace{-3mm} 
\item
The idea that galaxy interactions generally enhance the star formation rate (SFR) 
of galaxies  is well established 
from  observations (e.g., Larson \& Tinsley 1978; Joseph \& Wright 1985; 
Kennicutt et al. 1987; Barton et al 2003)  and simulations 
(e.g., Negroponte \& White 1983; Hernquist 1989; Barnes \& Hernquist 1991, 1996; 
Mihos \& Hernquist 1994, 1996;  Springel, Di Matteo \& Hernquist 2005b). 
However, simulations cannot uniquely predict the  factor by which  galaxy mergers enhance
the SF activity of galaxies over the last 7 Gyr, since both the SFR and properties 
of the remnants in simulations are highly sensitive to the  stellar feedback model,  
the bulge-to-disk ($B/D$) ratio, the gas mass fractions, and orbital geometry   
(e.g., Cox et al 2006; di Matteo et al. 2007).   
This motivates us  in $\S$~\ref{scsfr1} to empirically investigate  the  impact of 
interactions on the average UV-based and  UV+IR-based  SFR   of  high mass  
($M \ge$~$2.5 \times 10^{10}$ $M_{\odot}$) and intermediate-to-high mass 
($M \ge$~$1 \times  10^{9}$ $M_{\odot}$) galaxies over $z\sim$~0.24--0.80. 

\vspace{-3mm} 
\item
The SF properties of merging and non-interacting galaxies since $z< 1$ is of great 
astrophysical interest, given that the  
cosmic SFR density is claimed to decline by a factor of 4 to 10 since $z\sim$~1 (e.g., 
Lilly et al. 1996; Ellis et al 1996;  Flores \etal 1999; Haarsma \etal 2000; Hopkins 
2004; P{\'e}rez-Gonz{\'a}lez  et al.\ 2005;  Le Floc'h et  al.\ 2005).
In  $\S$~\ref{scsfr2}, we set quantitative limits on the  contribution of 
visibly merging systems to the UV-based and  UV+IR-based SFR density  
over $z\sim$~0.24--0.80. 
Our study covers a 4 Gyr interval  ($T_{\rm b}\sim$~3--7 Gyr or $z\sim$~0.24--0.80)  
and   extends the earlier studies carried out over a smaller 0.6 Gyr interval 
($T_{\rm b}\sim$~6.2--6.8 Gyr  or $z \sim$~0.65--0.75) by Wolf \etal (2005)  and
Bell \etal (2005) on the UV and IR luminosity density, respectively.
Our study also  complements  IR-based  studies by  Lotz et al. (2008), 
Hammer et al. (2005;  l95 galaxies at $z>0.4$ )  and  Melbourne et al. 
(2005; $\sim$~800 galaxies) in terms of better number statistics and the use of 
both UV-based and IR-based SFR indicators.


\end{enumerate}

\section{Dataset and Sample Selection}\label{scdatas}

This study uses data from the  Galaxy Evolution from Morphology and SEDS 
(GEMS; Rix \etal 2004)  survey,  which provides high resolution 
{\it Hubble Space Telescope (HST)} Advanced Camera for Surveys (ACS) images 
in  the F606W and F850LP filters  over  an 800 arcmin$^2$  ($\sim 28\arcmin \times 
28\arcmin$) field centered on the Chandra Deep Field-South (CDF-S).
Accurate spectrophotometric redshifts  [$\delta_{\rm z}$/(1 + $z$)~$\sim$~0.02 
down to  $R_{\rm Vega}$ = 24] and spectral energy distributions,  based  on 
5 broad bands ($UBVRI$) and 12 medium band filters, are available from the 
COMBO-17 project (Wolf \etal 2004).
The  ACS data reach a  limiting 5 $\sigma$ depth for point sources 
of  28.3  and 27.1 AB mag in F606W and F850LP, respectively (Rix \etal 2004; Caldwell et al. 2008).
The effective point spread function (PSF) in a single F606W image is 
$\sim$$0\farcs07$, corresponding to 260 pc at  $z \sim$~0.24 and 520 pc at 
$z \sim$~0.80. The PSF of combined drizzle images is $\sim$$0\farcs1$.
In addition to $HST$ ACS imaging, the GEMS field has panchromatic coverage 
which includes  $Spitzer$ (Rieke \etal 2004; Papovich \etal 2004) and  
$Chandra$  data (Alexander \etal 2003; Lehmer \etal 2005).

We use stellar masses  from  Borch \etal (2006).
We refer the reader to the latter publication for a detailed description and 
provide a  summary here. Using the 17-passband photometry from COMBO-17,
objects were classified as main sequence, stars,
white dwarfs, galaxies, and quasars using color indices and their photometric 
redshifts were estimated using simple dust-reddened single-burst SED  
templates  (Wolf \etal 2004).  
For galaxies and  quasars, the joint probability of a given redshift and a
given rest-frame SED is derived and this procedure  provides a minimum error
variance estimation of both the redshift and the SED template.
Once the redshfit has been estimated, the SEDs in 17 bands  were fitted with 
a new set of template SEDs with more plausible SF histories in order
to derive a stellar $M/L$ (Borch \etal 2006).  
The library of  SEDs is built using the PEGASE stellar population synthesis model 
(see Fioc \& Rocca-Volmerange 1997 for an earlier version of the model) and 
the  underlying SF histories are parameterized by the  three-component model, 
with a  Kroupa (Kroupa \etal 1993) initial mass function (IMF) adopted in the 
mass regime  0.1--120~$M_{\odot}$. 
We note that the  stellar masses are consistent within 10\% with the masses 
that would be estimated using a different  Kroupa (2001) or Chabrier (2003) IMF 
\footnote{We adopt a Chabrier (2003) IMF when exploring the contribution of 
mergers to the cosmic SFR  ($\S$~\ref{scsfr1} and $\S$~\ref{scsfr2}).}.
The reddest templates have smoothly-varying exponentially-declining star
formation episodes, intermediate templates have a contribution from a 
low-level constant level of star formation, while the bluer templates 
have a recent burst of star formation superimposed. 
Random stellar mass errors are $< 0.3$ dex on a
galaxy-by-galaxy basis, and systematic errors in the stellar masses were
argued to be at the 0.1 dex level (Borch \etal 2006). Bell \& de Jong (2001)
argued that galaxies with large bursts of recent star formation could 
drive down stellar M/L values by up to 0.5 dex at a given color, but 
we note that the Borch \etal (2006) templates do include bursts explicitly, 
thus compensating for the worst of the uncertainties introduced by bursting 
star formation histories.


We present the results based on the  visual classification and CAS 
(Conselice 2003)  parameters ($\S$~\ref{scmetho}) of GEMS  F606W, 
rather  than F850LP images, for the following reasons. 
The F606W images are $\sim$~1.2 AB magnitude deeper than the GEMS F850LP images 
and  allow more reliable characterization of morphological features in the presence of  
cosmological surface brightness dimming at the rate of (1+$z$)$^{-4}$ (e.g., Barden \etal 
2007). 
Furthermore, the low signal to noise in the F850LP  images leads to  large error bars
on the asymmetry $A$ and clumpiness $S$ parameters generated by the CAS code, 
effectively making it  impractical to use these values  in CAS merger diagnostics 
($\S$~\ref{sccas1}).
When using the F606W images,  we only include results over the redshift range
$z \sim$~0.24--0.80 in order to ensure that the rest-frame  wavelength $\lambda_{\rm rest}$
stays primarily in the optical band and does not shift well below the 4000 \AA \ break. 
In the fourth redshift bin  ($z\sim$~0.6 to 0.8)  $\lambda_{\rm rest}$ shifts to the  
violet/near-UV  (3700~\AA ~to 3290~\AA), but as we show in $\S$~\ref{scvctes}, 
this does not significantly impact the results.
We discard the last redshift bin  at $z>0.8$ where $\lambda_{\rm rest}$ shifts into 
the far-UV. 
These steps lead to  a sample  of 4766 galaxies selected down to $R_{\rm Vega} \le$ 24, 
over  $z\sim$~0.24--0.80 ($T_{\rm back}$~$\sim$~3--7 Gyr).

In this paper, we present  results for two samples of astrophysical interest, 
which are derived by applying stellar mass cuts to the above sample of $\sim$~4766
galaxies
The first sample (henceforth sample S1) focuses on  galaxies with 
high stellar mass  ($M \ge$~$2.5 \times  10^{10}$ $M_{\odot}$; Table~\ref{tvclas1}).
The sample size is originally 804 galaxies, out of which 798 (99.2\%) could be 
visually classified. For this stellar mass range,  the red sequence and blue cloud 
galaxies are both complete out to the highest redshift bin $z\sim$~0.62--0.80 
for our sample, and we have theoretical predictions for comparison 
(see $\S$ \ref{scmodel}) from  semi-analytical models 
(e.g, Somerville \etal  2008; Hopkins \etal 2007; Khochfar \& Burkert(2005); 
Bower \etal 2006),  $N$-body (D'Onghia \etal 2008), and hydrodynamical SPH  
simulations (e.g., Maller \etal 2006). Note that the survey has few galaxies 
above  $10^{11}$ $M_{\odot}$  (Fig.~\ref{fuvmas}),  and hence the high mass 
($M \ge$~$2.5 \times  10^{10}$ $M_{\odot}$) sample  primarily involves 
galaxies in the range $2.5 \times  10^{10}$ to  $10^{11}$ $M_{\odot}$.

We also present selected results for the sample S2 of $\sim$~3698 
galaxies with  $M \ge$~$1 \times  10^{9}$ $M_{\odot}$ and visual classes. 
Although this sample includes the  $\sim$~790 
high  mass galaxies in sample S1, it is dominated by systems of intermediate mass 
($ 10^{9} \le  M/M_{\odot} < 2.5 \times  10^{10}$). 
For the mass range $M \ge$~$1 \times  10^{9}$ $M_{\odot}$, 
the blue cloud is complete in our sample out to  $z\sim$~0.80, while the  red 
sequence  is incomplete in the higher redshift bins. Where appropriate, we 
will therefore present results for the blue cloud sample only 
(e.g., lower part of Table~\ref{tvclas2}).
The rest-frame $U-V$ color is plotted versus the stellar mass for the 
sample S2 in Figure \ref{fuvmas}.
The redshift interval is divided into four 1 Gyr bins.
The diagonal line  marks the separation of the red sequence 
and the blue cloud galaxies  (BCG) at 
the average redshift $z_{\rm ave}$ of the bin. We use the definition in Borch 
\etal (2006) and Bell \etal (2004) for CDF-S: 

\begin{equation}(U-V)_{\rm rest} > \rm 0.227 \ \rm log( M/M_{\odot}) - 1.26  - \rm 0.352 z
\end{equation}

\noindent 
The vertical lines on Figure \ref{fuvmas} marks the mass completeness limit 
(Borch \etal 2006) for the red sequence galaxies.  The blue cloud galaxies 
are complete well below this mass (Borch \etal 2006).

\section{Methodology: Identifying Mergers and Non-Interacting Galaxies}\label{scmetho}

\subsection{Overview of the Methodology}\label{sover}

Galaxy mergers with a  mass ratio $M1/M2 >$~1/10 can have a 
significant impact on galaxy evolution. 
According  to simulations, major mergers (defined as those with mass 
ratio 1/4 $< M1/M2 \le$ 1/1)   typically destroy stellar disks, 
transforming them  via violent relaxation, into systems with a steep
or r$^{1/4}$ de Vaucouleurs-type  stellar profile, such as ellipticals (e.g., 
Negroponte \& White 1983; Barnes \& Hernquist 1991; Mihos \& Hernquist 1996; 
Struck 1997; Naab \& Burkert 2001; but see Robertson et al. 2004). 
Simulations further suggest that recent major mergers at $z<1$ are 
typically associated with arcs, shells, ripples, tidal tails,  large tidal debris, 
highly asymmetric light distributions, double nuclei inside a common body, galaxies 
linked via tidal bridges of light, and galaxies enclosed within the same distorted 
envelope of light.

Minor mergers (defined as those with 1/10 $< M1/M2 \le$ 1/4) of two
spirals  will not destroy the disk of 
the larger spiral (e.g.,  Hernquist \& Mihos 1995; 
Smith \etal 1997; Jogee \etal 1999).
Typically, the smaller companion sinks via  dynamical friction,  may  excite 
warps, bars, spirals,  and other non-axisymmetric  perturbations,  and leads 
to vertical  heating,  arcs, shells, ripples,  tidal tails, tidal debris,  warps, 
offset rings, highly asymmetric light distributions, etc  (e.g., 
Quinn et al. 1993;  Hernquist \& Mihos 1995;  Mihos et al. 1995;  Quinn, Hernquist, \& 
Fullagar 1993; Smith \etal 1994; Jogee 1999; Jogee \etal 1998, 1999; review by Jogee 2006 and references 
therein).

One goal of this paper is to identify  systems whose 
morphology and other properties suggest they  they have recently experienced a 
merger  of mass ratio $M1/M2>$~1/10. 
We employ  two methods: a physically-driven visual classification system 
($\S$~\ref{scvis1} to $\S$~\ref{scvis2}) complemented with spectrophotometric redshifts  and 
stellar  masses, and a method based 
on quantitative  asymmetry ($A$), and clumpiness ($S$) 
parameters ($\S$~\ref{sccas1}) derived using the  CAS code (Conselice 2003). 
While many studies use only automated methods or visual classification, we  choose 
to use both methods in order to better assess the systematics, and to test the 
robustness of our results.

\subsection{Visual Classification of Mergers}\label{scvis1}

The visual classification system we adopted for identifying mergers 
was aimed at setting a procedural framework that allows merger 
fractions and rates from observations and the theoretical models (outlined in 
$\S$~\ref{scmodel})  to be defined in similar ways and to be readily compared. 

Specifically  the theoretical models described in $\S$~\ref{scmodel} 
track systems which have a stellar mass $M_{\ast} \ge M_{\rm cut}$  
 and have experienced a merger of a certain mass ratio
$M1/M2$ within  the last visibility timescale $t_{\rm vis}$, between times 
($t_{\rm obs}$ - $t_{\rm vis}$) and $t_{\rm obs}$.
Here $t_{\rm obs}$ is the time corresponding to the observed redshift $z$; 
$t_{\rm vis}$ is the timescale over which  morphological distortions remain 
visible  after a merger and we adopt a nominal value of 0.5 Gyr (see 
$\S$~\ref{scmfrac});  $M_{\rm cut}$ is  the cutoff mass for the merger, 
which is $2.5 \times  10^{10}$ $M_{\odot}$ for the high  mass sample S1 
and $1 \times  10^{9}$ $M_{\odot}$ for  the intermediate mass sample S2 
($\S$~\ref{scdatas}). In the models, major mergers are associated with
a mass ratio of 1/4~$< M1/M2 \le$~1/1, minor mergers with 1/10~$< M1/M2 \le$~1/4, 
and major+minor mergers with $M1/M2 >$~1/10.

Analogous to the theoretical models, the goal of the visual classification 
system is to identify systems with $M_{\ast} \ge M_{\rm cut}$, which show 
evidence of having  experienced  a merger of mass ratio $ >$~1/10 
within the last visibility timescale  $t_{\rm vis}$.
In this paper, we refer to such systems either as interacting galaxies or 
as mergers.
These mergers are denoted as having visual type `Int' in the figures of this 
paper and examples are shown in  Figure~\ref{fpec1}.  In practice, they 
consist of three types of mergers, which we denote as Int-1, Int-2a, and Int-2b, 
and define in $\S$~\ref{sint1} and $\S$~\ref{sint2}. These three types of 
systems encompass  young to advanced mergers, and are identified/handled 
in different ways, as described below.

\subsubsection{Mergers of type Int-1}\label{sint1}

Mergers of type  Int-1 primarily represent advanced mergers of mass 
$M_{\ast} \ge M_{\rm cut}$, which appear as a single system in ACS images, and are 
likely associated  with  a merger of mass ratio $>$~1/10 that occurred betwen 
times ($t_{\rm obs}$ - $t_{\rm vis}$) and $t_{\rm obs}$. 
Systems of type Int-1 are identified empirically based on the following criteria:               
\begin{enumerate}
\vspace{-3mm} 
\item 
    They have $M_{\ast} \ge M_{\rm cut}$ and 
    show morphological distortions, which are similar to those seen
     in simulations of mergers of mass ratio $>$~1/10, and  remain visible 
     over the visibility timescale $t_{\rm vis}$.   The distortions  include arcs, 
     shells, ripples, tidal tails, large tidal debris, highly asymmetric
     light distributions, double nuclei inside a common body, etc. The 
     presence of such distortions is considered indicative of a past 
     merger that occurred between  times  ($t_{\rm obs}$ - $t_{\rm vis}$) 
     and $t_{\rm obs}$. 
     We also make an extra test  to verify that the distortions are caused
     by a past merger, rather than a present tidal interaction, by verifying
     that such systems do not have a distorted companion  of similar
     spectrophotometric redshift within 40 kpc.
\vspace{-3mm} 
\item 
     They appear as a single distorted system, rather than 2 individually
     recognizable galaxies, in the ACS images of PSF~ $\sim$$0\farcs1$
     (corresponding to 380 pc at $z\sim$~0.24  and 750 pc at  $z\sim$~0.80). 
     This suggests the system is an advanced merger where the 2 progenitor
     galaxies have had time to  coalesce into a single ACS system by time 
     $t_{\rm obs}$. 

\end{enumerate}

Such advanced mergers have a single redshift (Wolf et al. 2004), stellar 
mass $M_{\ast}$ (Borch et al. 2006), and UV-based star formation rate 
(Bell et al. 2005, 2007) determined  from the COMBO-17 ground-based data of 
resolution $\sim$~$1\farcs5$. The lack of resolved COMBO-17 data for the individual 
progenitor galaxies that led to the remnant is not a problem because we are only 
concerned in the analysis with the  stellar mass and SFR of the merger. 
Furthermore, the condition $M_{\ast} \ge M_{\rm cut}$ is expected to apply
to the merger, and not to the progenitors, in both model and observations.

The majority of the mergers we identify are of type  Int-1. Examples in 
Figure~\ref{fpec1}  are cases 2, 3, 4, 5, 6, 7, 9, 10  11, and 12. 

For  mergers of type  Int-1 with a single redshift and stellar mass,
the evidence for a merger of mass ratio  $>$~1/10  does not come from a 
measured stellar mass ratio $M1/M2$, but instead is inferred from the presence 
of the afore-mentioned morphological distortions, which are seen in simulations 
of mass ratio $M1/M2 >$~1/10.

Without individual stellar masses $M1$ and $M2$, the further separation of 
these mergers into major and minor is not possible in every case, since the 
morphological disturbances induced depend not only on the mass ratio of the 
progenitors, but also on  the orbital geometry (prograde or retrograde), 
the gas mass fraction, and structural parameters (e.g., Mihos \&  Hernquist 
1996; Struck 1997; Naab \& Burkert 2001; Mihos et al. 1995, di Matteo et al. 
2007). Instead, we can only separate such mergers into three groups: clear 
major mergers clear  minor mergers, and ambiguous cases of `major or minor' 
merger as follows:
\begin{enumerate}
\vspace{-3mm} 
\item 
   The class of clear major mergers includes systems that host fairly
   unique tell-tale morphological distortions characteristic of a
    major merger, such as 2 tails of equal lengths, 2 nuclei of similar 
    luminosity, (e.g. case 6 in Fig.~\ref{fpec1}) and a train-wreck morphology
   (e.g.,case 12 in Fig.~\ref{fpec1}). 

\vspace{-3mm} 
\item  
   The class of clear  minor merger includes the merger systems where
   the outer disk (identified based on the presence of disk features, 
   such  as spiral arms and bars) has clearly survived the recent past 
   merger.  Examples include case 2  of a warped disk in Fig.~\ref{fpec1}. 
   The reasoning behind classifying such remnants as a minor merger is that 
   the outer disk of a  galaxy survives  a minor merger, but not a major
   merger. While  major mergers of extremely gas-rich disks with low star formation 
   efficiency  can lead to a remnant with an extended stellar disk (Robertson et  
   al. 2004), such mergers are unlikely to be relevant for our study, which 
   focuses on fairly massive systems  (with stellar masses $\ge 1 \times  
   10^{9}$ $M_{\odot}$)  at $z<1$.
  
   One further criterion is applied. When identifying minor mergers through
   the presence of a distorted surviving outer disk, we take care to check 
   that the candidate is not in the early phases of a merger, which would 
   destroy the disk in the near future, on a timescale $t_{\rm vis}$. We do 
   this by avoiding those galaxies, which have both a distorted outer disk 
   {\it and} a close companion of similar  redshift (within the photometric 
   redshift accuracy) and of mass ratio $>$~1/4.
 
\vspace{-3mm} 
\item  
   The class of ambiguous 'major or minor' merger is assigned to systems
   hosting  distortions, which could be due to both a major and a minor
   merger.  Examples in Fig.~\ref{fpec1} are cases 3, 4, 5, 7, 8, 10, and
   11.
\end{enumerate}

\subsubsection{Mergers of type Int-2}\label{sint2}

Mergers of type  Int-2 primarily represent  young mergers, which appear as 
very close pairs of overlapping galaxies (VCPOG) in ACS images, have 
$M_{\ast} \ge M_{\rm cut}$, and are likely associated  with  a merger of 
mass ratio $>$~1/10 that occurred betwen times ($t_{\rm obs}$ - $t_{\rm vis}$) 
and $t_{\rm obs}$. 

Systems of type  Int-2 are identified empirically based on the following criterion:                  
ACS images show 2 individually recognizable galaxies whose bodies 
overlap to form a common continuous envelope of light, and whose 
centers have a small separation $d<$~20 kpc. One or both of the galaxies
often have morphological distortions. These properties  suggest that
the 2 progenitor galaxies have recently merged, at a time close to 
$t_{\rm obs}$, but have not yet coalesced into a single ACS system 
of type  Int-1.

It is important to note that we are {\it not} concerned here with pairs of 
galaxies with a wide separation $d \ge t_{\rm vis} \times v$ (where $v$ is 
the relative speed between the 2 progenitor galaxies), because such systems 
represent potential {\it future} mergers that will occur in the next discrete 
time step of ($t_{\rm obs}$ + $t_{\rm vis}$). Instead,  we are only 
interested in very close overlapping pairs of galaxies with separation  
$d \ll t_{\rm vis} \times v$, which represent young mergers that have occurred  
at a past time close to $t_{\rm obs}$, and which will likely coalesce into 
a single ACS system well before the  next  discrete time step of  
($t_{\rm obs}$ + $t_{\rm vis}$). For $t_{\rm vis}$ of 0.5 Gyr and $v\sim$~100 km/s,
$d \ll t_{\rm vis} \times v$  translates to  $d \ll$~50 kpc. 
We thus only consider VCPOG where the galaxies overlap and have a 
separation $d<$~20 kpc, which  corresponds to  $d<$~$5\farcs3$ at 
$z\sim$~0.24  and $d<$~$2\farcs8$ at  $z\sim$~0.80.

One caveat in handling systems of type Int-2 is that some of the VCPOGs may 
be chance projections rather than real gravitationally bound mergers. 
However, this uncertainty does not significantly affect our result
as the vast majority ($>$~80\%) of the mergers in our study are of type
Int-1 rather than Int-2.  Furthermore, the likelihood of chance projection
is mitigated due to the fact that we are considering pairs of very small 
separation.

The COMBO-17 ground-based data of resolution $\sim$~$1\farcs5$ 
will resolve a fraction of the VCPOG that make up systems of type 
Int-2.  Thus, we divide the latter systems  into sub-classes 
Int-2a and Int-2b:

\begin{enumerate}
\vspace{-3mm} 
\item 
Mergers of type Int-2a: 
These very close pairs of overlapping galaxies  are not resolved by COMBO-17 data. 
Thus, only one stellar mass, redshift, and UV-based SFR  are available for the 
pair. The lack of resolved  COMBO-17 data for each galaxy in the the pair is not 
a problem because we are only concerned with the mass and  SFR of the progenitor.
Furthermore, the condition $M_{\ast} \ge M_{\rm cut}$ is expected to apply
to the merger, and not to the progenitors, in both model and observations.

Since only one mass is available for the entire merged system,  the evidence 
for a merger of mass ratio $>$~1/10  in systems of type Int-2a does not come from 
a measured mass ratio, but instead is deduced from the presence of 
morphological distortions, which are seen in simulations of mass ratio  
$M1/M2 >$~1/10. Essentially, the same approach described for systems of 
type Int-1 in $\S$~\ref{sint1} is used here.

\vspace{-3mm} 
\item 
Mergers of type Int-2b: 
These  VCPOG are resolved by COMBO-17 data such that stellar masses 
$M1$ and $M2$, as well as  redshifts and UV-based SFRs, are available 
for both galaxies in the pair. An example is case 1 in Fig.~\ref{fpec1}).
For systems of type Int-2b, the evidence for a merger of mass ratio 
$>$~1/10 comes directly from the measured mass ratio $M1/M2$.
The SFR  and mass of the merger is considered as the sum of that of its 
two progenitor galaxies

One caveat should be noted concerning the completeness of mergers of type Int-2b.
Strictly speaking, the criterion $M_{\ast} \ge M_{\rm cut}$  applies to 
the merger mass ($M1 + M2$) rather than to $M1$ or $M2$ individually. 
For $M_{\rm cut} \sim$~$2.5 \times  10^{10}$ $M_{\odot}$  counting all major 
mergers of type Int-2b with 1/4~$< M1/M2 \le$~1/1, requires our sample to 
be complete down to $1 \times  10^{10}$ $M_{\odot}$ for 1:1 major mergers, 
and down to $8 \times  10^{9}$ $M_{\odot}$ for  1:3 major mergers. Similarly,  
counting all minor mergers of type Int-2b with 1/10~$< M1/M2 \le$~1/4 requires 
our sample to be complete down to $6 \times  10^{9}$ $M_{\odot}$ for 1:4 minor 
mergers,  and down to $2.5 \times  10^{9}$ $M_{\odot}$ for  1:9 minor mergers. 
For the blue cloud, where we are complete down to  $1.0 \times  
10^{9}$ $M_{\odot}$, our count of mergers of type Int-2b is complete, but 
there will inevitably be incompleteness on the red sequence,
particularly among minor mergers.  The impact of this incompleteness on our results 
is mitigated by the fact that  most of the galaxies in our sample is on the blue 
cloud (see Fig.~\ref{fuvmas}) rather than on the red sequence, and the 
vast majority ($>$~80\%) of the mergers in our study are of type  Int-1 rather 
than Int-2.

\end{enumerate}

\subsection{Visual Classification of Non-Interacting Galaxies}\label{scvis2}

Systems that do not satisfy the criteria in $\S$~\ref{scvis1} and 
show no evidence of a recent merger of mass ratio~$>$~1/10 
are classified as non-interacting. These systems may harbor very subtle distortions, 
but none of the kind shown by the mergers of type `Int-1 and `Int-2'. 
The non-interacting systems are divided into  two sub-groups: 
non-interacting E-to-Sd  and non-interacting Irr1. These are shown in  
Figure~\ref{fmonta} and described below. 

\begin{enumerate}

\vspace{-3mm} 
\item `Non-Interacting Irr1': It is important to note that even non-interacting galaxies 
have some inherent level of small-scale asymmetries in optical light due to  star formation 
activity. In the case of low mass galaxies, further asymmetries may also arise due to the low ratio 
of rotational to random velocities, as is commonly seen in Im and Sm.  These 
{\rm internally-triggered} asymmetries  due to SF in non-interacting galaxies differ 
in scale (few 100 pc vs several kpc) and morphology from the {\rm externally-triggered} 
distortions typical of  the `Int-1' class.
We classify non-interacting galaxies with such {\rm internally-triggered} asymmetries 
as `Irr1'  (see Figure~\ref{fmonta}). Such systems may get erroneously classified as  
mergers in automated asymmetry-based codes (see $\S$~\ref{scasts}).

\vspace{-3mm}
\item  `Non-Interacting E-to-Sd':  Galaxies are assigned the `E to Sd' class if they are fairly symmetric,  
have Hubble types in the range E-to-Sd, and are not associated with any overlapping or 
contact companion.
\end{enumerate}

In this paper, we are primarily concerned about the differences between 
three groups: the  mergers  in class `Int',  the non-interacting E-to-Sd 
galaxies,  and the non-interacting Irr1 galaxies. 
The details of how E-to-Sd galaxies are further sub-divided
into  individual Hubble types  do not have any major impact on our 
main results. We nonetheless briefly  describe this  sub-classification
as it is of interest to other studies and   relevant for the test presented at the
end of  $\S$~\ref{scvctes}.
We use conventional definitions  (Binney \& Merrifield  1998) for  
individual Hubble types (E, S0, Sa, Sb-Sc, and Sd). 
We assign an elliptical (E) type if a galaxy exhibits 
a  smooth featureless appearance, shows no disk signatures, such as a bar or 
spiral arms, and appears to be a pure spheroid. 
We assign an S0 class if a galaxy hosts a smooth central brightness condensation, 
surrounded by an outer  component, which is relatively featureless (without 
spiral arms) and has a less steeply declining brightness profile.
We assign Sa, Sb-Sc, and Sd  types using primarily visual estimates of the 
$B/D$ ratio, and secondarily the smoothness/clumpiness of the disk.
At intermediate redshifts, where the faint smooth arms of Sa galaxies 
are not easily discernible, the distinction between E, S0, and Sa becomes 
blurred (see also $\S$~\ref{scvctes}).
However, this ambiguity between Es, S0s and Sas is not a problem for the
subsequent analyses in this paper, since  galaxies are grouped together 
either as    `E+S0+Sa' or  `E-to-Sd'.

The fraction of interacting systems (i.e., mergers), `non-interacting E-to-Sd' 
galaxies,  and  `non-interacting Irr1' galaxies is shown in   
Table~\ref{tvclas1} for the high mass sample, and in  Table~\ref{tvclas2}
for the intermediate mass sample.  
Further results and tests on the merger history from visual classes are 
presented in $\S$~\ref{scvctes} and $\S$~\ref{scmfrac}.

\subsection{CAS}\label{sccas1}

We derived the concentration $C$, asymmetry $A$, and clumpiness $S$  (CAS) parameters
by running the the CAS  code (Conselice 2003) on the F606W images.
As is standard practice,  the segmentation maps produced during the original source 
extraction (Caldwell \etal 2008) are used to mask neighbors on each ACS tile.
The CAS code derives the asymmetry index $A$  (Conselice 2003) by rotating a galaxy image by  
180 deg, subtracting the rotated image  from the original image, summing the absolute 
intensities of the residuals, and normalizing the sum to the  original galaxy flux. 
CAS improves the initial input center with the IRAF task `imcenter' and then performs
a further refinement within  a $3 \times~3$ grid, picking the center that minimizes $A$.
The CAS  concentration  index $C$ (Bershady \etal  2000) is proportional to  the logarithm 
of the ratio of the 80\% to 20\% curve of growth radii within 1.5 times the Petrosian 
inverted radius at $r$($\eta$~=~0.2), normalized by a logarithm 

\begin{equation}
C =  5 \times \rm log (r_{\rm 80\%}/r_{\rm 20\%})
\end{equation}

The clumpiness index $S$ (Conselice 2003)~is defined  as the ratio of the amount of light 
contained in high-frequency structures to the total amount of light in the galaxy. In order
to compute $S$, the CAS code first smooths the galaxy image with a filter of size equal 
to 1/6 of the Petrosian radius to produce a lower resolution image whose high-frequency 
structure has been washed out. The latter image is then subtracted from the original image  
to produce a residual map that 
contains only the high-frequency components of the galaxy's stellar light.
The flux of this residual light is then summed and divided by the sum of the original 
galaxy image flux to obtain a galaxy's clumpiness ($S$) value. 
Tests on the interaction history from CAS  are presented in $\S$~\ref{scasts}.

It has been argued that  the criterion  $A >$~0.35 and $A > S$  (henceforth referred 
to as the CAS  merger criterion) captures galaxies that exhibit large 
asymmetries produced by major mergers (Conselice 2003). We will assess 
this in $\S$~\ref{scasts}.

\section{Results and Discussion}\label{scresul}

\subsection{The merger fraction  from visual classes}\label{scvctes}

Fig~\ref{fvctst1} compares the  fraction $f$ of systems with evidence of a 
recent merger of mass ratio $>$~1/10, based on visual classification by 3 
classifiers (SJ, SM, KP). Results are shown for both the high mass  
($M \ge$~$2.5 \times  10^{10}$ $M_{\odot}$) sample S1 and the intermediate 
mass ($M \ge$~$1 \times  10^{9}$ $M_{\odot}$) sample S2. On this figure, the 
plotted error bar for the merger fraction $f$ only includes  the binomial 
term [$f$(1-$f$)/$N$]$^{1/2}$, for each bin of size $N$. 
The same trend is seen for all 3 classifiers and the maximum
spread $\delta_{\rm f}$/$f$  in the four bins is 
15\%, 17\%, 26\%, and 26\%, respectively.
In subsequent analyses, we  adopt a conservative error bar on $f$ that includes in  
quadrature both the binomial term and  a  dispersion of 26\% to capture 
the inherent subjectivity in the visual classification.

Another  key test is to assess the impact of redshift-dependent systematic 
effects, such as bandpass shifting.
When using the F606W filter  whose pivot wavelength is $\sim$~5915 \AA ,  
the rest frame wavelength ($\lambda_{\rm rest}$) corresponds
to the rest-frame optical at the mean redshift of  the first 3 bins, but shifts 
to the rest-frame violet/near-UV  (3700 \AA \ to 3290 \AA) in the last bin  
($z\sim$~0.6 to 0.8). 
Galaxies tend to look slightly more asymmetric at near-UV wavelengths due to 
the prominence of young stars. In order to quantitatively test the impact 
of bandpass shift on our visual classes, we use the  redder F850LP images 
from the GOODS survey, which overlaps with the central 20\% of the
GEMS survey area. The F850LP filter has a pivot wavelength 
of  9103 \AA  \ and traces the rest-frame optical (7340 \AA  \ to 5057 \AA)
in all four redshift bins out to $z\sim$~0.8. The F850LP images 
also have 5 times longer exposures  than the GEMS F850LP and F606W images.
Figure$~\ref{fvctst2}$ shows  GEMS F606W and GOODS F850LP images of 
typical  disturbed and normal galaxies in the last 2 redshift 
bins ($z\sim$~0.47 to 0.8). 

While the GOODS images have higher S/N, and trace redder older 
stars, they do not reveal dramatically different morphologies 
from those  in the GEMS F606W images (Fig$~\ref{fvctst2}$). 
Furthermore, the  855 intermediate mass ($M \ge$~$1 \times  10^{9}$ $M_{\odot}$) 
galaxies in the GEMS/GOODS overlap area, were classified using both 
GOODS  F850LP and GEMS F606W images by the 3 classifiers.
We find that the ratio of ($f_{\rm GEMS}$/$f_{\rm GOODS}$) ranges 
from 0.8 to 1.2 across the 3 classifiers (Table~\ref{tgoods1}), 
where $f_{\rm GEMS}$ and  $f_{\rm GOODS}$ are the fraction of 
merging systems based on the GEMS F606W and GOODS F850LP images, 
respectively. The mean $f$ changes by only 6\% (Table~\ref{tgoods1}). 
In effect, over $85\%$ of the systems  classified as  mergers 
(`Int') in the GEMS F606W images  retain the same visual class in the GOODS 
F850LP.   
Among the remaining objects, some classified as non-interacting
in GEMS F606W get reclassified as disturbed in GOODS F850LP, and vice-versa. 
The fact that $f$  does not change by a large amount between
GEMS F606W and GOODS F850LP is not surprising, since the 
rest-frame wavelength of GEMS F606W in the last bin shifts only  
to the violet/near-UV, rather than  to the far-UV, 
where morphological changes are more dramatic.
We conclude that our results are not highly impacted by bandpass
shifting, and any effect is accounted for by our error bars 
of $>$~$26\%$ in $f$.

Another redshift-dependent systematic effect is surface brightness dimming 
at the rate of (1+$z$)$^{-4}$ (e.g., Barden \etal 2008). 
This leads to surface brightness dimming by a factor of 
 1.0 to  2.5 magnitude over the redshift range 0.24 to 0.80.
This is mitigated in  part by  two factors:  galaxies are on average 
1.0 magnitude brighter in surface brightness by $z\sim$~0.8 (e.g., 
Barden \etal 2005), and  the average SFR rises by   a factor of $\sim$~4 out 
to $z\sim$~0.8  (e.g., see $\S$~\ref{scsfr1}).
Two approaches can be adopted to assess the impact of surface brightness dimming.
The first is to artificially redshift  disturbed galaxies in the lowest
redshift bin ($z\sim$~0.24) out to $z\sim$~0.8, either assuming passive evolution or 
adding in  a $\sim$~1 magnitude of brightening in surface brightness. 
However, this approach suffers from the limitation that it implicitly assumes 
that galaxies at $z\sim$~0.8 are similar to those at  $z\sim$~0.24 and evolve 
passively with time.
A better approach, which does not  make such assumptions,  is to repeat the 
analysis and visual classification using {\it deeper} images of the galaxies and 
assess the resulting change in  visual classes. 
The above-described test performed using the deep GOODS F850LP  image  
(Fig.~\ref{fvctst2})  is an example of such a test, and indicates that the 
eye-ball morphologies do not change within the error bars of $>$~$26\%$ in $f$.
We note however that  quantitative CAS parameters can change with the deeper
GOODS images (e.g., Conselice et al. 2008).

Finally, as an extra test, we checked the   distribution of S\'ersic indices $n$   
for single-component S\'ersic fits  (Barden \etal 2005) for the visual classes 
of the sample S2  of intermediate mass  $M \ge$~$1 \times  10^{9}$ $M_{\odot}$ galaxies 
(Fig.~\ref{fsersi}).  
Non-interacting disk-dominated systems are expected to have $n <$~2.5, while massive 
ellipticals and bulge-dominated systems typically have higher S\'ersic indices. 
We indeed find that  over 85\% of the systems visually classified as  
Sb-Sd  and Irr1 have  $n <$~2.5 in the  intermediate mass ($M \ge$~$1 \times  
10^{9}$ $M_{\odot}$) sample. Furthermore, as expected, the vast majority  of 
the systems typed as Sa have $n<$~4.  However, the systems typed as E and 
S0 span a broad range in $n$: most of them have $n>3$, but there is a tail 
of lower $n$ values. This is not surprising given the previously described  
difficulties ($\S$~\ref{scvis2})  in separating  E, S0, and Sa galaxies at  
intermediate redshifts.
However, this ambiguity between E, S0, and Sa systems is not a problem for the
subsequent analyses in this paper, since  galaxies are grouped together 
either as    `E+S0+Sa' or  `E-to-Sd'.
In fact, as stressed in $\S$~\ref{scvis2},  the main results presented 
in this paper depend only on the differences between three groups: 
mergers (`Int'), non-Interacting E to Sd galaxies, and non-interacting 
Irr1 galaxies.

\subsection{The   merger fraction  from  CAS}\label{scasts}

It has been argued that  the CAS merger criterion  ($A >$~0.35 and $A > S$)  
captures systems that exhibit large asymmetries produced by major 
mergers (Conselice 2003). This criterion is based on calibrations of the CAS system 
at  optical rest-frame wavelengths 
($\lambda$$_{\rm rest} > $~4500 \AA). \  However, there are several  caveats: 
a)~The CAS criterion ($A>$~0.35 and $A>S$) will 
miss out  interacting galaxies  where the morphological distortions
contribute  to less than 35 \% of the total galaxy flux.
(b)~Calibrations of $A$ with N-body simulations  (Conselice 2006) shows that 
during major mergers with mass ratios 1:1 to 1:3, the asymmetry oscillates with 
time.  Typically, it exceeds 0.35 for $\sim$ 0.2 Gyr in the early phases when the 
galaxies start to interact, falls to low values as the galaxies separate, 
rises for $\sim$~0.2 Gyr as they approach again for  the final merger, and 
eventually tapers down as the final remnant relaxes. On average, the $A >$~0.35
criterion is only satisfied for one third of the merger timescale in these
N-body simulations. 
For minor mergers of mass ratios 1:5 and below, the asymmetries are too low 
to satisfy  $A >$~0.35. 
(c)~To complicate matters, automated asymmetry parameters can also capture
non-interacting galaxies  whose visible light shows small-scale asymmetries 
due to star formation (e.g., Miller \etal  2008; Lotz \etal 2008).

Visual tests that verify  how well the  CAS criterion ($A>$~0.35 and $A>S$)  works 
at intermediate redshifts  have been performed  using spot checks and 
small-to-moderate samples  (e.g., Mobasher et al 2004;  Conselice 2003; 
Conselice  et al. 2003; Conselice et al. 2005). However, what has been missing
to date is a quantitative estimate, based on a large sample of  galaxies, of the 
recovery fraction  of CAS  (i.e.,  the fraction of visually-classified mergers 
that the CAS criterion picks up), and the contamination level of CAS  
(i.e.,  the fraction of visually-classified non-interacting  galaxies that 
the CAS criterion picks up). Both the recovery fraction and contamination level 
might be expected to depend on the rest-frame wavelength used, the mass and SFR
of the galaxies, etc.
In this paper, we perform one of the most extensive comparisons to date, 
at intermediate redshifts ($z\sim$~0.24 to 0.80),   between CAS-based 
and visual classification results  for both high mass  
($M \ge$~$2.5 \times  10^{10}$ $M_{\odot}$) and  intermediate mass 
($M \ge$~$1 \times  10^{9}$ $M_{\odot}$) galaxies. We  assess the 
effectiveness of the CAS merger criterion  ($A >$~0.35 and $A>S$) over this 
interval, where the  rest-frame wavelength $\lambda$$_{\rm rest}$ varies from 
4770 \AA \ to 3286 \AA. \  We note that the rest-frame wavelength range here 
extends to  somewhat bluer wavelengths than the range ($\lambda$$_{\rm rest} 
> $~4500 \AA) over which the CAS system was calibrated.  

Fig~\ref{fvctst1} compares the merger fractions that would be
obtained using the CAS criterion ($f_{\rm CAS}$), as opposed to  visual 
classification ($f$).
For the high mass  ($M \ge$~$2.5 \times  10^{10}$ $M_{\odot}$) galaxies, 
visually based and CAS-based merger fractions agree within a factor of two, with 
$f$ being  higher than  $f_{\rm CAS}$ at $z<0.5$, and being lower at $z>~0.5$ 
(top panel of Fig~\ref{fvctst1}). 
However, for the intermediate mass ($M \ge$~$1 \times  10^{9}$ $M_{\odot}$) 
galaxies (lower panel of Fig~\ref{fvctst1}), at $z>0.5$ the CAS-based  merger fraction 
can be systematically higher by a factor  $\sim$~3  than the visually based $f$.
The reason for this discrepancy, as we show below, is that at bluer rest-frame wavelengths 
(i.e., higher redshifts), the CAS criterion  picks up a significant number of 
non-interacting  dusty, star-forming galaxies.

Fig.~\ref{fcasvc} plots the CAS asymmetry $A$ and clumpiness $S$ parameter  for 
galaxies in  the four redshift bins covering  the interval  $z \sim$~0.24--0.80.
Galaxies satisfying the CAS  criterion  ($A >$~0.35 and $A > S$) lie in the upper  
left hand corner. 
One can see that while the CAS criterion captures a fair fraction of the 
mergers (coded as orange stars), 
it also picks up a large number of  non-interacting galaxies.
 
We define the  recovery fraction ($F_{\rm CAS-merger}$) of CAS as the fraction of  
visually-classified mergers (`Int'), which are 
picked up by  the CAS criterion  ($A >$~0.35 and $A > S$).  
For the high  mass sample ($M \ge$~$2.5 \times  10^{10}$ $M_{\odot}$) sample,
the recovery fractions in the four redshift bins are 
50\% (2/4),   14\% (1/7),   42\% (8/19),  and  56\% (20/36) respectively, 
with low number statistics dominating the first two bins.
For the intermediate mass ($M \ge$~$1 \times  10^{9}$ $M_{\odot}$) sample, 
the recovery fractions in the four redshift bins are
50\% (13/26),   69\% (30/43),   59\% (49/83),  and  73\% (85//116), respectively,
as illustrated in the top panel of   Fig.~\ref{fcasvc2}.
We inspected the visually-classified mergers  missed 
out by the CAS criterion  ($A >$~0.35 and $A > S$) 
and show typical cases in the top panel of  Fig.~\ref{fcasvc3}. The missed 
cases  include  galaxies where  tidal or accretion features  in the main disk  of 
a galaxy contribute  less than $35\%$  of the total light  
(e.g., case 3 in Fig.~\ref{fcasvc3});   galaxies with close double nuclei  
(e.g., case 2 in  Fig.~\ref{fcasvc3}) where CAS might refine the center to be between 
the two nuclei,  thereby leading to a  low $A < 0.35$;  and  pairs of 
fairly symmetric galaxies whose members have similar redshifts within the 
spectrophotometric error, appear connected via weak tidal features, and have a  stellar mass 
ratio $M1/M2 > 1/10$ (e.g., case 1 in Fig.~\ref{fcasvc3} where $M1/M2 \sim$~0.25).

We define the contamination fraction of CAS as the fraction of those systems 
which satisfy the CAS criterion ($A >$~0.35 and $A > S$) and are therefore considered
as likely major mergers by CAS, but  yet are visually classified as non-interacting. 
For the high  mass sample ($M \ge$~$2.5 \times  10^{10}$ $M_{\odot}$) sample,
the CAS contamination fractions in the four redshift bins are 
34\%, 75\%, 72\%, and 67\%  respectively, with low number statistics dominating 
the first two bins.
For the intermediate mass ($M \ge$~$1 \times  10^{9}$ $M_{\odot}$) sample, 
the corresponding CAS contamination fractions are 
 44\%, 53\%, 76\%, and 82\% respectively, as shown in the lower panel of  
Fig.~\ref{fcasvc2}. On the latter figure, $N_{\rm CAS}$ represents the total  number 
of galaxies  satisfying the CAS criterion  ($A >$~0.35 and $A > S$) in 
the four redshift bins. Plotted on the y-axis is the fraction 
$F_{\rm CAS-visual}$  of different visual types (mergers, non-interacting E-Sd, 
and non-interacting Irr1) among these  ``CAS mergers''. Across the four redshift bins, 
the non-interacting E-Sd and Irr1  make up  44\%, 53\%, 76\%, and 82\% of the 
CAS systems.  Typical  cases are shown  in the lower panel of Fig.~\ref{fcasvc3}. 
They include non-interacting  actively star-forming systems where SF induces small-scale 
asymmetries in the optical blue light  (e.g.,  cases 4 and 6 in  Fig.~\ref{fcasvc3});
systems where $A$ is high due to the  absence of a  clearly defined center  
(e.g., case 8  in Fig.~\ref{fcasvc3}) 
or due to the center being blocked by dust (e.g., cases 4 and 9  in Fig.~\ref{fcasvc3}); 
and compact or edge-on systems where the light profile is steep such that
small centering inaccuracies can lead to large $A$ (e.g., case 9 in Fig.~\ref{fcasvc3}).

In summary, we find that the CAS-based merger fraction agrees
within a factor of two with  visually based one for high mass  
($M \ge$~$2.5 \times  10^{10}$ $M_{\odot}$) galaxies, but can overestimate
the  merger fraction at $z>$~0.5 by a factor $\sim$~3  for  
intermediate mass ($M \ge$~$1 \times  10^{9}$ $M_{\odot}$) galaxies.
For the latter mass range, the systems counting toward $f_{\rm CAS}$ are a mixed bag: 
the CAS criterion misses about half of the visually-classified 
mergers, but picks up a dominant number  of non-interacting  dusty, 
star-forming galaxies.
We thus conclude that the CAS merger criterion  is ill-suited  for use  on  
$HST$ $V$-band images at $z >$~0.5,  where the rest frame wavelength falls 
below  $\lambda < 4000$ \AA,  particularly  in the case of intermediate
mass galaxies with significant SF, gas, and dust. Modified CAS criteria in the 
near-UV based on morphological k-corrections  (Taylor et al. 2007) might alleviate 
this problem.

\subsection{Interaction history of massive and intermediate mass galaxies}\label{scmfrac}

Based on the tests in $\S$~\ref{scvctes} and $\S$~\ref{scasts}, we decided to adopt  the 
mean merger fraction $f$ based on visual classes for our two samples of interest.
For the high mass  ($M \ge$~$2.5 \times  10^{10}$ $M_{\odot}$)  sample, which is 
complete on both the blue cloud and red sequence ($\S$~\ref{scdatas}), 
the results are shown on in   Table~\ref{tvclas1}.
The error bar shown on the merger fraction  $f$ in both tables now includes  the sum in quadrature of a binomial 
term [$f$(1-$f$)/$N$]$^{1/2}$ for each bin of size $N$, along with a fractional  
error  of $\pm$~26\% to capture the  dispersion between classifiers, 
and uncertainties due to bandpass shifting and surface brightness dimming.

From Table~\ref{tvclas1} and Fig~\ref{fvctst1}, it can be seen that
the merger fraction $f$  among
high  mass ($M \ge$~$2.5 \times  10^{10}$ $M_{\odot}$)  galaxies  does not show 
strong evolution over lookback times of 3--7 Gyr, ranging from  9\% $\pm$~5\% at 
$z \sim$~0.24--0.34, to 8\% $\pm$~2\%  at $z \sim$~0.60--0.80, as averaged over 
every Gyr bin. As discussed in $\S$~\ref{scvis1}, the merger fraction $f$ refers 
to systems with evidence  of a recent merger  of mass ratio $>$~1/10.

As outlined in $\S$~\ref{scvis1}, these mergers were further subdivided among 3 classes: 
clear major merger,    clear minor merger, and   ambiguous  `major or minor merger' cases.
The first two classes are  used to set  lower limits on  
the major and minor merger fraction.
The  lower limit on the major  ($M1/M2 >$ 1/4) merger  fraction, determined in this 
way, ranges from  1.1\%  to 3.5\%  over $z\sim$~0.24--0.80. 
The corresponding lower limit on the minor  (1/10 $\le M1/M2 <$ 1/4)  
merger fraction ranges  from  3.6\% to 7.5\%. 
To our knowledge, this is the first, albeit approximate, empirical 
estimate of the frequency of  minor mergers over the last 7 Gyr. 
The ambiguous cases of `major or minor merger' make up a fraction between 1.2\% to 2.0\%.

When converting  the observed fraction $f$  of galaxy mergers  into  
a merger rate $R$, we must  
bear in mind that in any observational survey of galaxies, 
mergers can only be recognized for a finite time $t_{\rm vis}$, which is the
timescale over which a merging galaxy will appear morphologically
distorted. This timescale depends on the mass ratio of the merger 
as well as the gas fraction of the progenitors: $t_{\rm vis} \sim 0.5$--0.8 
for gas-rich galaxies, and $t_{\rm vis} \sim 0.2$--0.4 Gyr for
gas-poor galaxies (T.J. Cox, private communication). This timescale
will also depend on many observational factors such as the method used
to identify mergers (e.g. visual classification vs. CAS or other
statistical methods) and the depth of the imaging used. We assume a
representative value of $t_{\rm vis} = 0.5$ Gyr here, but we must keep
in mind that there are at least factors of two uncertainty in this
number.  The merger  rate $R$ is  given by 
\begin{equation}
R  = \frac{n \ f}{t_{\rm vis}},
\end{equation}

\noindent
where $n$ is the comoving number density of galaxies above a certain mass limit 
in the redshift bin.

For the sample of high mass  ($M \ge$~$2.5 \times  10^{10}$ $M_{\odot}$) galaxies, 
our measured merger fraction $f$ and assumed value of $t_{\rm vis} \sim$~0.5 Gyr 
lead to a  corresponding  merger rate  $R$ of a 
few $\times 10^{-4}$ galaxies Gyr$^{-1}$ Mpc$^{-3}$.
Assuming a  visibility timescale of $\sim$~0.5 Gyr, it follows that on average, 
{\it  over $T_{\rm b}\sim$~3--7 Gyr, $\sim$~68\% of high mass systems have 
undergone a merger of mass ratio $>$~1/10. Of these, we estimate that 
$\sim$~16\%, 45\%, and 7\%  correspond respectively to clear major mergers, 
clear minor mergers, and ambiguous cases of `major or minor' mergers.}

At intermediate masses ($M \ge$~$1 \times  10^{9}$ $M_{\odot}$) where 
we are only complete in mass for the blue cloud ($\S$~\ref{scdatas}),  
we  consider $f$ to be meaningful only for the intermediate mass blue cloud sample.  
Results for this sample are shown  in the lower part of Table~\ref{tvclas2}.
The fraction of blue cloud galaxies 
having undergone recent mergers of mass ratio  $>$~1/10 ranges from  
7\% $\pm$~2\% to 15\% $\pm$~5\%  over $z \sim$~0.24--0.80.
The corresponding  merger rate $R$ ranges from  $8 \times 10^{-4}$  to
 $1 \times 10^{-3}$  galaxies  Gyr$^{-1}$ Mpc$^{-3}$ .
For an assumed visibility time of $\sim$~0.5 Gyr, we estimate that  on average, 
over $T_{\rm b}\sim$~3--7 Gyr, 84\% of  intermediate mass 
blue cloud galaxies have undergone  a merger  of mass ratio~$>$~1/10, 
with $\sim$~5\%, 22\%, and 57\% corresponding respectively to clear major 
mergers, clear minor mergers, and ambiguous cases of `major or minor' mergers


\subsection{Comparison with other studies}\label{sccompa}


When comparing our observed  merger fraction $f$  
in the high mass ($M \ge$~$2.5 \times  10^{10}$ $M_{\odot}$) 
sample over $z\sim$~0.24--0.80 with published studies, several caveats must be 
borne in mind. 
Many  studies have small samples and large error bars  at $z<0.8$ (e.g., 
Conselice 2003; Fig.~\ref{fcompa1}). Others 
focus on bright galaxies and luminosity-selected samples  (e.g.,  Lotz et al. 2008; 
Casatta et al. 2005 ) rather than stellar mass selected sample, because the data to 
derive stellar masses  were unavailable. 
Different studies target different systems,  ranging from morphologically
distorted systems to close pairs with  separation $d \sim$~5 to 40 kpc. Finally,  
many studies focus  only on major mergers, while the  interacting 
galaxies identified in our study are candidates for a merger of mass ratio 
$>$~1/10 ($\S$~\ref{scvis1}), and include both minor and major 
mergers.  Nonetheless, we attempt approximate comparisons.

Fig.~\ref{fcompa1} shows the merger fraction based primarily on  
morphologically distorted  galaxies  (filled circles), as well as 
the close pair fraction (open squares), as a function of redshift.  
The Lotz \etal (2008) study shows the fraction 
$f_{\rm Gini}$ of morphologically disturbed  systems based on Gini-M20 parameters  
among   $M_{\rm B} <$~-20.5 
and   $L_{\rm B}$~$>$~0.4 $L_{\ast}$ galaxies in the Extended Groth Strip.  
This study does  not present any results for a high mass sample, and thus we 
effectively are comparing their bright galaxies to our high mass galaxies. 
Over $z\sim$0.2--0.80, our results  are in very good  agreement, within a 
factor of less than two,   with $f_{\rm Gini}$. 
The CAS-based results from Conselice (2003) are derived from a  small sample in 
the Hubble Deep Field and have error bars  that are too large to set useful constraints at 
$z<1$ (Fig.~\ref{fcompa1}).
Our results of a fairly flat evolution of the merger rate out to $z\sim$~0.8 
also agree with the results of Cassata et al. (2005), which are based on 
both pairs and asymmetries.

Fig.~\ref{fcompa1} also shows  the result from  three studies based on close pairs.
The major merger fraction of  massive galaxies ($M_{\ast} \ge$~$2.5 \times  10^{10}$ $M_{\odot}$)
in close ($d <$~30 kpc) pairs,  based on  the 2-point correlation 
function in COMBO-17, is  5\%~$\pm$~1\%   averaged over  at 0.4~$< z<$~0.8 (Bell \etal  2006). 
This value is  lower than our merger fraction $f$ ($\sim 8\% \pm 2\%$), 
which represent likely mergers of mass ratio  $>$~1/10, and it is higher than 
the fraction of cases we see as clear major mergers ($\sim 1.3\% \pm 0.2\%$). 
The study of luminous  ($L_{\rm V}$~$>$~0.4~$L_{\ast}$)  pairs at projected 
separations of 5--20 kpc in the COSMOS field  (Kartaltepe \etal 2007) finds
a galaxy pair fraction  of $\sim$~1\%--3\% over $z\sim$~0.24--0.80, corresponding 
to  a galaxy merger fraction of $\sim$~2\%--6\% . 
Our observed  fraction $f$  of 9\% $\pm$~5 to 8\% $\pm$~2\%  over  $z\sim$~0.24--0.8 
is slightly higher and flatter than this study. 

The differences we see through these comparisons are already known (see  $\S$~\ref{scintro}).
Studies based on close pairs tend to show  moderate to fairly strong evolution in the major 
merger rate out to $z\sim$~1.2  (e.g., Kartaltepe et al 2007; Bell  et al 2006; Lin 2004), 
while studies based on asymmetries (e.g., Lotz et al. 2008; this study), and 
studies based on both pairs and asymmetries (Cassata et al. 2005) tend to report 
only mild evolution of the  merger rate  with redshift  up to  z$\sim$~1.  

It is not fully understood why  these  different methods yield  different results, but 
several factors likely play a part. 
First, it should be noted that  the claim of strong evolution in the close pair fraction out to 
$z\sim$~1.2  in the C0SMOS study by Kartaltepe et al. (2007) comes about when the low redshift 
$z\sim$~0 point from  the SDSS pair catalog (Allam et al. 2004)  is included in their 
analysis. The evolution within the internally consistent dataset from COSMOS over $z\sim$~0.15 
to 1.05 shows much weaker evolution  (Fig.~\ref{fcompa1}). 
The drop in   close pair fraction seems primarily to occur at $z<0.2$, but it is unclear
how reliable the  low $z\sim$~0  points are due to the small volume sampled and systematic 
effects between studies.
A further reason for the difference could be due to the fact that the methods used in these
studies trace different phases of an interaction, with the pair method tracing the potential 
pre-merger phase, while the method based on the distorted galaxies trace the later 
phases, including the merger and post-merger phases.

Another point is that both pair and asymmetry methods are imperfect ways of tracing 
the merger fraction. Methods tracing  morphologically  disturbed galaxies  may 
capture some fly-by tidal interactions rather than mergers, and this effect would 
cause the fraction of  interacting galaxies to overestimate the merger fraction.
However, this effect is not a dominant one due to the following reason:  interaction 
signatures typically persist for a visibility timescale of 0.5 Gyr ($T_{\rm vis}$),
and a fly-by companion causing the distortion would still be within  100 kpc 
of the disturbed galaxy, assuming an escape speed of 200 km/s. The distorted 
galaxies we identify do not typically have such a fly-by companion, of mass ratio
$>1/10$ and similar spectrophotometric redshift.
In studies based on close pairs, one source of uncertainty is that even pairs with 
members at the same redshift  may not become gravitationally bound in the future. 
This effect might cause pairs to overestimate the true major merger fraction. 
On the other hand, erroneous spectrophotometric redshifts  can cause 
us to either overestimate or  underestimate the true close pair fraction, with the 
latter effect being more likely.  Corrections for this effect are uncertain and depend 
on the shape of the spectrophotometric redshift errors  (e.g., see Bell et al. 2006 for 
discussion).

\subsection{Comparison of galaxy merger history with $\Lambda$CDM models}\label{scmodel}

We compare our empirical 
merger fraction $f$ (Fig.~\ref{fmodelf}) and merger rate $R$  (Fig.~\ref{fmodelr})  
to predictions from different theoretical models of galaxy
evolution in the context of a $\Lambda$CDM cosmology, including 
 the halo occupation distribution (HOD) models of Hopkins \etal (2007); 
semi-analytic models (SAMs) of Somerville \etal (2008), 
Bower et al. (2006), and Khochfar \&  Silk (2006); and the  cosmological 
smoothed particle hydrodynamics (SPH) simulations from Maller \etal (2006).
The models were provided to us directly by the authors or co-authors of 
these individual studies.

We first briefly describe the general problem of calculating galaxy merger
rates. Predicting the rate of mergers per comoving volume and per unit time
between \emph{isolated} dark matter (DM) halos within a $\Lambda$CDM model 
is relatively straightforward via semi-analytic methods or N-body simulations 
(e.g. Lacey \& Cole 1993; Gottl\"ober et al. 2001; Fakhouri \& Ma 2008; 
Neistein \& Dekel 2008; D'Onghia et al. 2008). 
However, making a direct prediction of the
\emph{galaxy} merger rates is more complicated due to a  number 
of factors, including  the difference between the galaxy and halo merger 
timescales,  tidal heating and stripping  of halos and sub-halos, the
effect of a dense core of baryons on merging satellites, and the non-linear
relation  at low mass between DM halo (or sub-halo) mass and galaxy mass 
(van den Bosch et al. 2007). 
Thus,  attempts to extract a \emph{galaxy} merger rate from $\Lambda$CDM
simulations also must attempt to model the relationship between dark
matter and galaxy properties. The three main methods for making this
connection are  HOD models, SAMs,  and hydrodynamic simulations.
We summarize below  how these three types of  models differ.

HOD models specify the probability that a DM halo of a given mass
$M$ harbors $N$ galaxies above a given mass or luminosity. The
parameters of this function are determined by requiring that
statistical observed quantities, such as galaxy mass or luminosity
functions and galaxy correlation functions, be reproduced. 
The merger rate of galaxies within their host halos is calculated
via standard or improved dynamical friction formulae.  
In the HOD models of  Hopkins \etal (2007) used here, different modified formulae 
can be used, which include the effect of a gravitational capture cross section, 
stripping of DM halos, and  calibration factors  from $N$-body simulations. 
The predicted model rate  can vary by a factor of $\sim$~two depending
on model assumptions for sub-halo structure and mass functions, 
the halo occupation statistics, and the dynamical friction formulae 
used.

In SAMS, merger trees of DM halos  are either extracted from cosmological 
N-body simulations or  derived  using analytic methods 
(e.g., Somerville \& Kolatt 1999).
Calibrated modified versions of the Chandrasekhar dynamical 
friction  approximation (e.g. Boylan-Kolchin et al. 2008) are used
to compute the galaxy merger rate. Simplified analytic formulae are used to model
the cooling of gas, star formation, supernova feedback, and more recently,
AGN feedback (e.g., Somerville \etal 2008; Bower \etal 2006; Croton et
al. 2006; Benson \etal 2005; Cole \etal 2000; Somerville \& Primack
1999). The free parameters in these  formulae are normalized to 
reproduce observations of nearby galaxies, such as the $z=0$ galaxy
mass or luminosity function. Fig.~\ref{fmodelr} shows results of three
independent SAMs from  Khochfar \& Silk (2006), Bower et al. (2006), and
Somerville \etal (2008). 

Cosmological hydrodynamic simulations attempt to model the detailed
physics of gas hydrodynamics and cooling as well as gravity by
explicitly solving the relevant equations for particles or grid cells.
SPH methods are most commonly used. SF and supernova feedback are
treated using empirical recipes. A drawback of this approach is that,
due to computational limitations, state-of-the-art simulations still
do not have the dynamic range to resolve the internal structure of
galaxies while simultaneously treating representative cosmological
volumes. 
Another well known problem  is  that cosmological SPH models, which
do not include some kind of suppression of cooling (e.g., due to AGN 
feedback)  in massive halos do not reproduce the observed number
density of galaxies on the mass scales of interest (few $\times
10^{10} M_{\odot}$). Thus, the simulations of  Maller et al. (2006) shown 
here, over-predict the number of high and low mass galaxies, while
galaxies at the bend of the Schechter mass function are a factor of 2
to 3 too massive.  In order to make the simulated mass function agree 
better with observations,  Maller et al. (2006) apply a correction factor 
of 2.75 for galaxies in the mass range $2 \times 10^{10}$~$ < 
M_{\ast}$/$M_{\odot} <$~$6 \times 10^{11}$.   This correction is 
already included in the model on  Fig.~\ref{fmodelr}.

When comparing the observations to the models, one must consider carefully 
how merger rates and fractions are determined in these simulations. Two approaches 
are used: one based  on simulation snapshots  and the other based  on a light cone. 
In the first approach,  simulation  outputs (``snapshots'') are stored at a 
sequence of redshifts. 
Two snapshots separated by a time $\Delta t$ are considered and modelers 
trace the merger history of galaxies whose final stellar mass $M_{\ast} $ 
is greater or equal to a given  mass cut  $M_{\rm cut}$. The same mass cut  
($M_{\ast} \ge 2.5 \times 10^{10}$) used in the data is applied to the 
simulations. In order to mimick the observations as closely as possible, 
the interval $\Delta t$ in the model should ideally be equal to 
the visibility timescale $t_{\rm vis}$\footnote{If 
 $\Delta t$  is larger than  $t_{\rm vis}$, a correction factor of order
 ($t_{\rm vis}$/$\Delta t$) needs to be applied to the model merger fraction 
$f_{\rm mod1}$}.
One then counts the number  $N_{\rm 1}$ of model galaxies with 
$M_{\ast} \ge M_{\rm cut}$, which have experienced a merger of 
mass ratio $M1/M2 > 1/10$  within the last $t_{\rm vis}$.  It is important
to note two points. Firstly, if a galaxy were to undergo multiple mergers
within a time $t_{\rm vis}$, these mergers would be counted only once 
in the model term $N_{\rm 1}$, analogous to the case of the data where 
one cannot discriminate between multiple mergers within the time  $t_{\rm vis}$ 
over which a galaxy appears distorted due to a merger. Secondly, the fact that 
$t_{\rm vis}$ is so short makes it very unlikely for a model galaxy to 
undergo more than one merger over this timescale. As a result the model term 
$N_{\rm 1}$  is essentially  tracing the number  $N_{\rm mrg1}$ of mergers 
within the last $t_{\rm vis}$. One can then determine the merger rate 
using  $R_{\rm mod1}$~=~$N_{\rm mrg1}/(\Delta t \ V)$, where $V$ is the 
comoving volume of the simulation box.  
The merger fraction is $f_{\rm mod1}$~=~$N_{\rm  mrg1}/N_{\rm gal1}$, 
where $N_{\rm gal1}$ is the total number of galaxies above the relevant mass limit.
Except for Somerville et al. (2008), all the models presented on  Fig.~\ref{fmodelf} 
and Fig.\ref{fmodelr}  derive the merger fraction $f$ and rate $R$  using  the above
approach, based on  simulation snapshots.


For the Somerville \etal (2008) models, the simulation analysis was
carried out in a way that is closer to the observations.  We construct
a light cone with a geometry that is equivalent to three GEMS fields
(2700 arcmin$^2$ from $0.1 < z < 1.1$). We then divide the galaxies
into redshift bins, exactly as in the observational analysis, and 
we count the number $N_{\rm mrg2}$ of galaxies that have had a merger within a 
time $t_{\rm vis}$ in the past.  These galaxies  appear as 
morphologically distorted  mergers in the observations.  The model  merger 
fraction is  $f_{\rm mod2} = N_{\rm  mrg2}/N_{\rm gal2}$, where $N_{\rm gal2}$  is 
the total number of galaxies above the relevant mass limit in this light cone
or redshfit bin. 
The model merger  rate is calculated exactly as in the data 
using  $R_{\rm mod2}= f_{\rm mod2} \  n_{\rm mod2}/t_{\rm vis}$, where 
 $n_{\rm mod2}$ is the comoving  number density of galaxies above a certain 
mass limit in the redshift bin. 
Note that  $N_{\rm mrg2}$ and $f_{\rm mod2}$ are quite sensitive 
to  $t_{\rm vis}$, while $R$ is independent of  $t_{\rm vis}$.

In  Fig.~\ref{fmodelf},  we compare the empirical merger fraction $f$
to the corresponding model predictions $f_{\rm mod1}$ and  $f_{\rm mod2}$.
The comparison between the empirical merger rate $R$ (equation (3) in $\S$~\ref{scmfrac}) 
and the model predictions $R_{\rm mod1}$ and  $R_{\rm mod2}$  is in Fig.~\ref{fmodelr}.
In both data and models, major and minor mergers  are defined as those 
with mass ratio ($M1/M2 >$ 1/4), and (1/10 $\le M1/M2 <$ 1/4), respectively.  
The only slight exception is in the case of  Maller et al (2006) model where the 
extracted major mergers were defined with a slightly lower mass cutoff ($M1/M2 >$ 1/3).  
The dotted lines on  Fig.~\ref{fmodelf} and Fig.~\ref{fmodelr}  show 
the major merger rate  for all the models. 
The solid line show the  (major+minor) merger rate,  in other words,  the rate of 
mergers with mass ratio  ($M1/M2 >$ 1/10).   This is shown  for all models except  the 
Maller et al (2006)  SPH simulations,  where the limited dynamic range of the current 
simulations  only allows predictions for major mergers. The solid line  can be directly 
compared to our empirical merger fraction  $f$ or rate $R$, 
since the latter refer to mergers of mass ratio $M1/M2 >$ 1/10.

{\it
We find  qualitative  agreement  between the observations and models, such 
that the (major+minor) merger fraction (Fig.~\ref{fmodelf}) and 
merger rate (Fig.~\ref{fmodelr}) from different models (solid lines) bracket
the corresponding empirical estimates (stars) and show a factor of five  dispersion.}
One can now anticipate that in the near future, improvements in both the observational 
estimates and model predictions will start to rule out certain merger scenarios and refine
our understanding of the merger history of galaxies.

\subsection{The impact of galaxy mergers on the average SFR 
over the last 7 Gyr}\label{scsfr1}

Both observations (e.g., Larson \& Tinsley 1978; Joseph \& Wright 1985; 
Kennicutt et al. 1987;  Barton et al 2003) and simulations (e.g., 
Negroponte \& White 1983; Hernquist 1989; Barnes \& Hernquist 1991, 1996; 
Mihos \& Hernquist 1994, 1996;  Springel, Di Matteo \& Hernquist 2005b) 
suggest that galaxy interactions and mergers trigger star formation. 
However, simulations cannot uniquely predict the  factor by which mergers enhance
the SF activity of galaxies over the last 7 Gyr, since the star formation rate 
in simulations is highly sensitive to the  stellar feedback model,  
the bulge-to-disk ($B/D$) ratio, the gas mass fractions, and orbital geometry   
(e.g., Cox et al 2006; di Matteo et al. 2007).   Thus, we explore here the  impact of 
interactions on the average UV-based and  UV+IR-based  SFR   of  intermediate-to-high 
mass ($M \ge$~$1 \times  10^{9}$ $M_{\odot}$) galaxies over $z\sim$~0.24--0.80.

We adopt the SFRs  in  Bell \etal (2005, 2007), based on 
COMBO-17 UV  data (Wolf \etal 2004) and deep Spitzer 24~$\mu$m 
observations with a  limiting flux of $\sim$~83~$\mu$ Jy ($5\sigma$) from  
the  Spitzer Guaranteed Time Observers (Papovich \etal 2004;  Gordon \etal  2005). 
The unobscured SFR based on the directly observable   UV light from young stars 
was computed using 
SFR$_{\rm UV}$~=~$9.8 \times 10^{-11}$ (2.2 $L_{\rm UV}$), where  
$L_{\rm UV}$~=~1.5$\nu$$l_{\nu,2800}$ is a rough estimate of the total 
integrated 1216--3000  \AA \  UV luminosity, derived using the 2800 
\AA \ rest-frame luminosity from COMBO-17 $l_{\nu,2800}$. 
The factor of 1.5 used in converting the 2800 \AA  \ luminosity to  total UV luminosity 
accounts for  the UV spectral shape of a 100 Myr-old population with constant SFR.  
The factor of 2.2 corrects for the  light emitted longward of 3000 \AA \ and shortward 
of 1216 \AA. \  The SFR calibration  is  derived from P\'egase assuming a 100 Myr old stellar 
population with constant SFR  and a Chabrier (2003) IMF. 

The obscured SFR  can be calculated from  dust-reprocessed IR  emission using the 
expression  SFR$_{\rm IR}$~=~$9.8\times 10^{-11}$ $L_{\rm IR}$, where  
$L_{\rm IR}$ is the total IR luminosity (TIR) over  8--1000~$\mu$m  (Bell et al. 2007). 
 $L_{\rm IR}$  is  constructed from the
observed  24~$\mu$m  flux  (corresponding to rest-frame wavelengths of 19--13~$\mu$m over 
$z\sim$~0.24--0.80)  using the method outlined in  Papovich \& Bell (2002), based on 
an average  Sbc template from  the Devriendt \etal (1999) SED library. In converting
from  $L_{\rm IR}$  to SFR$_{\rm IR}$, Bell \etal (2007) assume that the bulk of the 
24~$\mu$m emission comes from SF, and not from AGN activity, based on 
the statistical result that less than 15\% of the total 24~$\mu$m 
emission at $z <$~1 is in X-ray luminous AGN (e.g.,  Silva \etal 2004;  Bell \etal 
2005; Franceschini \etal 2005; Brand \etal 2006).
Uncertainties in these SFR estimates are no less than a factor of 2 for individual 
galaxies  while  the systematic uncertainty in the overall SFR scale is 
likely to be less than a factor of 2  (Bell \etal 2007).


We investigate the star formation properties of the sample S1 of $\sim$~789 high mass   
($M \ge$~$2.5 \times  10^{10}$ $M_{\odot}$) and the sample S2 of $\sim$3698  
intermediate mass ($M \ge$~$1.0 \times 10^{9}$ $M_{\odot}$) galaxies. As described 
in $\S$~\ref{scdatas}, the high mass sample S1 is complete for both the red sequence and blue cloud, 
while the intermediate mass sample S2  is only complete for the blue cloud and suffers from 
incompleteness on the red sequence in the highest redshift bins. 
However, since most of the SFR density  originates 
from the blue cloud, this incompleteness does not have any major impact on the 
results.  Fig.~\ref{fsfrms} shows the UV-based SFR  plotted {\it versus}  
the stellar mass in each redshift bin. The  UV-based SFR ranges from  $\sim$~0.01 
to 25 $M_{\odot}$ yr$^{-1}$,  with most galaxies having a rate below 5  
$M_{\odot}$ yr$^{-1}$.

While it is desirable to use the  Spitzer 24~$\mu$m data in order to account 
for obscured star formation, only  $\sim$~24\%  ($\sim$~878 galaxies) of 
the 3698 galaxies in our intermediate mass sample have a Spitzer 24~$\mu$m  
detection, although over 86\%  of the sample  is covered 
by the Spitzer  observations down to  a  limiting flux of $\sim$~83~$\mu$ Jy.  
The detected galaxies  yield  a median ratio of (SFR$_{\rm IR}$/SFR$_{\rm UV}$) of  
$\sim$~3.6, indicative  of  a sustantial amount of obscured star formation. 
Three of the interacting  galaxies in the first redshift bin had anomalously high 
SFR$_{\rm UV +IR}$  ($\sim$~41, 18, and 15 $M_{\odot}$  yr$^{-1}$). Two of 
these turned to have  infrared spectra consistent with an AGN and were removed 
before computing the  IR-based SF properties  shown in Fig.~\ref{favesf} to Fig.~\ref{fcasf4}.

The  average UV-based SFR (based on 3698 galaxies)  and   UV+IR-based SFR (based on only 
the 876  galaxies with  24um detections)  are plotted in  the top 2 panels of 
Fig.~\ref{favesf} for three groups of intermediate mass galaxies: mergers, 
non-Interacting  E-Sd, and  non-Interacting Irr1. 
The corresponding plot for the high mass sample is  in Fig.~\ref{favesf2}.
It can be seen (Fig.~\ref{favesf} and Fig.~\ref{favesf2}) that over 
$z \sim$~0.24--0.80,  the average UV-based and  UV+IR-based SFR 
of  mergers  (in the phase where they are recognizable as mergers)
{\it are only modestly enhanced},  at best by a factor of a few, 
compared to the  non-interacting galaxies.  This result applies to both high 
mass ($M \ge$~$2.5 \times  10^{10}$ $M_{\odot}$) galaxies and intermediate mass
($M \ge$~$1.0 \times 10^{9}$ $M_{\odot}$) blue cloud galaxies.  A similar 
result is also found by 
Robaina et al. (in preparation) in high mass systems.  This modest 
enhancement is consistent with the  recent statistical 
study   of di Matteo et al. (2007),  who find from  numerical simulations 
of several hundred galaxy collisions that the maximum SFR  in galaxy mergers 
is typically only a factor of 2-3 larger than that of corresponding 
non-interacting  galaxies.
Their results suggest that the results of some early simulations  (e.g., Mihos \& 
Hernquist 1996; Hernquist \& Mihos 1995),  where mergers converted 
50 to 80 per cent of their original gas mass into stars, may not represent the typical 
situation at $z<1$.

In order to further test the robustness of our result, we used
 the stacking procedure described in Zheng \etal (2006) 
to get a more representative measure of the IR-based SFR for the following  
three groups of intermediate mass systems: mergers, non-interacting  E-Sd, 
and  non-interacting Irr1 galaxies. 
For every group, the individual galaxies were cross-correlated with the 
Spitzer 24~$\mu$m  catalog in order to identify detected and undetected objects.
Then the PSF-removed 24~$\mu$m images  for the undetected objects  were stacked,
and a mean flux was derived from the average/median stacked image.
An average 24um luminosity was determined from the individually-detected fluxes 
and individually-undetected fluxes estimated by stacking. The  3215 intermediate 
mass galaxies in the Spitzer field were used in this process, giving a more 
representative 24um luminosity  than the  mere 878 galaxies with detections.
A final uncertainty 
can be obtained by combining background error and bootstrap error in quadrature.
The IR-based SFR was estimated from the 24um luminosity using the procedure described
above, and combined with the  UV-based SFR to estimate the total SFR.
The average UV+IR-stacked SFR is plotted in the bottom panel of Fig.~\ref{favesf} and  Fig.~\ref{favesf2}: again, only a modest enhancement 
is seen in the average SFR of  mergers (in the 
phase where they are recognizable as mergers), compared to  non-interacting galaxies.

\subsection{The contribution of interacting galaxies to the cosmic 
SFR density over  the last 7 Gyr}\label{scsfr2}

Over the last 8 Gyr  since  $z\sim$~1, the cosmic SFR density is observed to  
decline by a factor of 4 to 10 (e.g., Lilly et al. 1996; Ellis et al 1996;  Flores \etal 1999; 
Haarsma \etal 2000; Hopkins 2004; P{\'e}rez-Gonz{\'a}lez  et al.\ 2005;  Le Floc'h et  al.\ 2005).
Earlier GEMS studies  by Wolf \etal (2005)  and Bell \etal (2005)   over a 0.6 Gyr interval
($z \sim$~0.65--0.75 or $T_{\rm b}\sim$~6.2--6.8 Gyr)  showed that the  UV and IR 
luminosity density over this interval  are  dominated by non-interacting galaxies.
Here, we extend the  earlier GEMS studies to  cover a  six-fold larger time interval 
of 4 Gyr ($z\sim$~0.24--0.80 or $T_{\rm b}\sim$~3--7 Gyr), and set quantitative limits 
on  the  contribution of  merging systems to the UV-based and  
UV+IR-based SFR density. We use  the sample S2 of $\sim$3698  
intermediate mass ($M \ge$~$1.0 \times 10^{9}$ $M_{\odot}$) galaxies.
Our study also  complements the  IR-based  studies by  Hammer et al. (2005;  l95 
galaxies at $z>0.4$ ) and  Melbourne et al. (2005; $\sim$~800 galaxies) 
 in terms of  sample size or/and   SFR indicators.

Fig.~\ref{fsfhis1} shows the SFR density for intermediate mass  mergers, 
non-interacting  E-Sd galaxies, and  non-interacting Irr1 galaxies over  $z\sim$~0.24--0.80.
The top panel shows the   UV-based SFR density from the full sample. The 
middle panel show the UV+IR-based SFR density  from the 878 galaxies with  
individual  24um detections. 
Finally, the bottom panel shows the UV+IR-stacked SFR 
density determined via the stacking of 3215 galaxies with Spitzer coverage, as outlined
in $\S$~\ref{scsfr1}.
In all three panels, one finds that {\it interacting galaxies only account for 
a small fraction ($<$ 30\%) of the cosmic SFR density over $z\sim$~0.24--0.80, corresponding to  
lookback times of 3--7 Gyr (Fig.~\ref{fsfhis1}).}
The same results hold for the sample of high mass ($M \ge$~$2.5 \times  10^{10}$ $M_{\odot}$) 
galaxies, as illustrated in Fig.~\ref{fsfhis2}.

Thus, our results  suggest   that the behavior  of the cosmic SFR density
over $z \sim$~0.24--0.80 is  predominantly shaped by non-interacting galaxies. 
A similar result is found  by Lotz et al. (2008).
Our result is a direct consequence of the fact that  the merger 
fraction $f$ (Table~\ref{tvclas2}; $\S$~\ref{scmfrac}), as well as the
enhancement in the average  SFR from interactions ($\S$~\ref{scsfr1}), are 
both modest.
Our results   agree remarkably well with  models for the self-regulated growth of 
supermassive black holes in mergers involving gas-rich galaxies (Hopkins \etal 2006). 
These models  predict that galaxy mergers contribute only $\sim$~20\%  of the SFR 
density  out to  $z\sim$~1, and even out to  $z\sim$~2.

It is legitimate to ask whether the results hold despite the uncertainties 
in identifying  mergers. 
We first note that based on the tests of $\S$~\ref{scvctes}, we have already
included a large fractional error term on $f$ to account for 
the binomial standard deviation, the dispersion  between classifiers, and 
the  effect of moderate bandpass shifting, and surface brightness  dimming.
Therefore, the results presented here already  take into account 
at least some of these sources of uncertainties.

Another source of uncertainty might be that some of the galaxies, which we 
have classified as  non-interacting Irr1 under the assumption that  
their small-scale asymmetries   are likely caused by SF rather than 
interactions ($\S$~\ref{scvis2}), may  be  borderline cases of mergers or 
interacting systems. However, it can be seen from Fig.~\ref{fsfhis1}, that even if we were
to add the SFR density of {\it all} the   non-interacting Irr1 galaxies 
to that of the mergers, the sum would still be significantly lower than the 
contribution of  non-interacting E to Sd galaxies. Thus, the results would
be largely unchanged.

Another test is  to repeat the analyses using the CAS merger criterion 
($A >$~0.35 and $A > S$) to identify  mergers. 
The limited recovery rate (50\% to 73\%) and large level of  contamination 
impacting the CAS criterion  ($\S$~\ref{sccas1}) make it more difficult to 
interpret the SF properties of systems identified 
as mergers or non-interacting  with CAS.
For both the intermediate mass sample (Fig.~\ref{fcasf1}) and 
the high mass sample  (Fig.~\ref{fcasf2}),  the average SFR of  
CAS mergers is only  modestly enhanced 
compared to CAS non-interacting galaxies, in agreement with the 
results from $\S$~\ref{scsfr1}.

Furthermore, for the intermediate mass sample, Fig.~\ref{fcasf3}   shows that 
CAS mergers contribute only 16\% to 33\% of the UV SFR density and 
22\% to 38\% of the UV+IR SFR density.   While the upper limits of these 
values are slightly  higher than those based on the visual types, it is 
nonetheless reassuring  that  CAS non-interacting galaxies dominate the
SFR density. Similar results hold for the sample of 
high mass ($M \ge$~$2.5 \times  10^{10}$ $M_{\odot}$) galaxies, 
as illustrated in Fig.~\ref{fcasf4}.


For intermediate mass ($M \ge$~$1.0 \times 10^{9}$ $M_{\odot}$)  galaxies, 
we find  that the  cosmic  SFR density  declines by a factor of $\sim$~3   from 
$z \sim$~0.80 to 0.24  (lookback time $\sim$~7 to 3 Gyr).
Since  non-interacting galaxies dominate the cosmic SFR density in every redshift 
bin, it follows that this decline
is largely the result of a shutdown in the SF of   non-interacting 
galaxies.
The question of what drives this shutdown will be addressed in detail in a future paper,
and is only considered briefly here. 
One possibility is the depletion of the internal  cold gas supply of galaxies by 
star formation, or the reduction in the accretion rate of gas 
from cosmological filaments.  Future facilities like ALMA will 
be instrumental in exploring this issue further.
Another related possibility  is that over time, most of the SFR is shifting to 
lower stellar masses. 
High mass systems are associated with a lower SSFR  (Cowie 
\etal  1996; Brinchmann \etal 2004;  Brinchmann \& Ellis 2000;  
Fontana et al. 2003;  Bauer \etal 2005; Zheng \etal 2007; Fig. 1 of Noeske  et al. 2007a),  
consistent with the idea that they  have experienced the  bulk of their stellar mass  growth 
at  earlier epochs ($z>1$). In staged SF models (Noeske et al. 2007b)  
the SF history of low mass systems is consistent with  exponential SF models 
associated with a late onset and a long duration.

\section{Summary and Conclusions}\label{scsumm}

We have performed a comprehensive observational estimate of the galaxy  merger fraction 
over  $z \sim$~0.24--0.80 (lookback times of 3--7 Gyr), and explored  the 
impact of mergers on the star formation  of  galaxies over
this interval. Our study is based on  $HST$ ACS, COMBO-17, and Spitzer 24~$\mu$m data 
from the GEMS survey. 
We use a large sample of $\sim$~3600 ($M \ge$~$1 \times  10^{9}$ $M_{\odot}$)  galaxies 
and  $\sim$~790 high mass  ($M \ge$~$2.5 \times 10^{10}$ $M_{\odot}$) galaxies 
($\S$~\ref{scdatas}). We primarily identify mergers  using a visual classification 
system, which is based on visual morphologies, spectrophotometric redshifts, and stellar
masses  ($\S$~\ref{scvis1} to $\S$~\ref{scvis2}), and  identifies 
systems that show evidence of having  experienced  a merger of mass ratio  
$M1/M2 >$~1/10 within the last visibility timescale. 
While many earlier studies focused only on major mergers  (defined as mergers with $M1/M2 >$ 1/4),  
we also attempt to constrain the frequency of minor mergers  (defined as mergers 
with 1/10 $ < M1/M2 \le$ 1/4), since they dominate the merger rates in $\Lambda$CDM models.
Below is a summary of our  results: 

\begin{enumerate}

\item
For the high mass  ($M \ge$~$2.5 \times  10^{10}$ $M_{\odot}$)  sample of 
$\sim$~790 galaxies, which is  complete on both the blue cloud and red sequence, 
we find the following. {\it \rm The fraction $f$  of  visually-classified 
systems that show evidence of a recent merger of mass ratio $>$~1/10,
does not show strong evolution over lookback times of 3--7 Gyr, and 
ranges from  9\% $\pm$~5\% at 
$z \sim$~0.24--0.34, to 8\% $\pm$~2\%  at $z \sim$~0.60--0.80 
(Table~\ref{tvclas1}; Fig~\ref{fvctst1}).}
These mergers are further subdivided into three categories: clear `major merger',  
clear `minor merger', and  ambiguous  `major or minor merger' cases.
The first two classes are used to set  lower limits on the major and minor 
merger fraction. 
The  lower limit on the major  merger  fraction, determined in this 
way, ranges from  1.1\%  to 3.5\%  over $z\sim$~0.24--0.80  (Table~\ref{tvclas1}).
The corresponding lower limit on the minor 
merger fraction ranges  from  3.6\% to 7.5\%.  This is the first, albeit approximate, 
empirical estimate of the frequency of  minor mergers over the last 7 Gyr.

For an assumed visibility timescale of $\sim$~0.5 Gyr, it follows 
that {\it \rm  over $T_{\rm b}\sim$~3--7 Gyr, $\sim$~68\% of high mass systems have 
undergone a merger of mass ratio $>$~1/10, with $\sim$~16\%, 45\%, and 7\%  of 
these corresponding respectively to major, minor, and ambiguous `major or minor' 
mergers.} The corresponding merger rate $R$ is a few $\times 10^{-4}$ galaxies 
Gyr$^{-1}$ Mpc$^{-3}$.

\item
At intermediate masses ($M \ge$~$1 \times  10^{9}$ $M_{\odot}$), 
we are only complete in mass for the blue cloud.
{\it \rm
Among $\sim$~2840 blue cloud galaxies of mass $M \ge$~$1.0 \times 10^{9}$ $M_{\odot}$, 
the fraction of visually-classified systems that  show evidence of a recent merger of 
mass ratio $>$~1/10, ranges from  
7\% $\pm$~2\% to 15\% $\pm$~5\%  over $z \sim$~0.24--0.80 (Table~\ref{tvclas2}).}

For an assumed visibility time of $\sim$~0.5 Gyr, we estimate  that  on average, 
over $T_{\rm b}\sim$~3--7 Gyr, 84\% of  intermediate mass blue cloud galaxies 
have undergone  a merger  of mass ratio~$>$~1/10, with $\sim$~5\%,  22\%, and 57\% 
corresponding respectively to major, minor, and ambiguous `major or minor' mergers.
The corresponding  merger rate $R$ ranges from  $8 \times 10^{-4}$  to 
 $1 \times 10^{-3}$  galaxies  Gyr$^{-1}$ Mpc$^{-3}$ .

\item
We  compare our visual mergers to those identified using the widely used 
CAS merger criterion  ($A >$~0.35 and $A > S$), based 
on  CAS  asymmetry $A$ and clumpiness $S$ parameters ($\S$~\ref{scasts}).
The  merger fraction based on  the CAS merger criterion 
agrees within a factor of two with  the visually based merger fraction for high mass  
($M \ge$~$2.5 \times  10^{10}$ $M_{\odot}$) galaxies. However,  for  
intermediate mass ($M \ge$~$1 \times  10^{9}$ $M_{\odot}$) galaxies, CAS  can overestimate
the merger fraction at $z>$~0.5 by a factor $\sim$~3.  In effect, over $z \sim$~0.24--0.80, 
$\sim$~50\% to 70\% of  the galaxies visually classified as mergers satisfy the  
CAS criterion, but the latter also picks up a dominant  number of non-interacting  dusty, 
star-forming galaxies (Fig.~\ref{fcasvc} and Fig.~\ref{fcasvc2}).  These non-interacting 
systems  make up as much as $\sim$~45\% to 80\% of the systems picked up the CAS criterion. 
We thus conclude that the traditional CAS merger criterion  is ill-suited  for use  on  
$HST$ $V$-band images at $z >$~0.5  where the rest frame wavelength falls 
below  $\lambda < 4000$ \AA,  particularly  in the case of intermediate
mass galaxies with significant SF, gas, and dust. Modified CAS criteria in the 
near-UV based on morphological k-corrections  (Taylor et al. 2007) might alleviate 
this problem.

\item
We compare our empirical merger fraction $f$ and merger rate $R$ for high mass 
($M \ge$~$2.5 \times  10^{10}$ $M_{\odot}$) galaxies 
to predictions from different  $\Lambda$CDM-based simulations  of galaxy evolution,
including  the halo occupation distribution (HOD) models of Hopkins \etal (2007); 
semi-analytic models (SAMs) of Somerville \etal (2008), 
Bower et al. (2006), and Khochfar \&  Silk (2006); and smoothed 
particle hydrodynamics (SPH) cosmological simulations from Maller \etal (2006) 
with a corrected stellar mass function (see $\S$~\ref{scmodel}). 
To our knowledge, such extensive comparisons have not been attempted to date, and 
are long overdue.
{\it 
We find  qualitative  agreement  between the observations and models, such 
that the (major+minor) merger fraction (Fig.~\ref{fmodelf}) and 
merger rate (Fig.~\ref{fmodelr}) from different models bracket
the corresponding empirical estimates and show a factor of five  dispersion.}
One can now anticipate that in the near future, improvements in both the observational 
estimates and model predictions will start to rule out certain merger scenarios and refine
our understanding of the merger history of galaxies.

\item
We explore the impact of galaxy mergers on  the SF activity of  galaxies since $z<0.8$. 
In the sample of $\sim$~789 high mass ($M \ge$~$2.5 \times  10^{10}$ $M_{\odot}$) galaxies, 
as well as the sample of $\sim$~3600 intermediate mass ($M \ge$~$1.0 \times 10^{9}$ $M_{\odot}$) 
galaxies, we find that 
{\it 
the average SFR  of visibly merging galaxies is  
only modestly enhanced compared to non-interacting galaxies over $z \sim$~0.24--0.80 
(Fig.~\ref{favesf})}. 
This result is found for SFRs based on UV, UV+IR, as well as UV+stacked-IR data.
This modest enhancement is consistent with  the  empirical results of Robaina 
et al. (in preparation),  and the  recent  statistical study  of di 
Matteo et al. (2007) based on numerical simulations of several hundreds of 
galaxy collisions.

\item
Among  both high mass and intermediate mass  galaxies, 
our results  of a modest merger fraction $f$ and a modest 
enhancement in the average SFR due to mergers,  culminate in our finding that 
{\it visibly merging systems only account for a small fraction ($<$ 30\%) of 
the cosmic SFR density over lookback times of $\sim$~3--7 Gyr ($z\sim$~0.24--0.80; 
Fig.~\ref{fsfhis1} and Fig.~\ref{fsfhis2})}. 
Our result complements  that of Wolf et al. (2005) 
over a smaller lookback time interval of  $\sim$~6.2--6.8 Gyr. In effect, our result 
suggests that {\it the behavior  of the cosmic SFR density over the  last 7 Gyr is  
predominantly shaped by non-interacting galaxies, rather than  mergers and interacting 
galaxies.} 
We suggest that our observed decline in the  cosmic  SFR density  by 
a factor of $\sim$~3  since $z \sim$~0.80  is largely the result of a  shutdown 
in the SF of relatively non-interacting galaxies. This shutdown may be driven by the 
depletion of the internal  cold gas supply of galaxies, the reduction in the accretion 
rate of gas from cosmological filaments, and  the transition of  SF activity to lower 
mass systems. 


\end{enumerate}

\acknowledgments
S. J. thanks  Phil Hopkins, Sadegh Khochfar, Andrew Benson, Andi Burkert, 
and Ari Maller for useful discussions.
S. J. acknowledges support from the National Aeronautics and Space
Administration (NASA) LTSA grant NAG5-13063, NSF grant AST-0607748, 
and $HST$ grants G0-10395 from STScI, which is operated by 
AURA, Inc., for NASA, under NAS5-26555. 
E.\ F.\ B., K. J., and A.\ R.\ R.\ acknowledge support from the Deutsche 
Forschungsgemeinschaft through the Emmy Noether Programme.
X.\ Z.\ Z. acknowledges support from NSFC through grant 10773030,10833006.
D. H. M. acknowledges support from NASA  LTSA grant NAG5-13102 issued
through the office of Space Science.
C. Y. P. is grateful for support provided through STScI and NRC-HIA 
Fellowship. C. W. acknowledges support from  an STFC Advanced Fellowship. 
Support for  GEMS was provided by NASA through $HST$ grant GO-9500 from
the Space Telescope Science Institute, which is operated by 
AURA, Inc., for NASA, under NAS5-26555.
This research has made use of NASA's Astrophysics Data System Service.


{}


\clearpage
\setcounter{figure}{0}
\begin{figure}
\epsscale{0.9}
\plotone{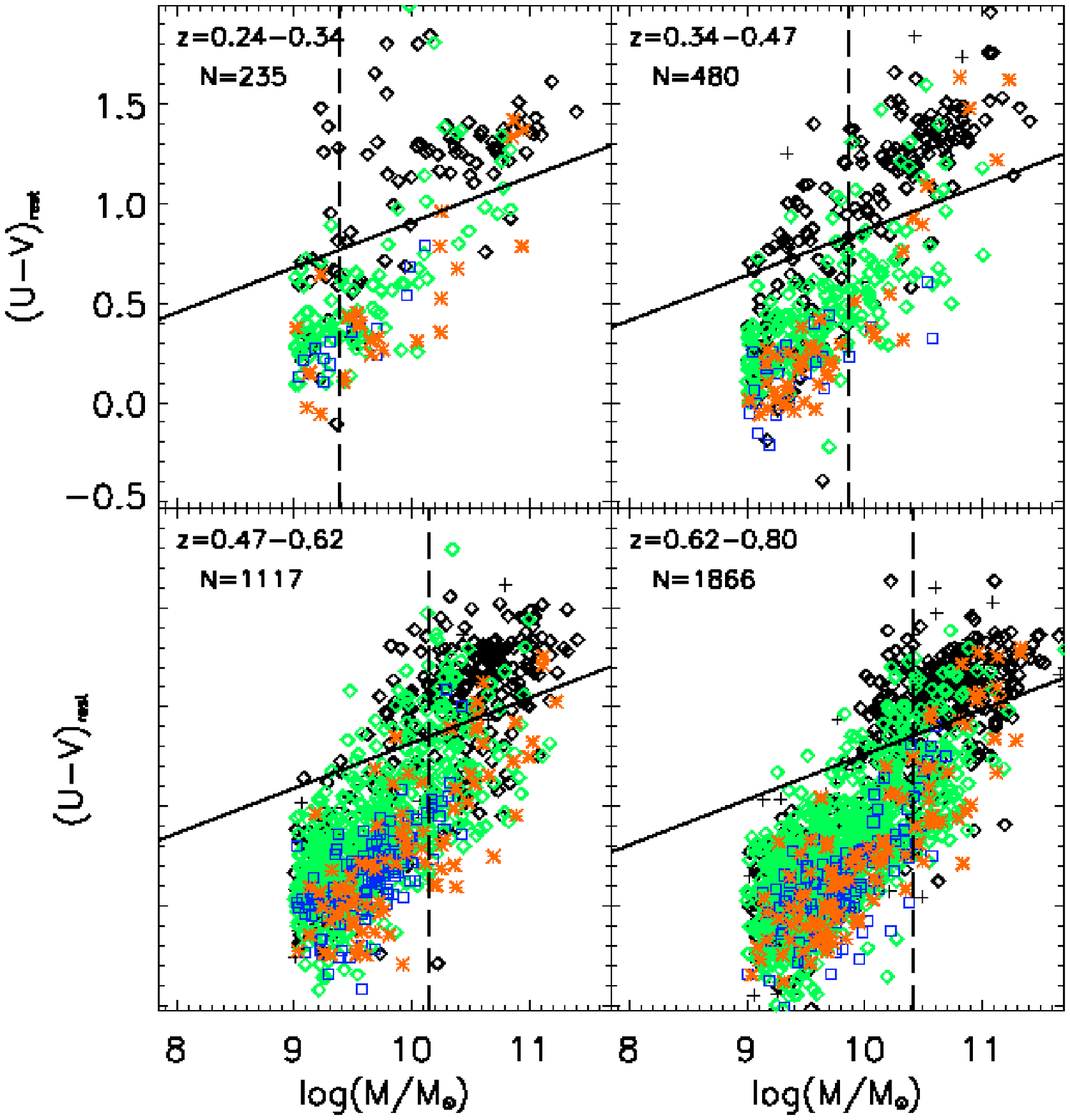}
\caption{
The rest-frame $U-V$ color is plotted {\it versus} the stellar mass 
over $z \sim$~0.24--0.80 for the sample S2 of $\sim$~3698 galaxies with  
$M \ge$~$1 \times  10^{9}$ $M_{\odot}$. 
The four panels denote the four redshift bins, which span 1 Gyr each, and 
cumulatively cover  the interval $z \sim$~0.24--0.80 ($T_{\rm back}$~$\sim$~3--7 Gyr). 
$N$ denotes the number of galaxies in each bin. 
The diagonal line  marks the separation of the red sequence 
galaxies  and the blue cloud galaxies at the average redshift $z_{\rm ave}$ 
of the bin. The vertical lines  marks the mass completeness 
limit (Borch \etal 2006) for the red sequence galaxies, while 
the blue cloud galaxies are complete well below this mass.  
For the mass range $M \ge$~$1 \times  10^{9}$ $M_{\odot}$, 
the blue cloud is complete in our sample out to  $z\sim$~0.80, while the  red 
sequence  is incomplete in the higher redshift bins.
Galaxies are coded according to their visual type (VT) in the F606W band: 
`Mergers' (orange stars), `Non-Interacting E+S0+Sa' (black diamonds), 
`Non-Interacting Sb-Sc + Sd' (green diamonds), and  
`Non-Interacting Irr1' (blue squares).
\label{fuvmas}} 
\end{figure}

\clearpage
\begin{figure}
\epsscale{0.80}
\plotone{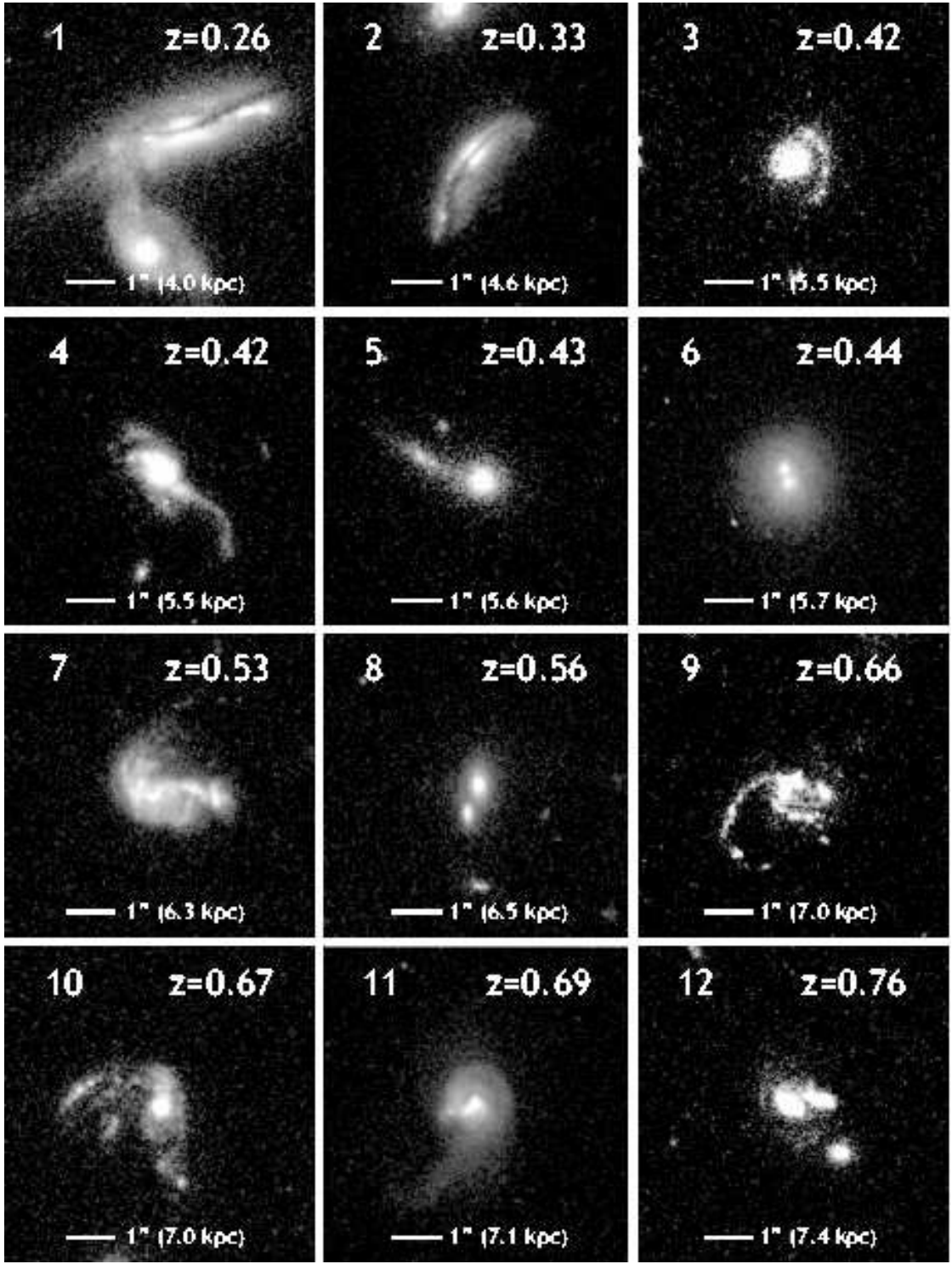}
\caption{
This montage show examples of mergers, namely  systems that 
show evidence of having  experienced  a merger of  mass ratio $>$~1/10  
within the last visibility timescale, as described in $\S$~\ref{scvis1}. 
The evidence includes morphological distortions 
similar to those seen in simulations of mergers of mass ratio $>$~1/10, 
such as  multiple nuclei (e.g., case 6, 8); components (e.g., case 12) connected  
by a bridge or common envelope; tidal tails and asymmetric features 
(e.g., cases 3, 4, 5, 9, 11);  and warped disks (e.g., case 2).
Systems classified as `Int-1' mergers primarily represent advanced mergers, 
which appear as single systems in ACS images (e.g., cases 2, 3, 4, 5, 6, 
7, 9, 10  11, and 12). Conversely, `Int-2' mergers (e.g., case 1) primarily 
represent  young mergers, which appear as very close pairs of overlapping 
galaxies  in ACS images.
The mergers can be divided into 3 groups: clear major mergers (cases 1, 6, 12),
clear minor mergers (cases 2, 9), and ambiguous `major or minor mergers' 
(cases 3, 4, 5, 7, 8, 10, 11).
%
\label{fpec1}}
\end{figure}

\clearpage
\begin{figure}
\epsscale{0.9}
\plotone{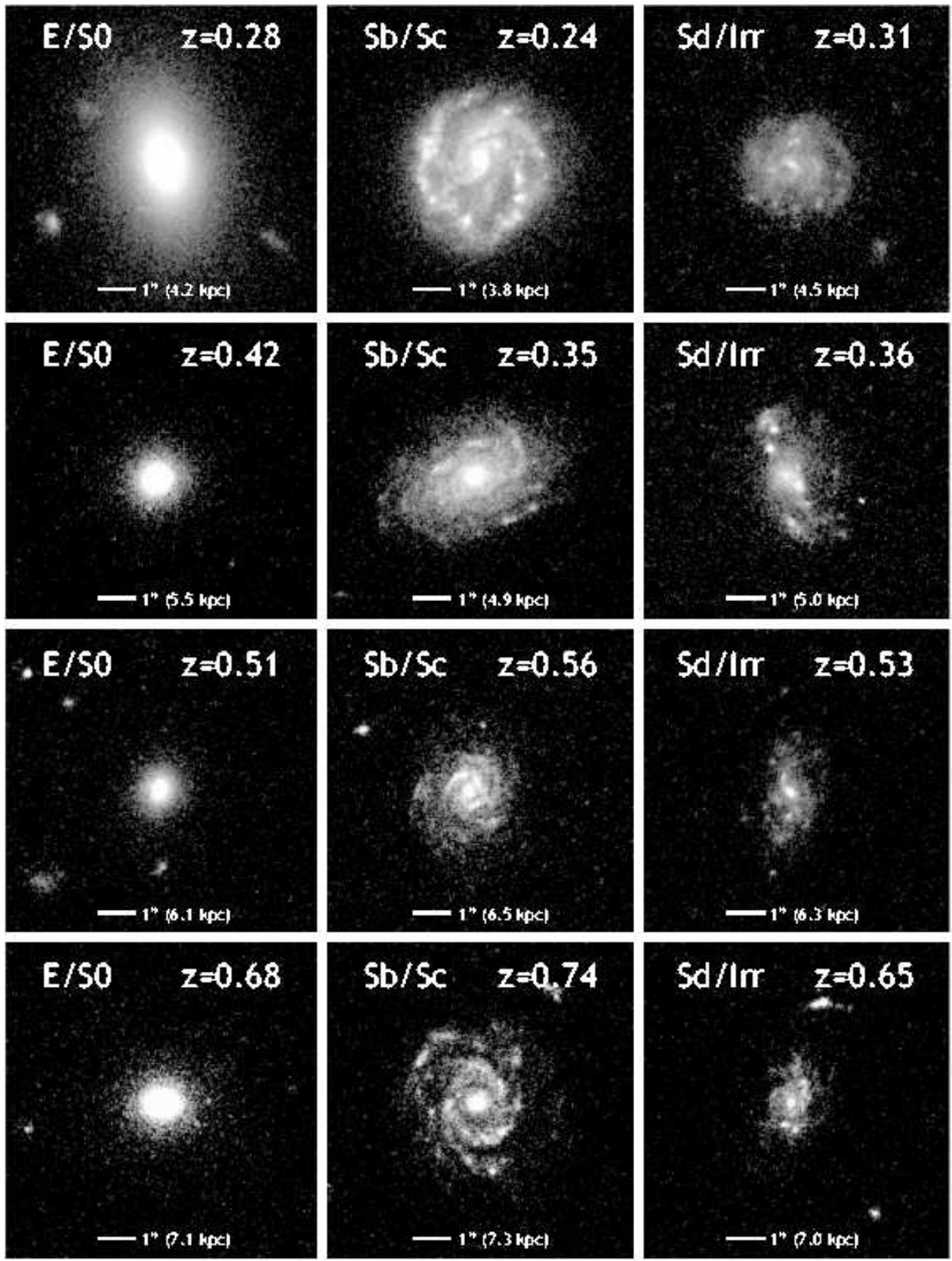}
\caption{
This montage shows  examples of galaxies classified as  non-Interacting E-to-Sd and  
non-interacting Irr1 galaxies, according to the criteria in $\S$~\ref{scvis2}. 
Within  the broad class of non-interacting E-to-Sd, galaxies have Hubble types E, S0, Sa,
Sb--Sc, and Sd, as shown in Fig.~\ref{fsersi}.
\label{fmonta}}
\end{figure}

\clearpage
\begin{figure}
\epsscale{1.0}
\plotone{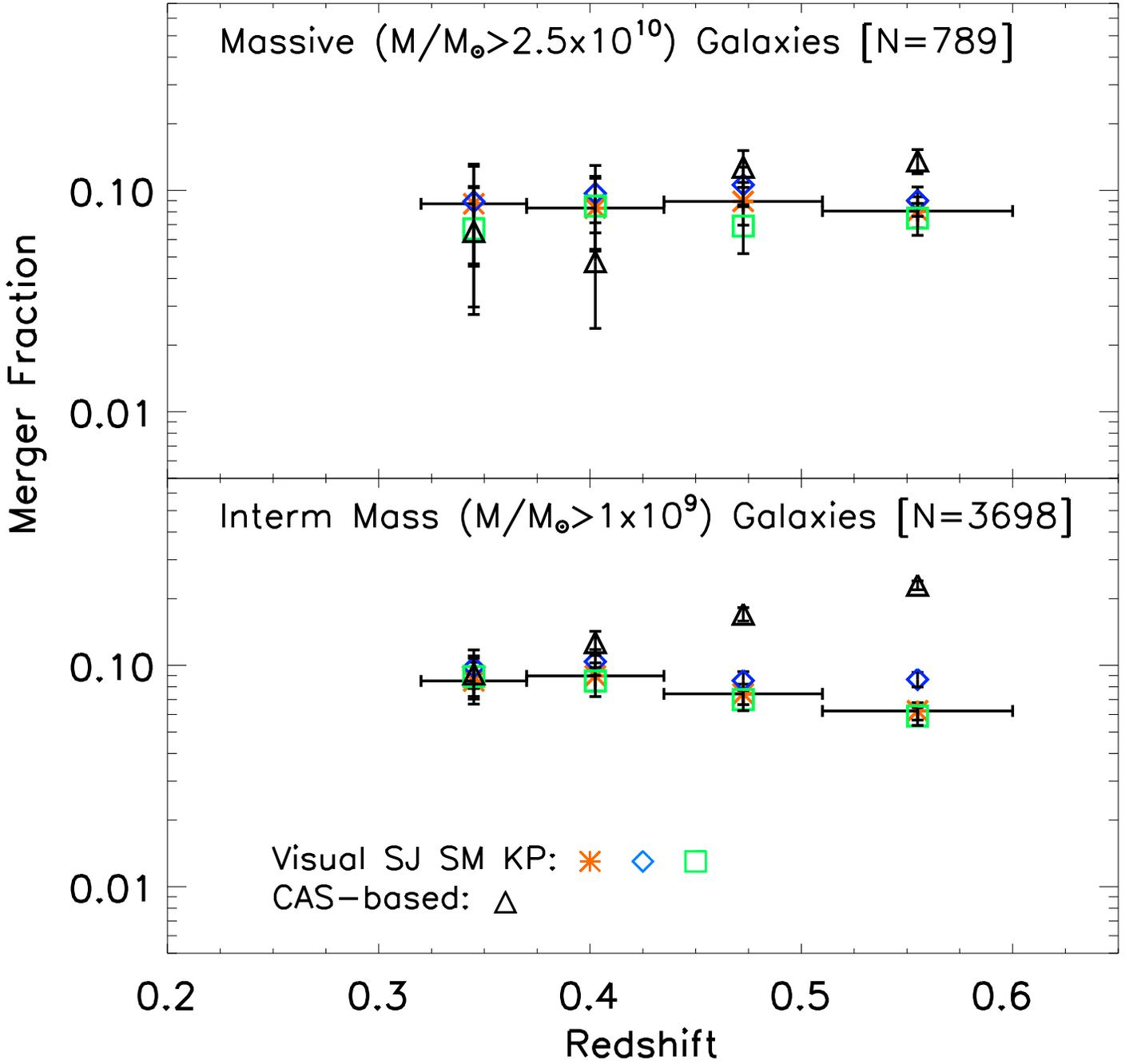}
\caption{
This figure compares the  the merger fraction ($f$)  based on visual classification 
by 3 classifiers (SJ, SM, KP), 
to  merger fraction ($f_{\rm CAS}$) that would be
obtained using the CAS criterion  ($A>$~0.35 and $A>S$).  
The results are shown for  both high mass  ($M \ge$~$2.5 \times  10^{10}$ $M_{\odot}$; 
top panel) and  intermediate mass ($M \ge$~$1 \times  10^{9}$ $M_{\odot}$; bottom 
panel) sample.  The plotted error bar for $f$, at this stage,  only includes  
the binomial term [$f$(1-$f$)/$N$]$^{1/2}$, for each bin of size $N$. 
The same trend is seen for all 3 classifiers and the maximum
spread $\delta_{\rm f}$/$f$  in the 4 bins is $\sim$ 26\%.
$f_{\rm CAS}$ agrees within a factor of two with  the visually based $f$ 
merger fraction for high mass   galaxies. However,  for  
intermediate mass galaxies, CAS  can overestimate
the merger fraction at $z>$~0.5 by a factor $\sim$~3, as it 
picks up a significant number of non-interacting  dusty, star-forming galaxies 
(see $\S$~\ref{scasts}).
\label{fvctst1}}
\end{figure}

\clearpage
\begin{figure}
\epsscale{1.0}
\plotone{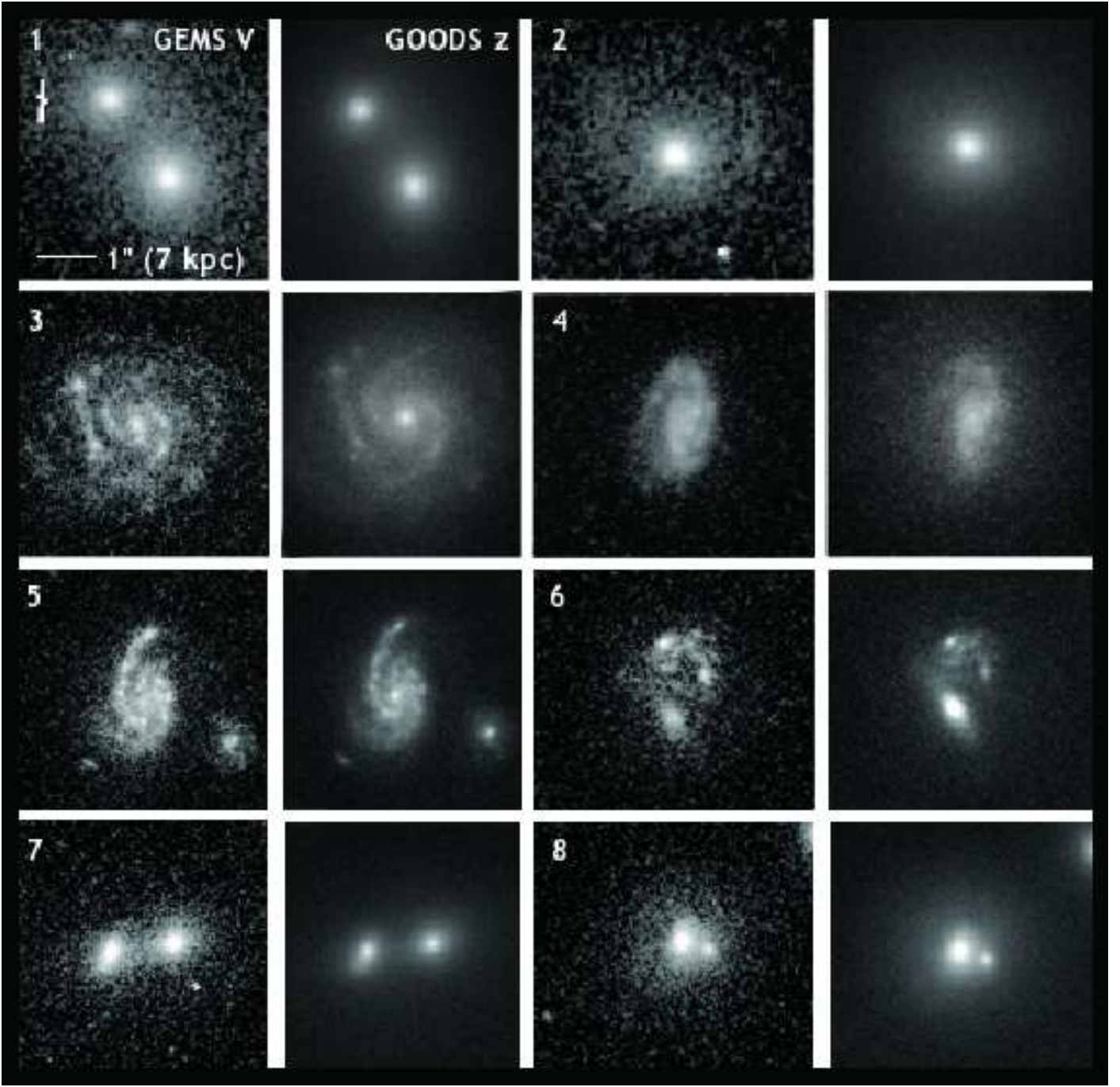}
\caption{
This montage illustrates a test for bandpass shift 
and surface brightness dimming. It  compares the bluer shallower 
GEMS F606W images ($V$ band; pivot $\lambda \sim$~5915 \AA) and deeper 
redder GOODS F850LP  ($z$ band; pivot $\lambda \sim$~9103 \AA) images 
of typical interacting and non-interacting 
galaxies in the last redshift bin ($z\sim$~0.60 to 0.80), where 
bandpass shift and surface brightness dimming are 
expected to  be most severe. 
In this redshift bin, the rest-frame wavelength traced by the 
GEMS images shift from optical to violet/near-UV  (3700 \AA \ to 
3290 \AA).  However, while the GOODS images have higher S/N, and trace 
redder older  stars, they do not reveal dramatically different morphologies 
from from those  in the GEMS F606W images. Cases 1, 5, 6, 7, and 8 are 
interacting systems. A statistical analysis is shown in Table~\ref{tgoods1}.
\label{fvctst2}}
\end{figure}

\clearpage
\begin{figure}
\epsscale{1.0}
\plotone{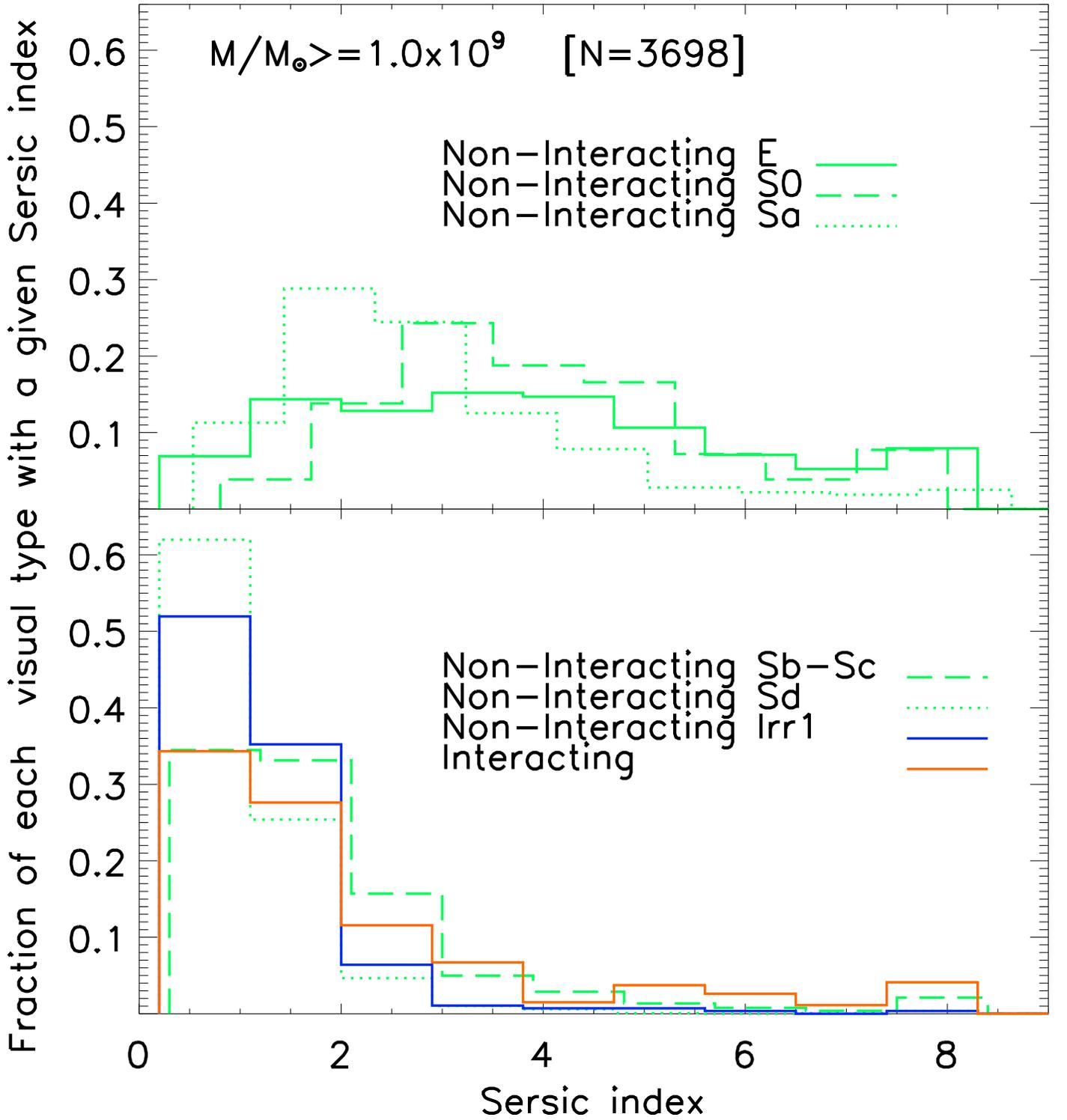}
\caption{
For intermediate mass ($M \ge$~$1 \times  10^{9}$ $M_{\odot}$) galaxies, 
the distribution of  S\'ersic index $n$ from single-component  
S\'ersic fits  is plotted for  mergers, non-interacting  Irr1, and 
non-interacting E-to-Sd systems.  The latter class  is further 
subdivided as  E, S0, Sa, Sb-Sc, and Sd.
The majority  of systems visually classified  as non-interacting Sb-Sc, 
Sd, and Irr1  have  $n <$~2.5, as expected for disk-dominated systems. 
Most of the systems visually typed as Sa have $n<$~4. Those galaxies typed 
as E and S0  primarily have $n>$~3, but some have low $n$, reflecting the 
inherent difficulty in separating E, S0, and Sa at intermediate
redshifts.
\label{fsersi}}
\end{figure}

\clearpage
\begin{figure}
\epsscale{1.0}
\plotone{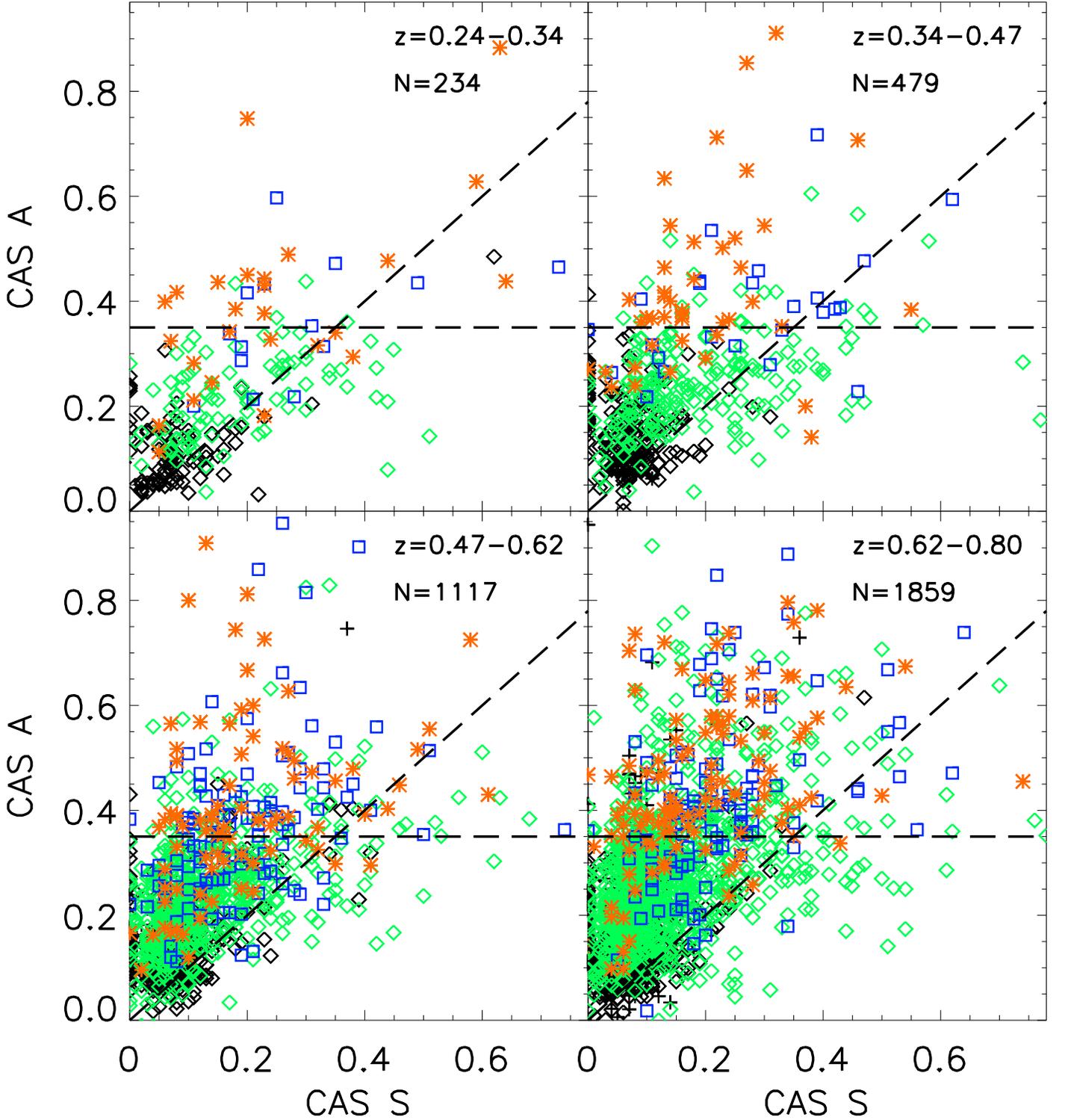}
\caption{
The CAS asymmetry $A$ and  clumpiness $S$ parameters are plotted  for  
intermediate mass ($M \ge$~$1 \times  10^{9}$ $M_{\odot}$) galaxies in the  
four redshift bins of Fig.~\ref{fuvmas}, using the same color coding. 
Galaxies  satisfying the CAS  criterion  ($A >$~0.35 and $A > S$) lie in the upper 
left hand corner, bracketed by the $A$~=~$S$ diagonal line and the $A$~=~0.35 horizontal 
line. The CAS criterion captures a fair fraction  of the  galaxy mergers, 
but it also picks up ``contaminants'' in the form of   non-interacting galaxies.  
This is  further illustrated 
in  Fig.~\ref{fcasvc2}.
\label{fcasvc}}
\end{figure}

\begin{figure}
\epsscale{1.0}
\plotone{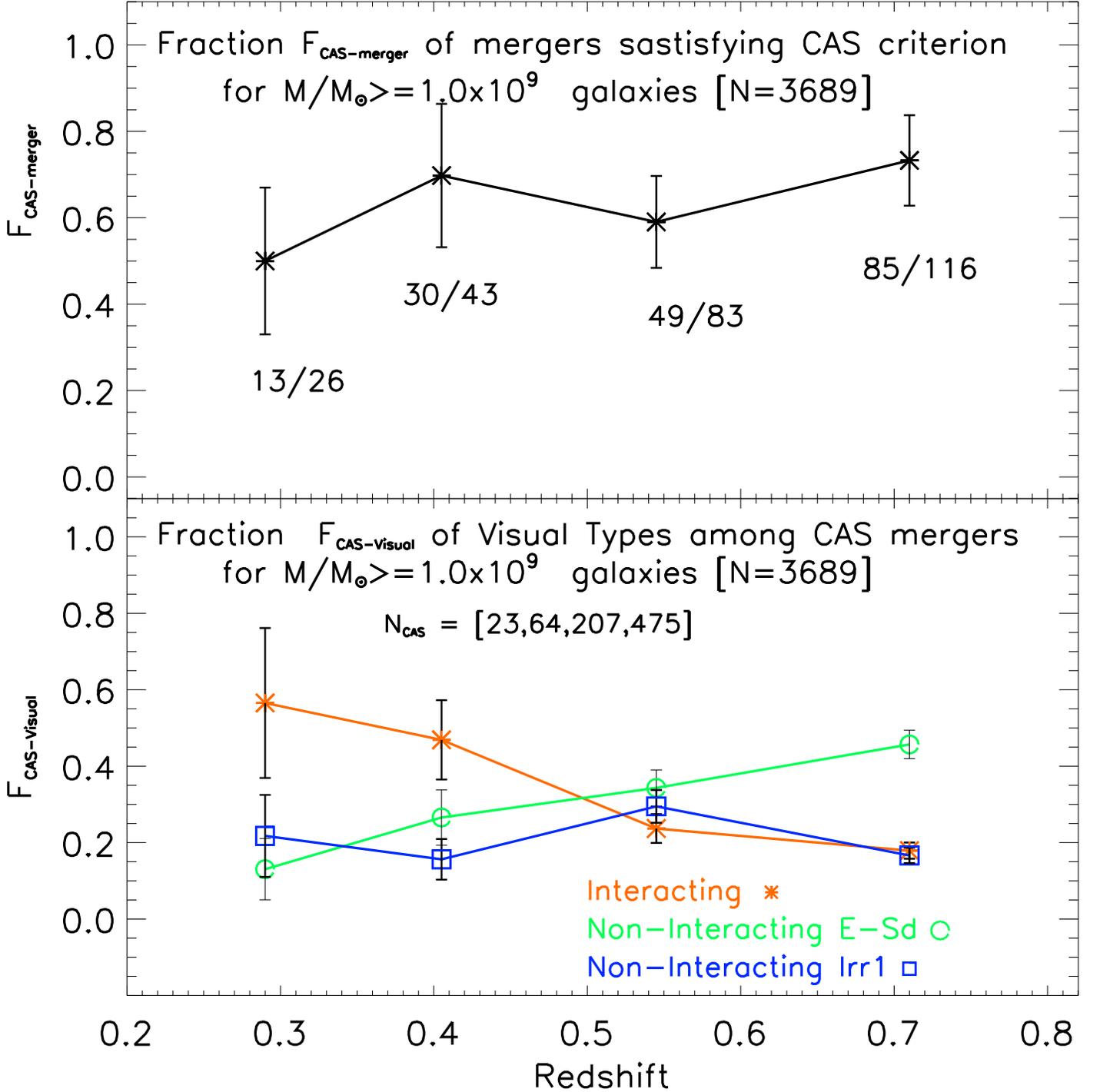}
\caption{
For galaxies with $M \ge$~$1.0 \times 10^{9}$ $M_{\odot}$, the top panel shows  that 
the fraction $F_{\rm CAS-merger}$ of visually-classified mergers, which 
satisfy the CAS  criterion  is $\sim$~ 50\% to 70\%  across the four redshift bins.
The bottom panel shows the degree to which non-interacting galaxies contaminate the 
systems  picked up by CAS. $N_{\rm CAS}$ represents the total number of galaxies, which satisfy 
the CAS criterion and are considered as ``CAS mergers'' across the four redshift bins.  
The fraction $F_{\rm CAS-visual}$  of different visual types among these 
 ``CAS mergers'' is plotted on the y-axis. At $z>0.5$, the vast majority 
(44\% to 80\%) of the systems considered as mergers by the CAS criterion turn out to be 
 non-interacting [E-Sd and Irr1] systems. 
\label{fcasvc2}}
\end{figure}

\begin{figure}
\epsscale{0.9}
\plotone{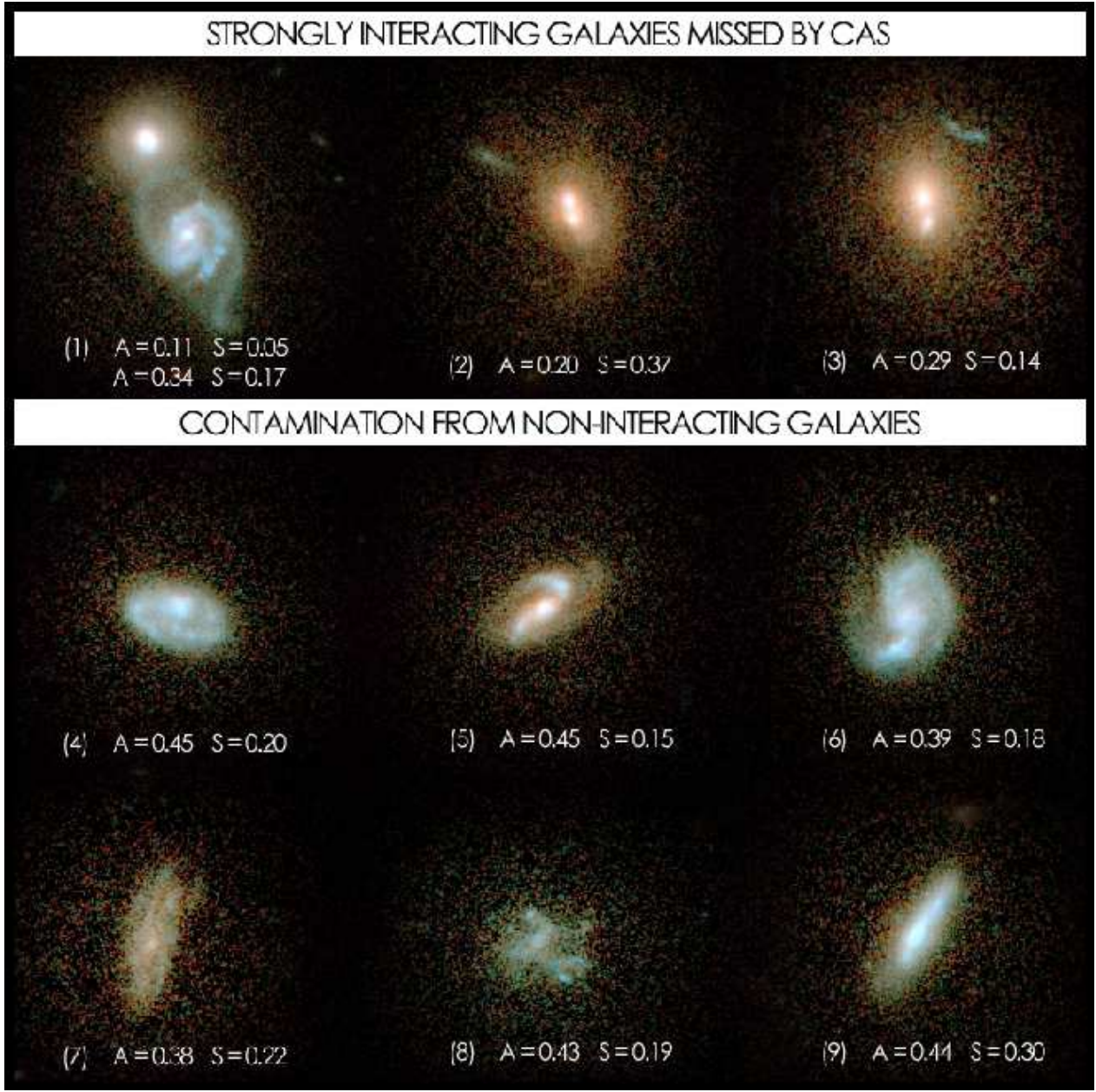}
\caption{
The montage shows typical systems where the  CAS  criterion  ($A >$~0.35 and $A > S$) fails 
(see $\S$~\ref{scasts} for details).
Cases  1-3 are systems, which are visually classified as mergers, but are 
missed by the CAS criterion.  They include systems with  tidal debris (e.g., case 3) 
that may contribute  less than $35\%$  of the total light;  systems  with close 
double nuclei (e.g., case 2) where CAS might refine the center to be between  
the two nuclei,  thereby leading to a  low $A < 0.35$; and  pairs of fairly symmetric 
galaxies whose members have similar redshifts within the spectrophotometric error, 
appear connected via weak tidal features, and have a stellar mass ratio $M1/M2 > 1/10$ 
(e.g., case 1 where $M1/M2 \sim$~0.25).
Conversely, cases  4-9 are systems, which are visually classified as non-interacting 
galaxies, but are  picked  by the CAS criterion. 
They include non-interacting, actively star-forming systems with 
small-scale asymmetries in the optical blue light  (cases 4 and 6);
systems where $A$ is high due to the absence of a clear center (case 8) or 
due to the center being blocked by dust (case 4, 9); 
edge-on systems and compact systems, where the light profile is steep such that
small centering inaccuracies lead to large $A$ (case 9).
\label{fcasvc3}}
\end{figure}

\clearpage
\begin{figure}
\epsscale{1.0}
\plotone{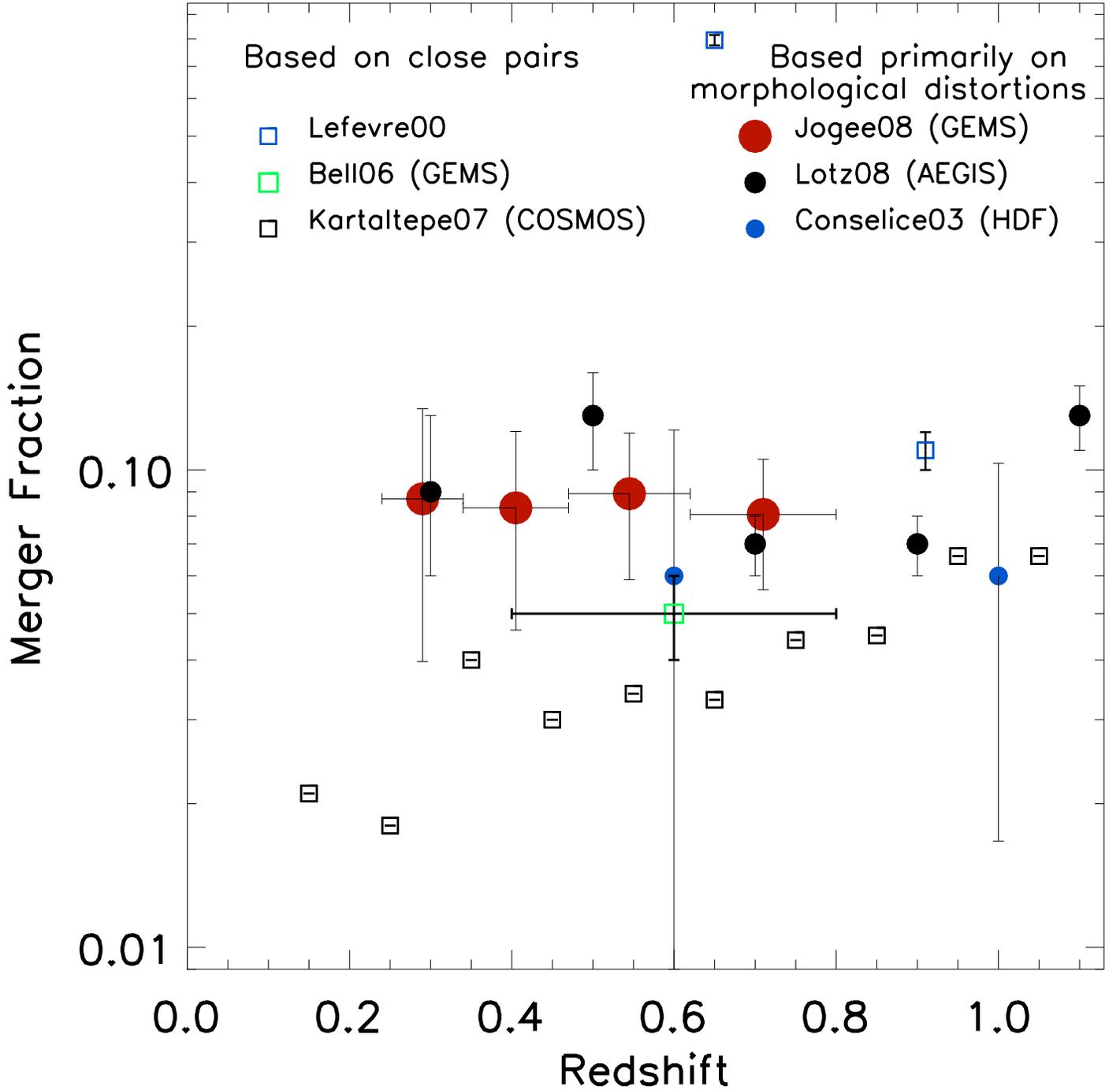}
\caption{
The observed  merger fraction $f$   in the high 
mass ($M \ge$~$2.5 \times  10^{10}$ $M_{\odot}$) sample is compared to other studies, noting 
the caveats outlined in $\S$~\ref{sccompa}. Shown here are  the merger fraction based primarily 
on  morphologically distorted galaxies  (filled circles: this study;  Lotz \etal 2008;  
Conselice 2003), and the close pair fraction (open squares: Le Fevre \etal 2000; Bell \etal  
2006; Kartaltepe \etal 2007) as a function of redshift.  See   $\S$~\ref{sccompa} for details.
\label{fcompa1}}
\end{figure}


 \clearpage
\begin{figure}
\epsscale{1.0}
\plotone{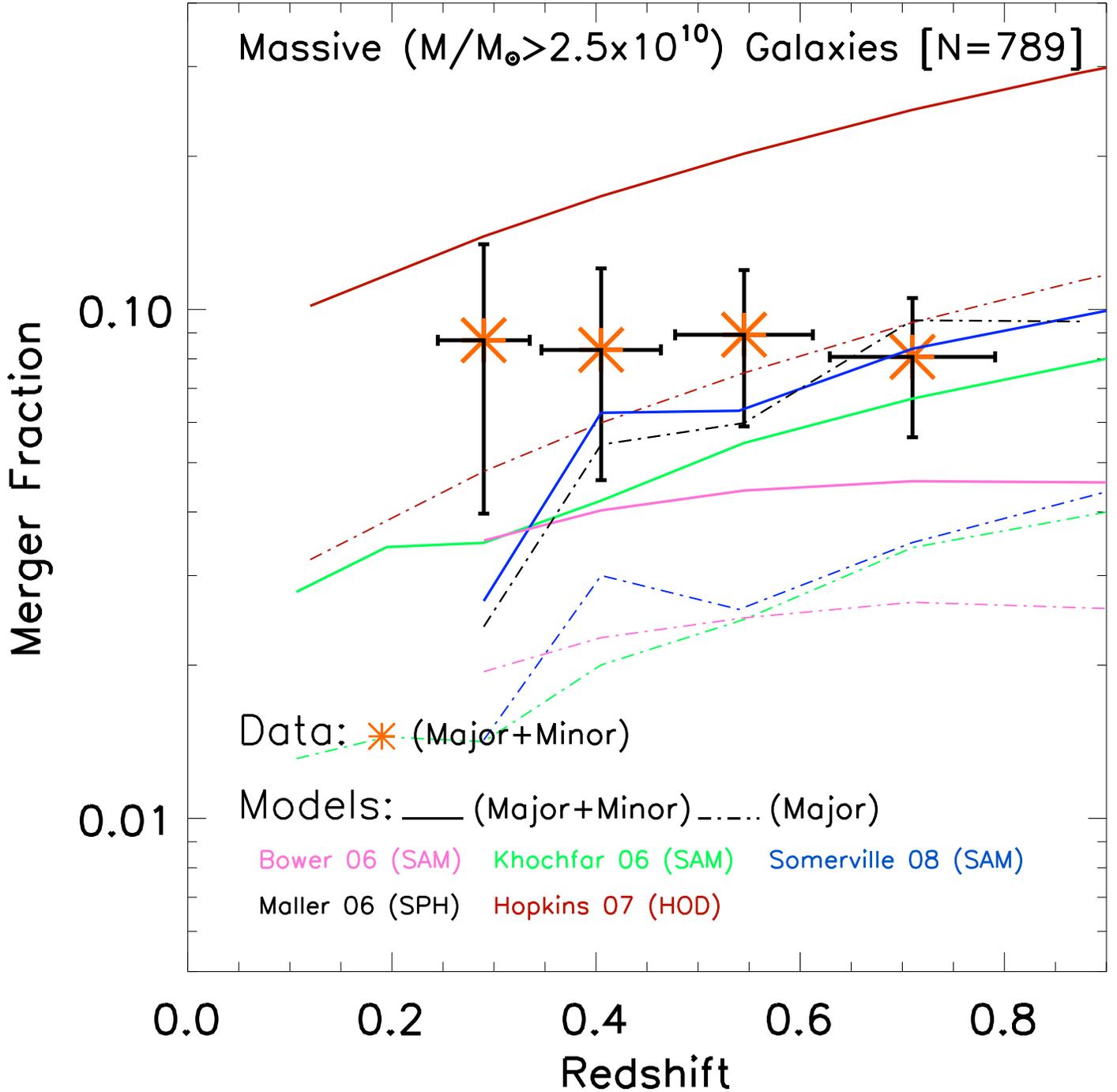}
\caption{
The empirical merger fraction  $f$ (orange stars) for mergers with mass ratio $M1/M2>$~1/10 
among   high mass galaxies   is compared to the fraction of (major+minor)  
mergers (solid lines; stellar mass ratio $M1/M2>$~1/10)  predicted   
by different  $\Lambda$CDM-based simulations  of galaxy evolution,
including the halo occupation distribution (HOD) models of Hopkins \etal (2007); 
semi-analytic models (SAMs) of Somerville \etal (2008), 
Bower et al. (2006), and Khochfar \&  Silk (2006); and smoothed 
particle hydrodynamics (SPH) cosmological simulations from Maller \etal (2006) 
(see $\S$~\ref{scmodel}). 
\label{fmodelf}}
\end{figure}

 \clearpage
\begin{figure}
\epsscale{1.0}
\plotone{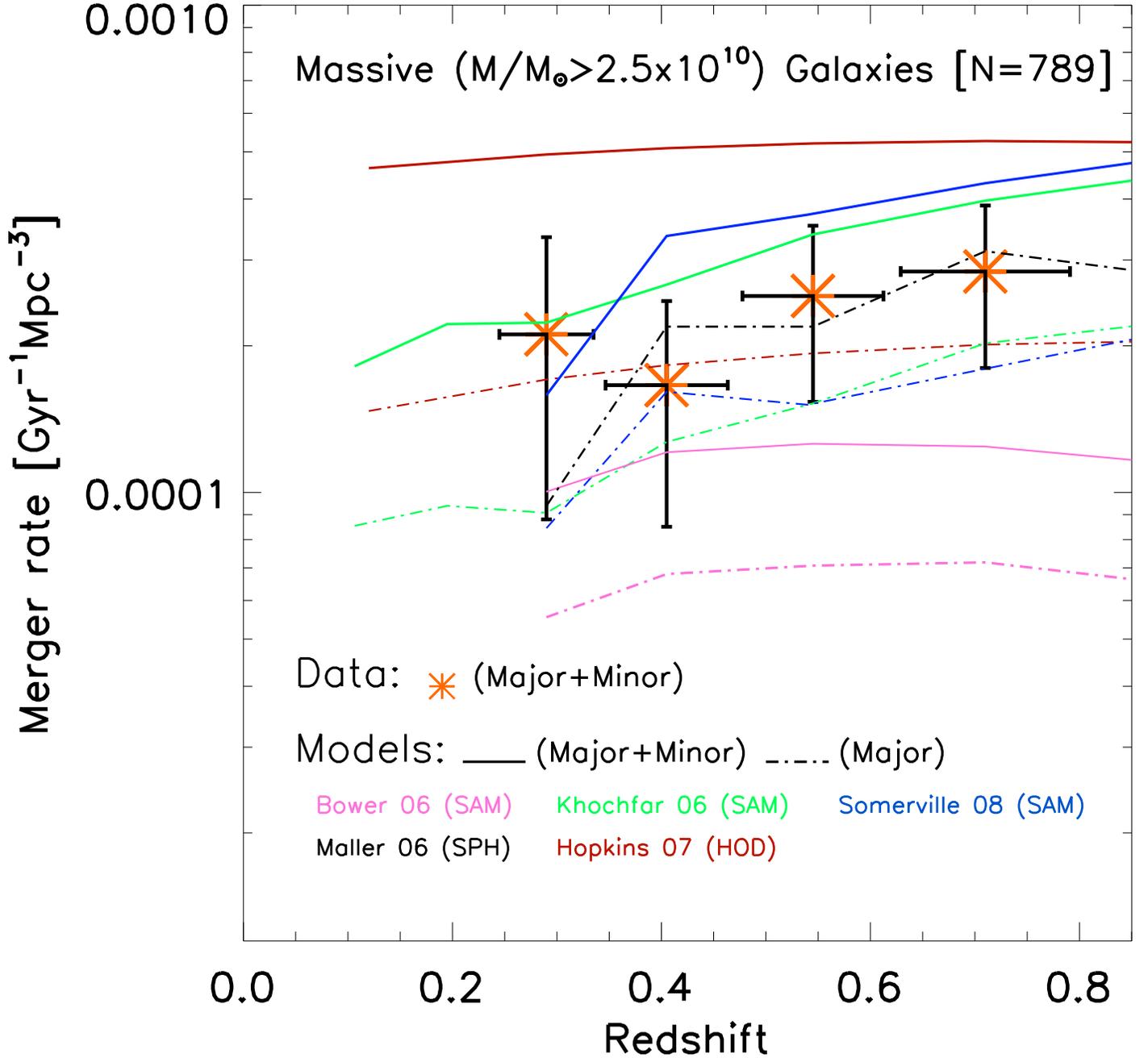}
\caption{
As in Figure 11, but now comparing the  empirical rate  $R$ (orange stars) of 
mergers with mass ratio $M1/M2>$~1/10 among   high mass galaxies  to the  
rate of (major+minor)  mergers (solid lines; stellar mass ratio $M1/M2>$~1/10)  
predicted   by different  $\Lambda$CDM-based simulations  of galaxy evolution.
\label{fmodelr}}
\end{figure}

\clearpage
\begin{figure}
\epsscale{1.0}
\plotone{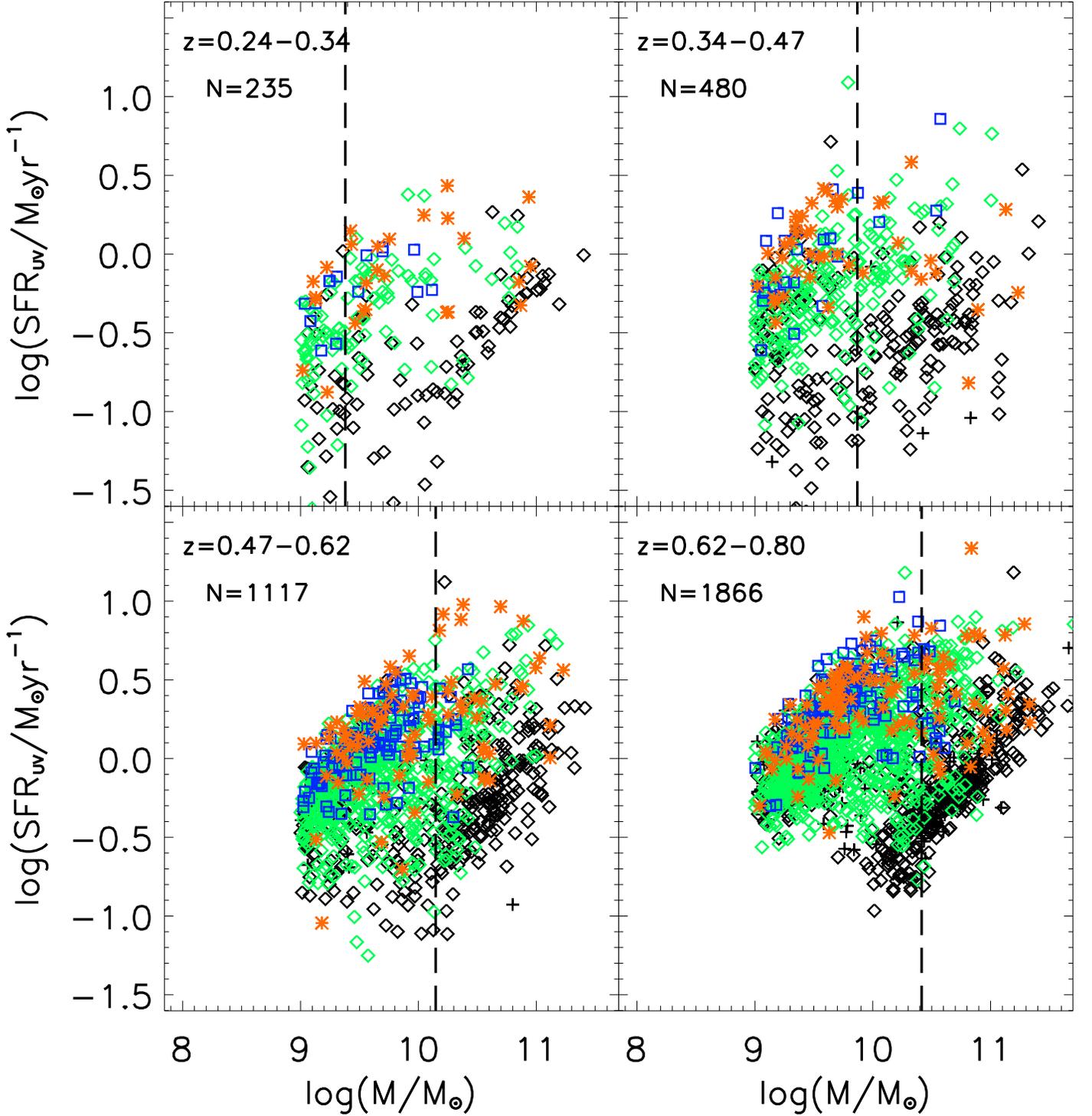}
\caption{
The  UV-based star formation rate  is plotted {\it versus}  the stellar mass 
over $z \sim$~0.24--0.80 
for the sample S2 of galaxies with  $M \ge$~$1 \times  10^{9}$ $M_{\odot}$. 
The four panels show the four redshift bins, which span 1 Gyr each, and 
cumulatively cover  the interval $z \sim$~0.24--0.80 ($T_{\rm back}$~$\sim$~3--7 Gyr). 
$N$ denotes the number of galaxies plotted in each bin.
Galaxies are coded  as in   Fig.~\ref{fuvmas}, with  merging systems 
denoted by orange stars.
The average SFR and total SFR density  in both the UV and the IR, are  
further illustrated  in  Fig.~\ref{favesf} and  Fig.~\ref{fsfhis1}.
\label{fsfrms}}
\end{figure}

\clearpage
\begin{figure}
\epsscale{1.0}
\plotone{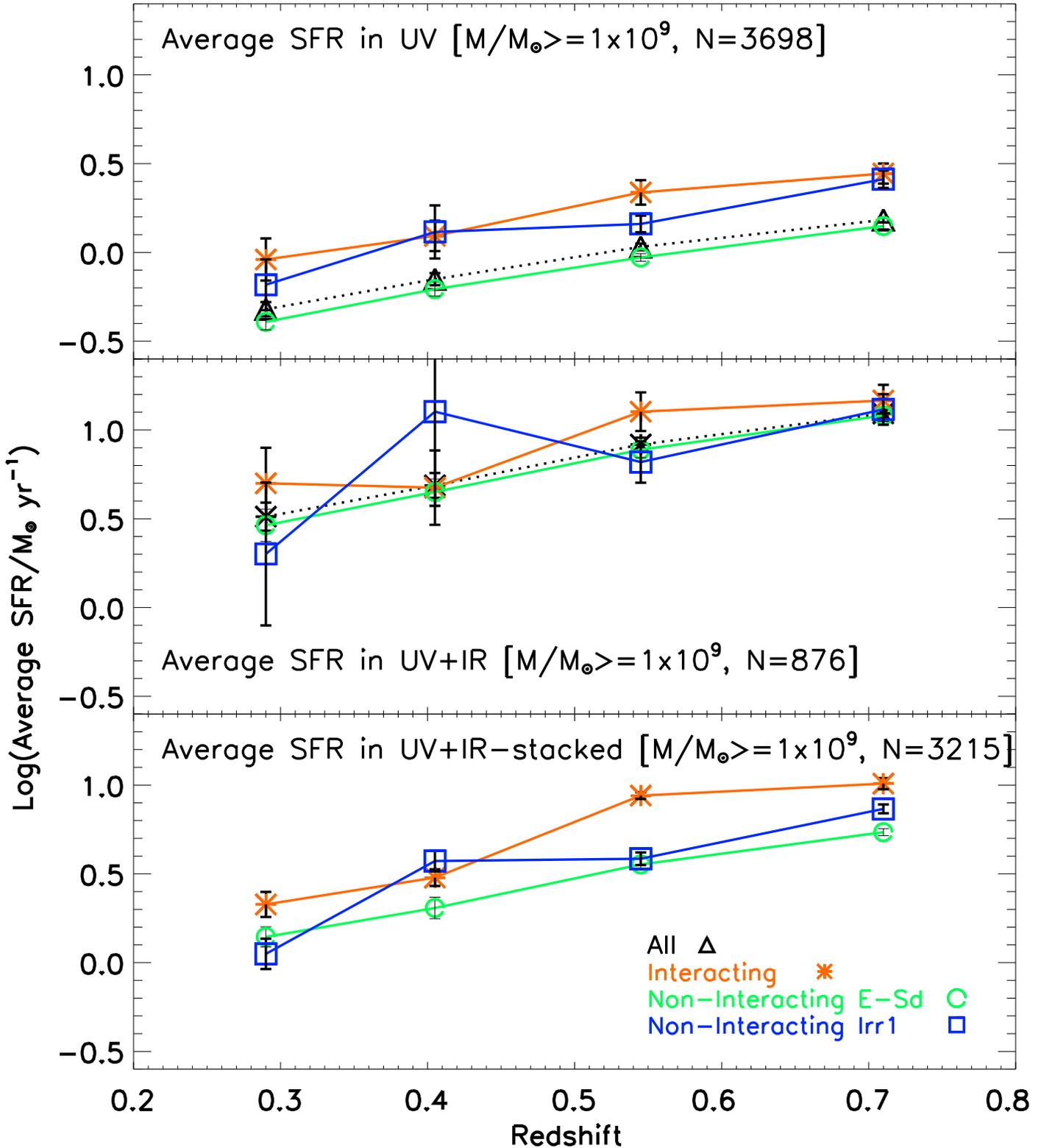}
\caption{
For the sample S2 of  galaxies with  $M \ge$~$1 \times  10^{9}$ $M_{\odot}$,
the average SFR of merging systems, non-interacting  E-Sd galaxies, and  
non-interacting Irr1 galaxies are compared over $z \sim$~0.24--0.80. $N$ 
denotes the number of galaxies used.
The average UV-based SFR (top panel; based on 3698 galaxies),  average UV+IR-based SFR
(middle panel; based on only the 876 galaxies with  24um detections), and  average
UV+IR-stacked SFR  (based on 3215 galaxies with  24um coverage) are shown.
In all there cases, the average SFR  of  visibly merging systems is only modestly
enhanced 
compared to non-interacting galaxies  over $z \sim$~0.24--0.80 (lookback time
$\sim$~3--7 Gyr). See $\S$~\ref{scsfr1} for details.
\label{favesf}}
\end{figure}

\clearpage
\begin{figure}
\epsscale{1.0}
\plotone{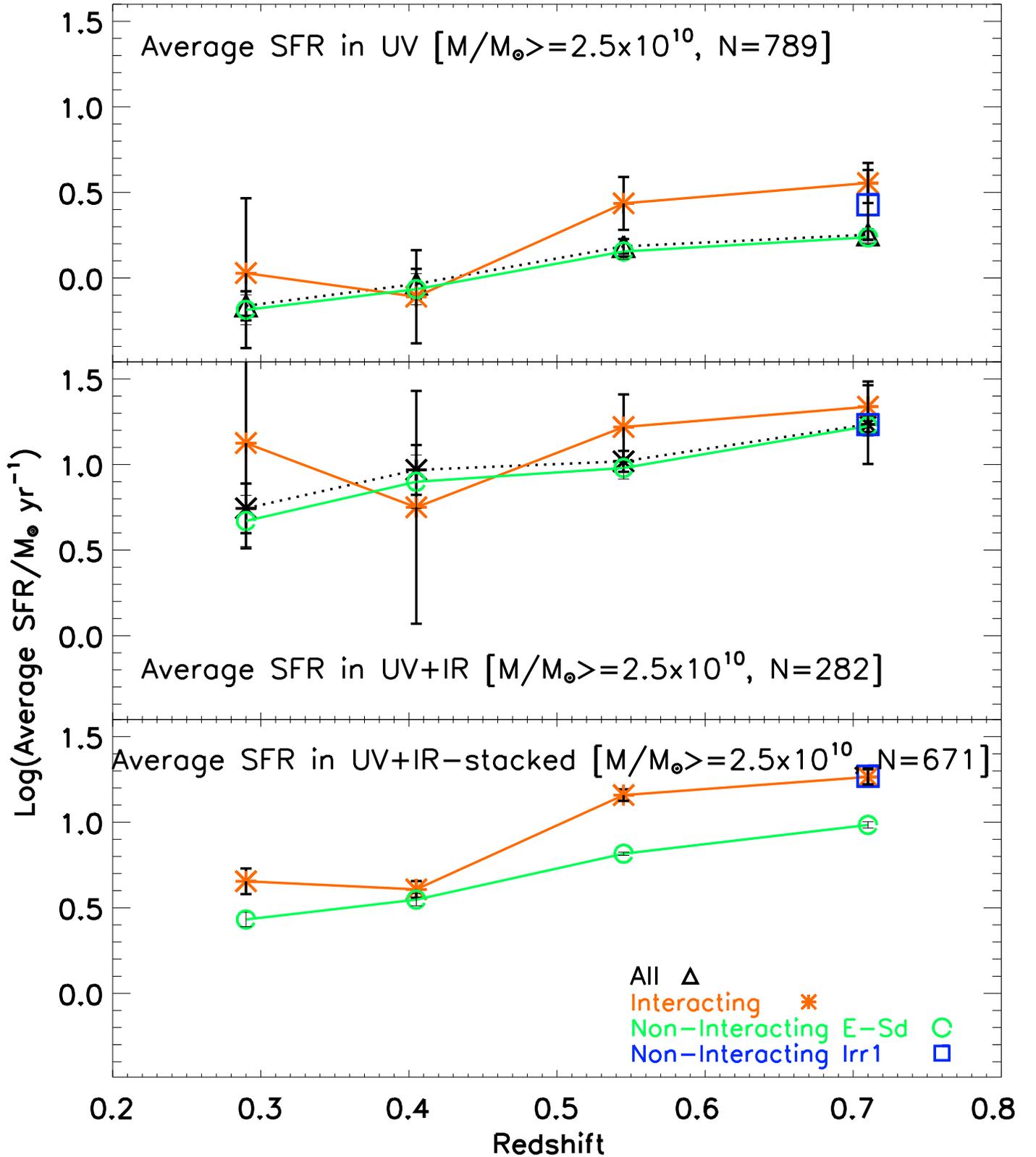}
\caption{
As in Fig.~\ref{favesf}, but for  the sample S1 of high mass ($M \ge$~$2.5 \times 10^{10}$ $M_{\odot}$) 
galaxies.  Only data points with at least 3 galaxies are shown. 
Again, the average SFR  of visibly merging galaxies is only modestly
enhanced compared to non-interacting galaxies  over $z \sim$~0.24--0.80 (lookback time
$\sim$~3--7 Gyr).
\label{favesf2}}
\end{figure}

\clearpage
\begin{figure}
\epsscale{1.0}
\plotone{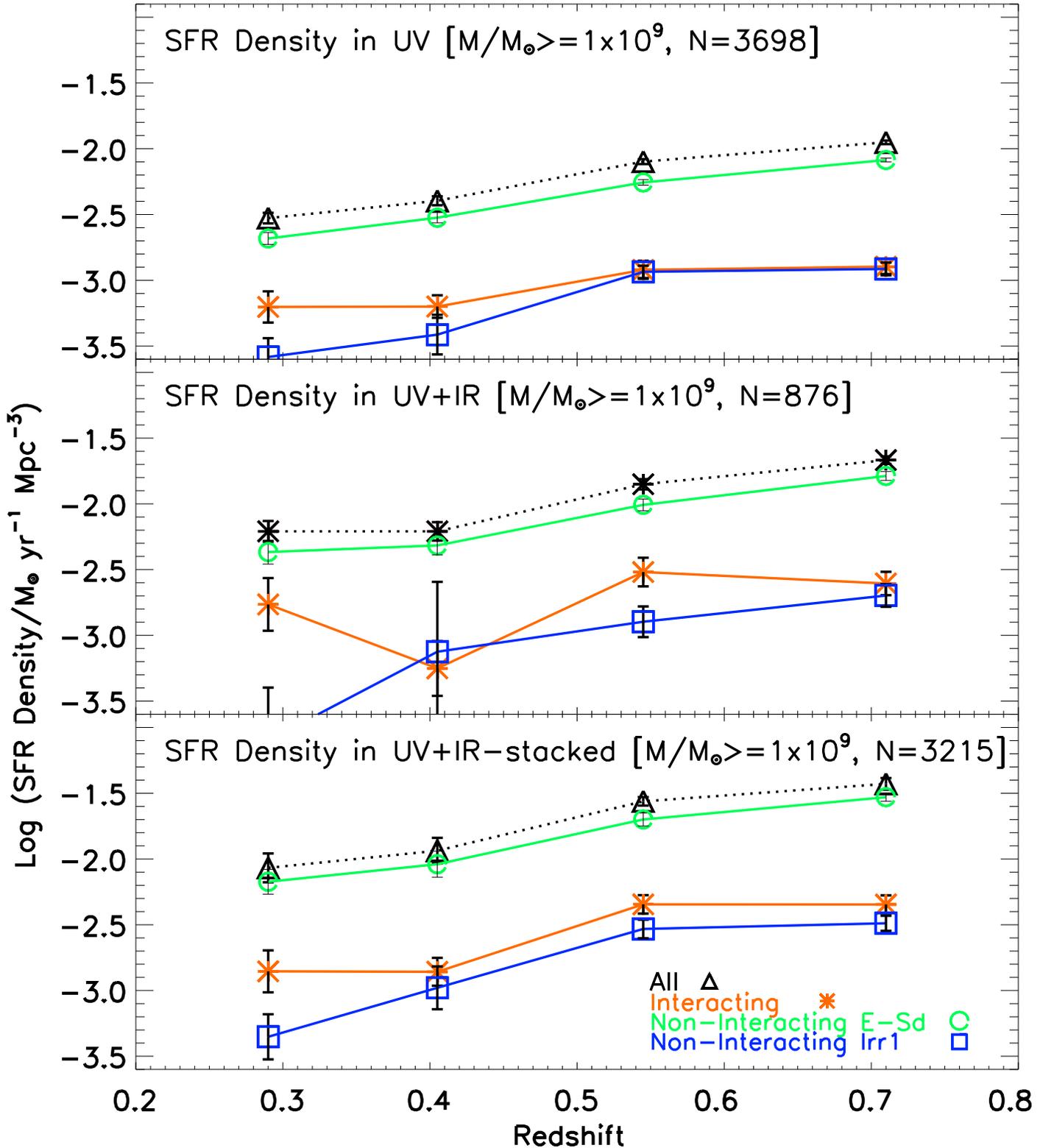}
\caption{
For the sample S2 of galaxies with  $M \ge$~$1 \times  10^{9}$ $M_{\odot}$,
the SFR density of  merging systems, non-interacting  E-Sd galaxies, and  
non-interacting Irr galaxies are compared over $z \sim$~0.24--0.80. 
Results based on UV (top panel), UV+IR (middle panel), as well as 
UV+stacked-IR data (bottom panel), are shown in the top, middle, and bottom panels.
In all bins, visibly merging  systems only contribute a small 
fraction (typically below 30\%) of the total SFR density.
In effect, the behavior  of the cosmic SFR density over the
last 7 Gyr is  predominantly shaped by non-interacting E-Sd  galaxies rather 
than visibly merging galaxies.
\label{fsfhis1}}
\end{figure}

\clearpage
\begin{figure}
\epsscale{1.0}
\plotone{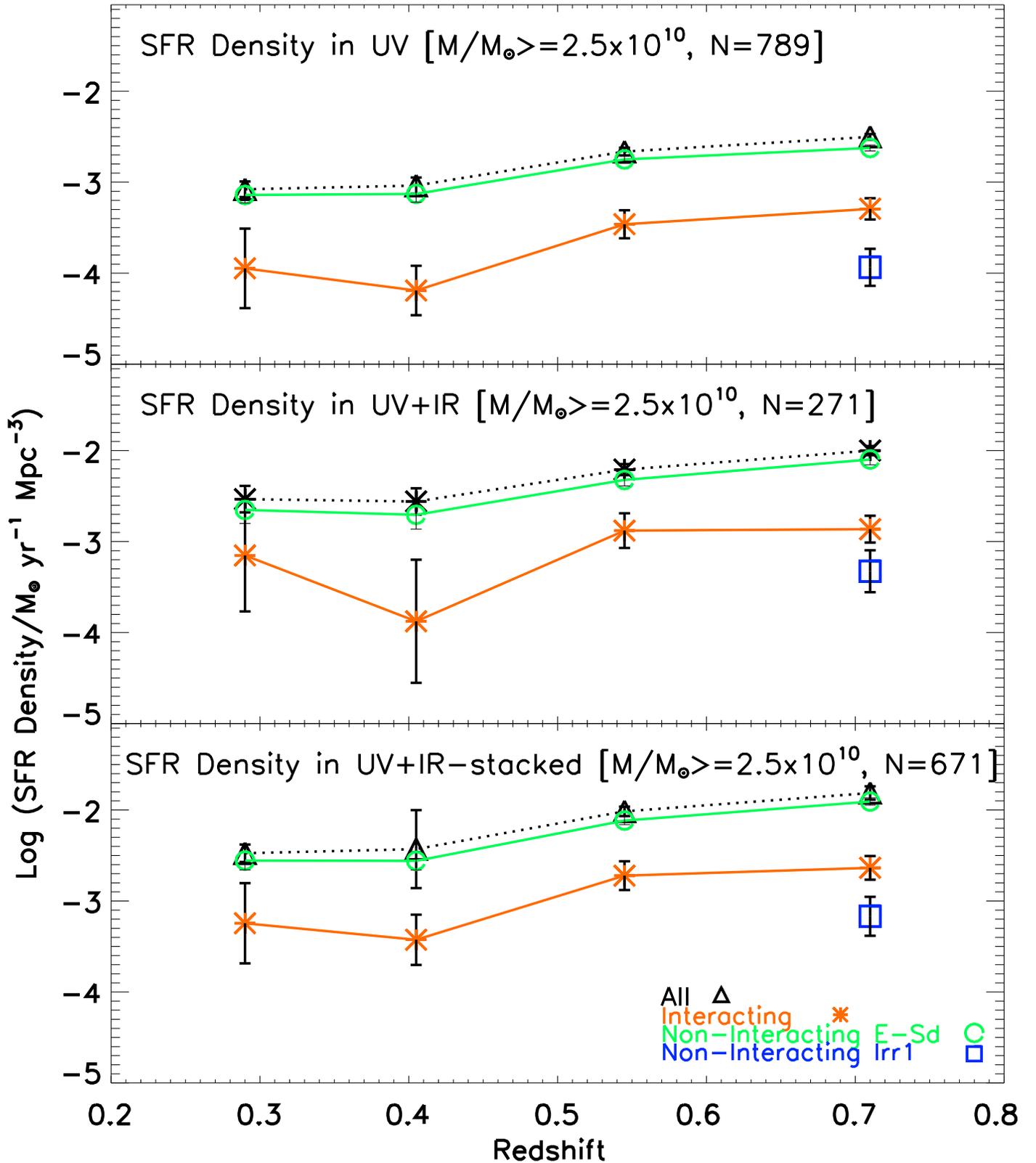}
\caption{
As in Fig.~\ref{fsfhis1}, but for  the sample of high mass ($M \ge$~$2.5 \times 10^{10}$ $M_{\odot}$) 
galaxies.   Only data points with at least 3 galaxies are shown. 
The same conclusion holds: the cosmic SFR density over the last 7 Gyr is  predominantly shaped 
by  non-interacting E-Sd  galaxies rather than  visibly merging galaxies.
\label{fsfhis2}}
\end{figure}

\clearpage
\begin{figure}
\epsscale{1.0}
\plotone{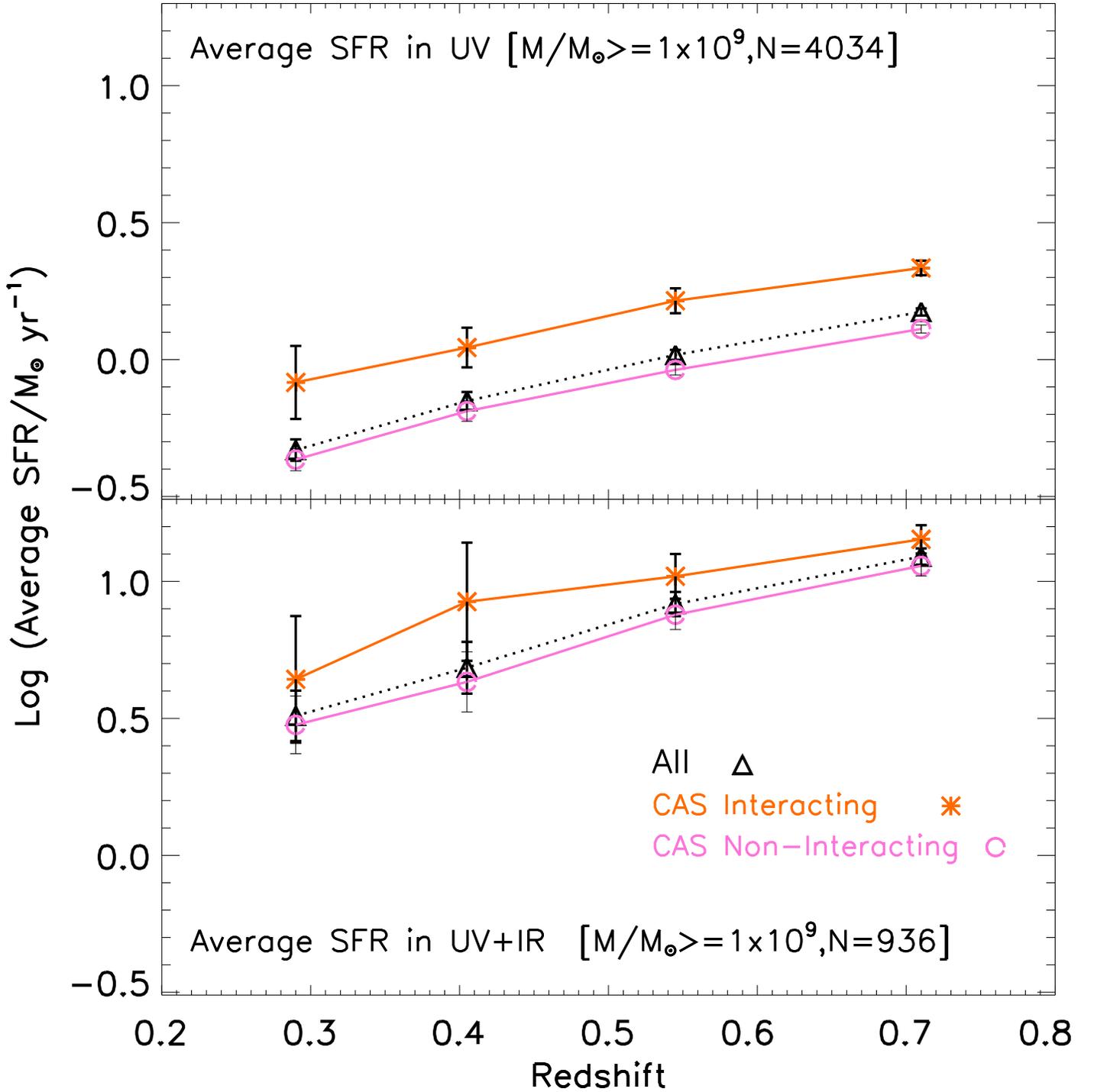}
\caption{
Same as in   Fig.~\ref{favesf}, but using the CAS  criterion 
($A >$~0.35 and $A > S$) to identify  interacting systems  in the sample of 
intermediate mass ($M \ge$~$1.0 \times 10^{9}$ $M_{\odot}$) galaxies.
The average SFR of `CAS-interacting' galaxies  
is only  modestly enhanced compared to `CAS non-interacting' galaxies,
in agreement with the results from $\S$~\ref{scsfr1}.
\label{fcasf1}}
\end{figure}

\clearpage
\begin{figure}
\epsscale{1.0}
\plotone{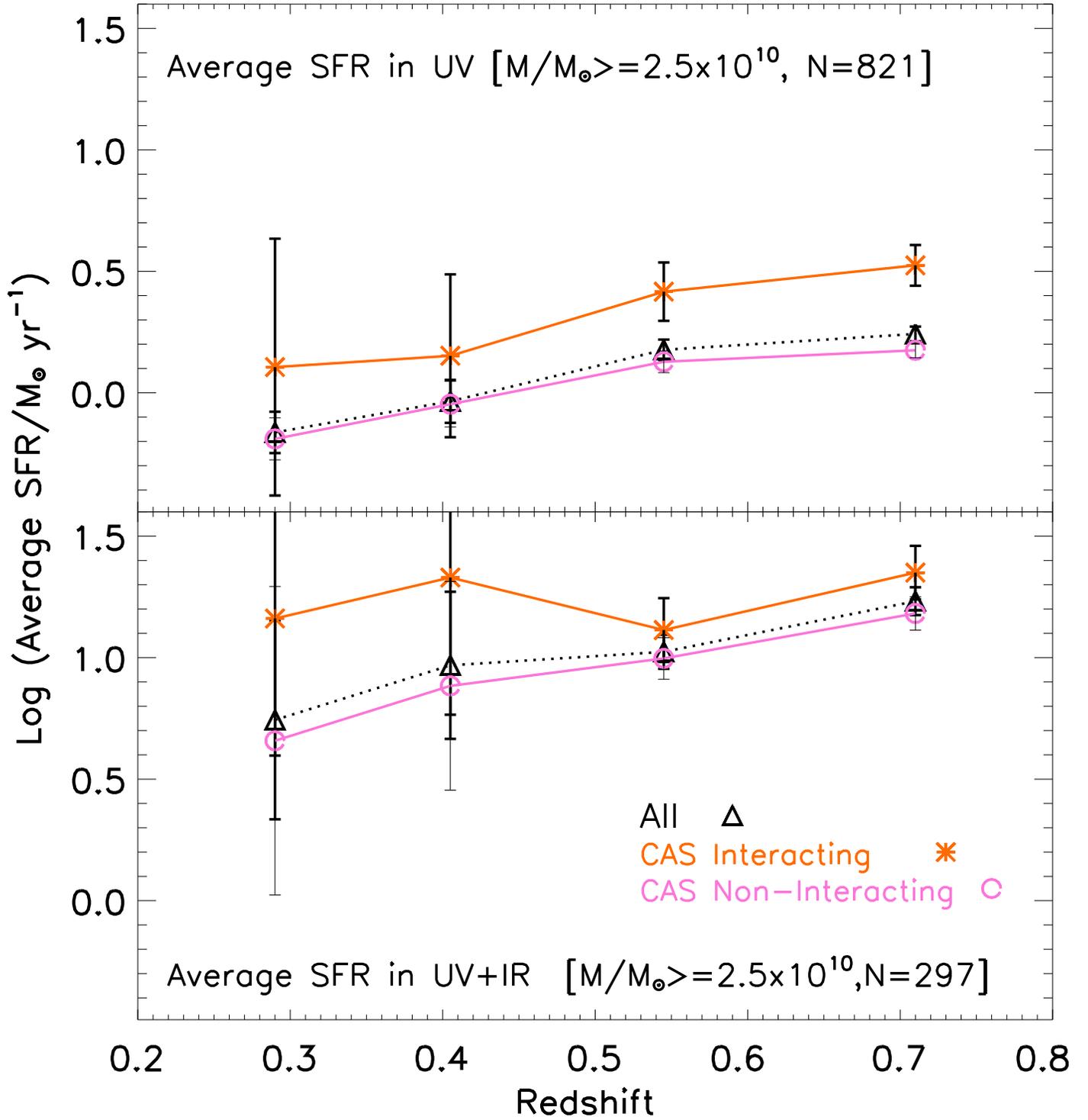}
\caption{
As in Fig.~\ref{fcasf1}, but for  the sample of high mass ($M \ge$~$2.5 \times 10^{10}$ $M_{\odot}$) 
galaxies.  The same conclusion holds: the average SFR of `CAS-interacting' galaxies  
is only  modestly enhanced compared to `CAS non-interacting' galaxies.
\label{fcasf2}}
\end{figure}

\clearpage
\begin{figure}
\epsscale{1.0}
\plotone{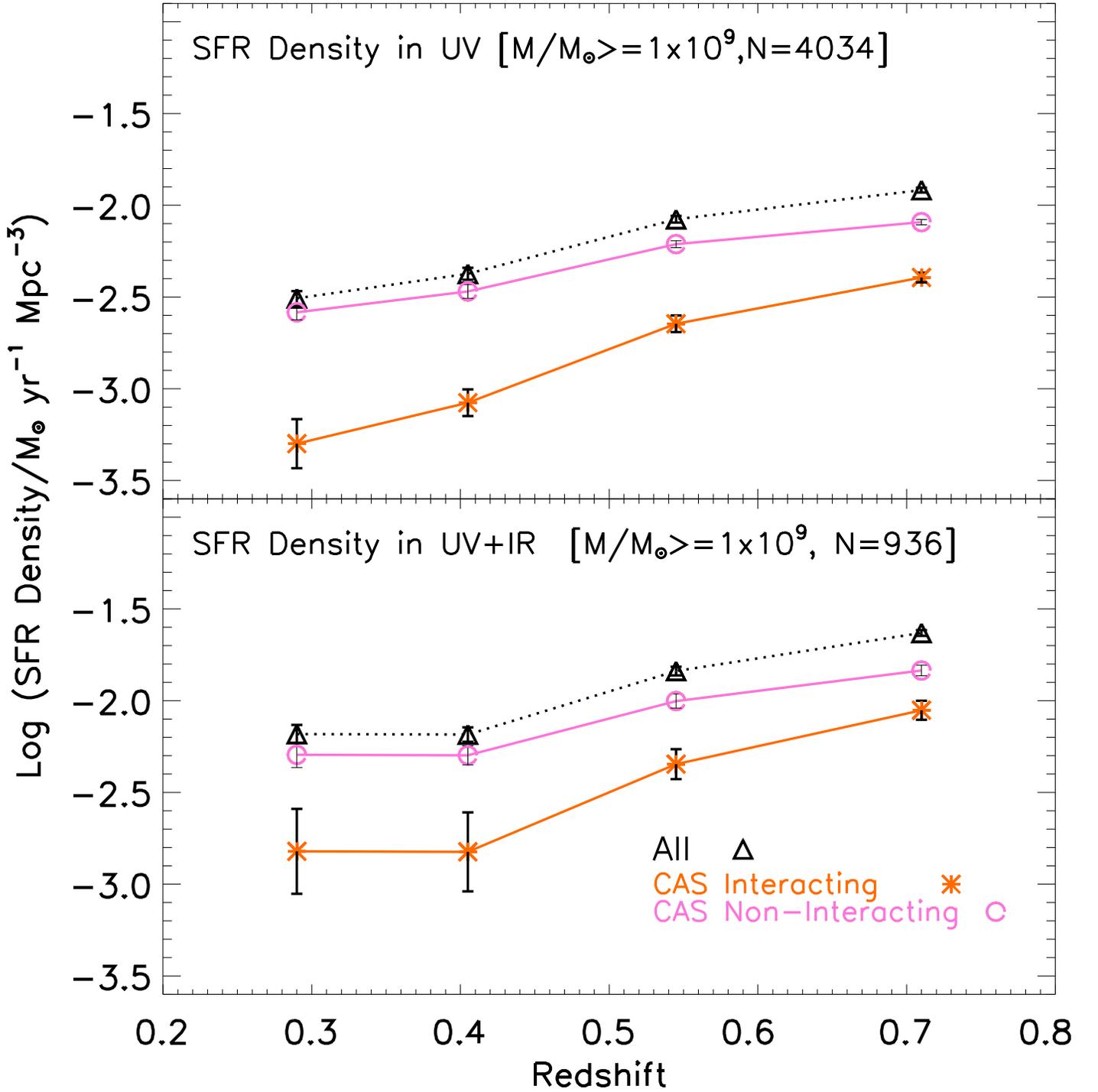}
\caption{
Same as in   Fig.~\ref{fsfhis1}, but using the CAS merger criterion 
($A >$~0.35 and $A > S$) to identify  interacting galaxies in the sample of 
intermediate mass ($M \ge$~$1.0 \times 10^{9}$ $M_{\odot}$) galaxies. 
The `CAS-interacting' galaxies  contribute only 16\% to 33\% of the UV SFR density and 
22\% to 38\% of the UV+IR SFR density.  While the upper limits of these 
values are slightly  higher than those based on the visual types 
(Fig.~\ref{fsfhis1}), the  `CAS non-interacting' galaxies' clearly dominate the
SFR density.
\label{fcasf3}}
\end{figure}

\begin{figure}
\epsscale{1.0}
\plotone{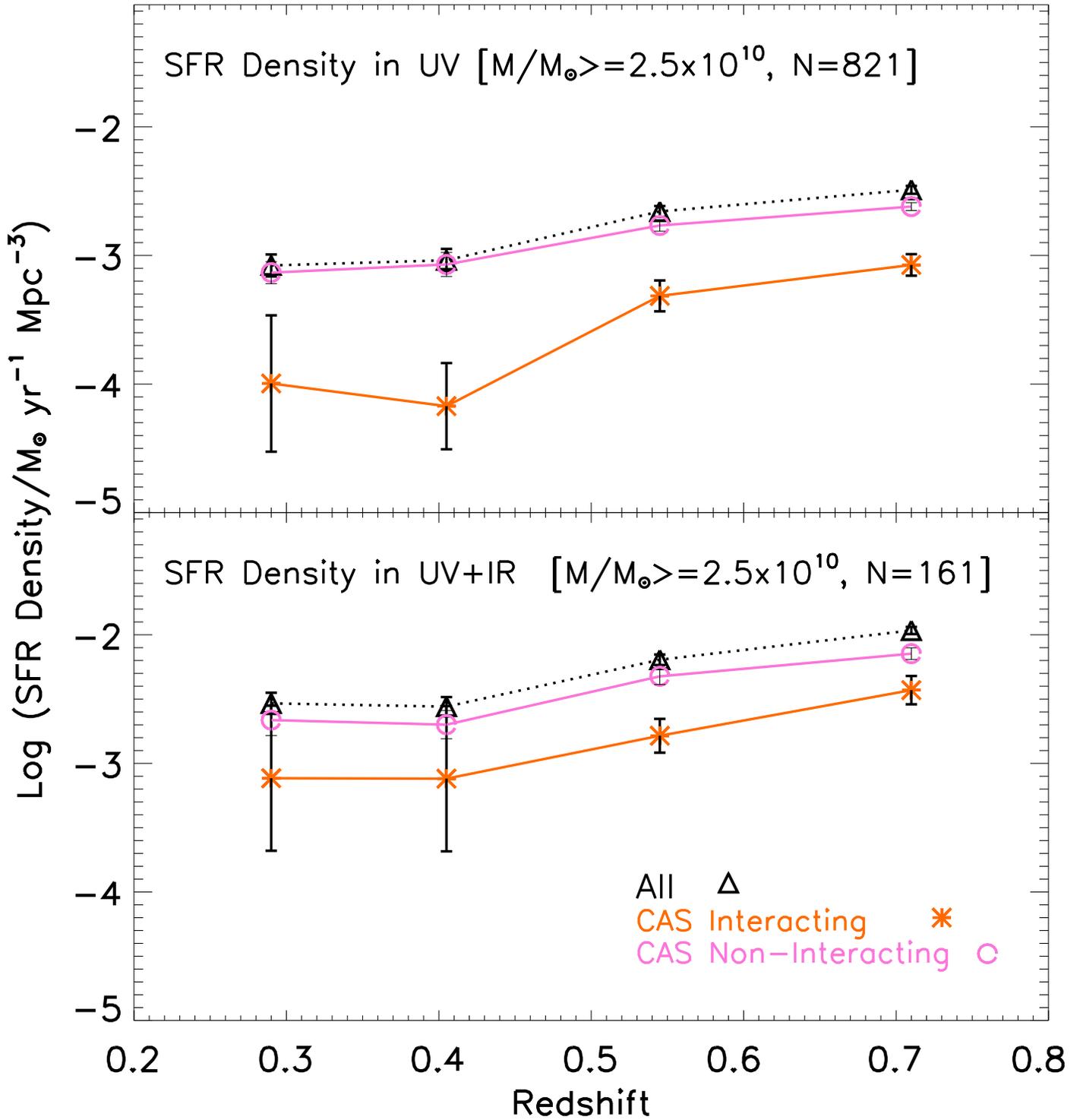}
\caption{
As in Fig.~\ref{fcasf3}, but for  the sample of high mass ($M \ge$~$2.5 \times 10^{10}$ $M_{\odot}$) 
galaxies.  The same conclusion holds.
\label{fcasf4}}
\end{figure}



\clearpage
\begin{table}
\begin{center}
\caption{Visual Merger Fraction for High Mass ($M \ge$~$2.5 \times  10^{10}$ $M_{\odot}$) 
Sample S1 [N=789]}
\begin{tabular}{cccccc}
\tableline
     &    &   &   &   &  \\   
(1)  & Redshift bin                              &  1               &  2                &     3            & 4            \\
(2)  & Redshift range                            &  0.24--0.34      &  0.34--0.47       &  0.47-0.62       & 0.62--0.80   \\
(3)  & Lookback time [Gyr]                       &  3.0--4.0        &  4.0--5.0         &  5.0--.6-0       & 6.0--7.0     \\
(4)  & $\lambda$$_{\rm rest}$ in F606W  [\AA]    & 4470--4414       & 4414--4023        &  4023--3651      & 3651--3286   \\
(5)  & Total no of  galaxies                     &   46             &   84              &  213             &   446 \\
(6) & Fraction  of mergers                & 0.087$\pm$0.047  &  0.083$\pm$0.037  & 0.089$\pm$0.030  & 0.0807$\pm$0.025 \\
(6a) & Lower limit on major merger fraction      & 0.022$\pm$0.021  &  0.035$\pm$0.022  & 0.014$\pm$0.009  & 0.011$\pm$0.006 \\
(6b) & Lower limit on minor merger fraction      & 0.065$\pm$0.040  &  0.036$\pm$0.022  & 0.075$\pm$0.027  & 0.049$\pm$0.016  \\
(6c) & Fraction of ambiguous minor/major mergers  & 0.00             &  0.012$\pm0.012$  &  0.00            & 0.020$\pm$0.008 \\
(7)  & Fraction of non-interacting E-Sd          & 0.913$\pm$ 0.241 & 0.869$\pm$0.229   & 0.878$\pm$0.229  & 0.785$\pm$0.205 \\
(8)  & Fraction of non-interacting Irr1          & 0.000            & 0.024$\pm$0.018   & 0.009$\pm$0.007  & 0.025$\pm$0.010  \\
(9)  & Fraction of compact                       &  0.000            & 0.024$\pm$0.018   & 0.023$\pm$0.012  &  0.11$\pm$0.032 \\
\tableline
\end{tabular}
\tablecomments{ 
Rows are~:
(1) Redshift bin. (2) Range in redshift covered by the bin; 
(3) Range in lookback time covered by the bin;
(4) Range in rest-frame wavelength traced by the F606W filter over the bin, assuming a pivot wavelength of 5915\AA; 
(5)  Total number of high mass galaxies per bin;
(6) Fraction of systems with evidence of a recent merger of mass ratio $>$~1/10.
    These include both major ($M1/M2 \ge 1/4$) and minor  (1/10$<M1/M2\le$~1/4) mergers;
(6a)  Lower limit on the fraction of galaxies undergoing  major mergers; 
(6b)  Lower limit on the fraction of galaxies undergoing  minor mergers; 
(6c)  Remaining fraction of galaxies that could be either major or minor mergers; 
(7-9) Fraction of  non-interacting E-Sd, non-interacting Irr1, and  compact systems.}
\label{tvclas1}
\end{center}
\end{table}

\clearpage
\begin{table}
\begin{center}
\caption{Visual Merger Fraction for $M \ge$~$1.0 \times  10^{9}$ $M_{\odot}$ Sample S2 [N=3698]}
\begin{tabular}{cccccccc}
\tableline
    &    &   &   &   &  &    &   \\
(1)  & Redshift bin                              &  1               &  2                &     3            & 4            \\
(2)  & Redshift range                            &  0.24--0.34      &  0.34--0.47       &  0.47-0.62       & 0.62--0.80   \\
(3)  & Lookback time [Gyr]                       &  3.0--4.0        &  4.0--5.0         &  5.0--.6-0       & 6.0--7.0     \\
(4)  & $\lambda$$_{\rm rest}$ in F606W  [\AA]    & 4470--4414       & 4414--4023        &  4023--3651      & 3651--3286   \\
\tableline
      &                       &            &         &           &        \\
\multicolumn{6}{c}{All [N=3698]} \\
(5) & Total no of  galaxies                      &     235          &    480           &    1117          &    1866\\
(6) & Fraction of mergers                     & 0.111$\pm$0.035  & 0.090$\pm$0.027  & 0.074$\pm$0.021  & 0.062$\pm$0.017 \\
(6a) & Lower limit on major merger fraction      & 0.013$\pm$0.008  & 0.006$\pm$0.004  & 0.004$\pm$0.002  & 0.009$\pm$0.003 \\
(6b) & Lower limit on minor merger fraction      & 0.038$\pm$0.016  & 0.021$\pm$0.008  & 0.021$\pm$0.007  & 0.017$\pm$0.005  \\
(6c) & Fraction of ambiguous minor/major mergers  & 0.060$\pm$0.022  & 0.062$\pm$0.020  & 0.048$\pm$0.014  & 0.036$\pm$0.010 \\
(7) & Fraction of non-interacting E-Sd           & 0.850$\pm$0.220  & 0.846$\pm$0.220  & 0.796$\pm$0.207  & 0.793$\pm$0.206 \\
(8) & Fraction of non-interacting Irr1           & 0.064$\pm$0.023  & 0.052$\pm$0.017  & 0.108$\pm$0.030  & 0.064$\pm$0.018 \\
(9) & Fraction of compact                        & 0.000            & 0.012$\pm$0.006  & 0.021$\pm$0.007  & 0.080$\pm$0.022 \\
\tableline
      &                       &            &         &           &        \\
\multicolumn{6}{c}{Blue Cloud [N=2844] } \\
(10) & Total no of  galaxies                     &    154          &    332            &    876          &    1482\\
(11) & Fraction of mergers                   & 0.149$\pm$0.048  & 0.114$\pm$0.034  & 0.088$\pm$0.025  & 0.069$\pm$0.019 \\
(11a) & Lower limit on major merger fraction     & 0.013$\pm$0.009  & 0.00             & 0.005$\pm$0.003  & 0.008$\pm$0.003 \\
(11b) & Lower limit on minor merger fraction     & 0.046$\pm$0.020  & 0.024$\pm$0.010  & 0.023$\pm$0.008  & 0.016$\pm$0.005  \\
(11c) & Fraction of ambiguous minor/major mergers & 0.091$\pm$0.033  & 0.090$\pm$0.028  & 0.060$\pm$0.018  & 0.046$\pm$0.013 \\
(12)  & Fraction of non-interacting E-Sd         & 0.753$\pm$ 0.199 & 0.801$\pm$0.209  & 0.753$\pm$0.196  & 0.784$\pm$0.204 \\
(13)  & Fraction of non-interacting Irr1         & 0.097$\pm$0.035  & 0.075$\pm$0.024  & 0.136$\pm$0.037  & 0.080$\pm$0.022 \\
(14)  & Fraction of compact                      &  0.000           & 0.009$\pm$0.006  & 0.023$\pm$0.008  &  0.067$\pm$0.018 \\
\tableline
\end{tabular}
\tablecomments{ Rows are~:
(1) to (9): As in Table 1, but for intermediate mass ($M \ge$~$1.0 \times  10^{9}$ $M_{\odot}$) galaxies. 
However, note that the intermediate mass sample is incomplete for the red sequence.
(10) to (14);  Ditto, but for the blue cloud, where the sample is complete.
}
\label{tvclas2}
\end{center}
\end{table}

\clearpage
\begin{table}
\begin{center}
\caption{Merger fraction in GEMS F606W ($V$) and GOODS F850LP ($z$) [$N$=855]}
\begin{tabular}{llccccc}
\tableline\tableline
   &   &           &            &          &           &            \\    
   &   & GEMS $V$  & GOODS $z$ & GOODS $z$ & GOODS $z$ & GOODS $z$  \\
   &   & Average   & Average   &  SJ       &  SM       & KP         \\
\tableline
    &                                    &                    &                  &                 &                  &                \\   
(1) & Fraction $f$  of merging galaxies  &  0.046$\pm$0.007   & 0.049$\pm$0.007  & 0.051$\pm$0.007 & 0.057$\pm$0.008  & 0.038$\pm$0.006 \\
(2) & Ratio of $f$ in GOODS $z$ to GEMS $V$  &       -            & 1.06$\pm$0.22    & 1.10 $\pm$0.23  & 1.20$\pm$0.25    & 0.83 $\pm$0.18 \\ 
\tableline
\end{tabular}
\tablecomments{ 
As a test for bandpass shift and surface brightness dimming, the table shows a comparison of the 
fraction of visual mergers based on  GEMS F606W images and deeper redder GOODS F850LP images. 
The sample consists of the 855  intermediate mass ($M \ge$~$1.0 \times  10^{9}$ $M_{\odot}$) 
galaxies at $z\sim$0.24 to 080, which are 
common to both GEMS F606W and GOODS F850LP surveys. Columns are: 
(2) Fraction  $f$  of mergers in GEMS F606W. The error bar only includes the 
binomial term [$f$(1-$f$)/$N$]$^{1/2}$;
(3) Average fraction  $f$  of mergers in  GOODS F850LP based on results 
by  classifiers (SJ,SM,KP), shown in columns 4--6.
}
\label{tgoods1}
\end{center}
\end{table}

\end{document}